\documentclass[iop,revtex4]{emulateapj}
\usepackage{graphics,amsmath}

\newcommand{\sdss}{{\small {SDSS}}}
\newcommand{\sdssi}{{\small {SDSS-I}}}
\newcommand{\sdssii}{{\small {SDSS-II}}}
\newcommand{\sdssiii}{{\small {SDSS-III}}}

\newcommand{\snls}{{\small {SNLS}}}
\newcommand{\essence}{{\small {ESSENCE}}}
\newcommand{\boss}{{\small {BOSS}}}
\newcommand{\sn}{{\small {SN}}}
\newcommand{\ccsn}{{\small {CC~SN}}}

\newcommand{\snia}{{\small {SN~I}}a}
\newcommand{\snibc}{{\small {SN~I}}b/c}
\newcommand{\snii}{{\small {SN~II}}}

\newcommand{\sed}{{\small {SED}}}
\newcommand{\dmB}{$\Delta m_{15}(B)$}

\newcommand{\apo}{{\small {APO}}}

\newcommand{\mjd}{{\small {MJD}}}

\newcommand{\hst}{{\it {\small {HST}}}}

\newcommand{\mdm}{{\small {MDM}}}
\newcommand{\wht}{{\small {WHT}}}
\newcommand{\het}{{\small {HET}}}
\newcommand{\eso}{{\small {ESO}}}
\newcommand{\ntt}{{\small {NTT}}}
\newcommand{\nnot}{{\small {NOT}}}
\newcommand{\tng}{{\small {TNG}}}
\newcommand{\mlcs}{{\small {MLCS}}2k2}
\newcommand{\salt}{{\small {SALT}}}
\newcommand{\saltii}{{\small {SALT2}}}

\newcommand{\signoi}{{\small {S/N}}}

\newcommand{\psf}{{\small {PSF}}}

\newcommand{\id}{{\small {ID}}}

\newcommand{\ab}{{\small {AB}}}
\newcommand{\pt}{{\small {PT}}}
\newcommand{\dr}{{\small {DR}}}
\newcommand{\sfr}{{\small {SFR}}}
\newcommand{\ssfr}{s{\small {SFR}}}

\newcommand\pia{$P_{\mathrm{Ia}}$}
\newcommand\pibc{$P_{\mathrm{Ibc}}$}
\newcommand\pii{$P_{\mathrm{II}}$}

\newcommand\chir{$\chi^2_{r}$}

\newcommand\pfit{$P_{\mathrm{fit}}$}
\newcommand\pnnia{$P_{\mathrm{NN,Ia}}$}
\newcommand\pnnibc{$P_{\mathrm{NN,Ibc}}$}
\newcommand\pnnii{$P_{\mathrm{NN,II}}$}
\newcommand\trest{$T_{\mathrm{rest}}$}
\newcommand\av{$A_V$}
\newcommand\tmax{$T_{\mathrm{max}}$}
\newcommand\zspec{$z_{\mathrm{spec}}$}

\newcommand\zhost{$z_{\mathrm{host}}$}

\newcommand\psnid{{\small{PSNID}}}
\newcommand\psnidnn{{\small{PSNID/NN}}}
\newcommand\psnia{p{\small {SN~I}}a}
\newcommand\zsnia{z{\small {SN~I}}a}
\newcommand\psnibc{p{\small {SN~I}}bc}
\newcommand\zsnibc{z{\small {SN~I}}bc}
\newcommand\psnii{p{\small {SN~II}}}
\newcommand\zsnii{z{\small {SN~II}}}

\newcommand\snana{{\small{SNANA}}}

\newcommand\smp{{\small{SMP}}}

\newcommand\iau{{\small IAU}}
\newcommand\des{{\small DES}}
\newcommand\lsst{{\small LSST}}
\newcommand\panstarrs{Pan-{\small STARRS}}

\newcommand{\dlr}{{\small {DLR}}}
\newcommand{\ddlr}{$d_{\mbox{\scriptsize DLR}}$}
\newcommand\fsps{{\small {FSPS}}}
\newcommand\PEGASE{{\small {P\'EGASE.2}}}

\newcommand\ncand{10,258} 
\newcommand\nsncand{4607} 

\newcommand\nspec{1360} 
\newcommand\nsnspec{889} 

\newcommand\nunknown{1584}   
\newcommand\nvariable{3225}  
\newcommand\nagn{906}   
\newcommand\nslsn{3}
\newcommand\nssnia{500}
\newcommand\nssniap{36}
\newcommand\nssnibc{19}  
\newcommand\nssnib{8}
\newcommand\nssnic{11}   
\newcommand\nssnii{62}
\newcommand\nsccsn{81}   
\newcommand\npsnia{677}  
\newcommand\nzsnia{907}  
\newcommand\npsnibc{24}
\newcommand\nzsnibc{62}
\newcommand\npsnii{1841} 
\newcommand\nzsnii{411}  

\newcommand\nsnzfixed{1443} 
\newcommand\nsnzfit{677}    

\newcommand\snanaversion{version 10.31b}

\newcommand{\hostless}{hostless}

\begin{document}

\title{The Data Release of the Sloan Digital Sky Survey-II Supernova Survey}

\author{
Masao~Sako\altaffilmark{1},
Bruce~Bassett\altaffilmark{2,3,4},
Andrew~C.~Becker\altaffilmark{5},
Peter~J.~Brown\altaffilmark{6},
Heather~Campbell\altaffilmark{7,8},
Rachel~Cane\altaffilmark{1},
David~Cinabro\altaffilmark{9},
Chris~B.~D'Andrea\altaffilmark{7},
Kyle~S.~Dawson\altaffilmark{10},
Fritz~DeJongh\altaffilmark{11},
Darren~L.~Depoy\altaffilmark{6},
Ben~Dilday\altaffilmark{12},
Mamoru~Doi\altaffilmark{13,14},
Alexei~V.~Filippenko\altaffilmark{15},
John~A.~Fischer\altaffilmark{1},
Ryan~J.~Foley\altaffilmark{16,17},
Joshua~A.~Frieman\altaffilmark{11,18,19},
Lluis~Galbany\altaffilmark{20,21},
Peter~M.~Garnavich\altaffilmark{22},
Ariel~Goobar\altaffilmark{23,24},
Ravi~R.~Gupta\altaffilmark{1,25},
Gary~J.~Hill\altaffilmark{26},
Brian~T.~Hayden\altaffilmark{22,27},
Ren\'ee~Hlozek\altaffilmark{28},
Jon~A.~Holtzman\altaffilmark{29},
Ulrich~Hopp\altaffilmark{30,31},
Saurabh~W.~Jha\altaffilmark{32},
Richard~Kessler\altaffilmark{18,19},
Wolfram~Kollatschny\altaffilmark{33},
Giorgos~Leloudas\altaffilmark{23,34},
John~Marriner\altaffilmark{11},
Jennifer~L.~Marshall\altaffilmark{6},
Ramon~Miquel\altaffilmark{20,35},
Tomoki~Morokuma\altaffilmark{13},
Jennifer~Mosher\altaffilmark{1},
Robert~C.~Nichol\altaffilmark{7},
Jakob~Nordin\altaffilmark{27,36},
Matthew~D.~Olmstead\altaffilmark{10},
Linda~\"Ostman\altaffilmark{23},
Jose~L.~Prieto\altaffilmark{37},
Michael~Richmond\altaffilmark{38},
Roger~W.~Romani\altaffilmark{39},
Jesper~Sollerman\altaffilmark{23,40},
Max~Stritzinger\altaffilmark{41},
Donald~P.~Schneider\altaffilmark{42,43},
Mathew~Smith\altaffilmark{44},
J.~Craig~Wheeler\altaffilmark{45},
Naoki~Yasuda\altaffilmark{14}, and
Chen~Zheng\altaffilmark{46}
}

\lastpagefootnotes
\altaffiltext{1}{Department of Physics and Astronomy, University of
  Pennsylvania, 209 South 33rd Street, Philadelphia, PA 19104, USA}
\altaffiltext{2}{African Institute for Mathematical Sciences,
  Muizenberg, 7945, Cape Town, South Africa}
\altaffiltext{3}{South African Astronomical Observatory, Cape Town,
  South Africa}
\altaffiltext{4}{Department of Mathematics and Applied Mathematics,
  University of Cape Town, Cape Town, South Africa}
\altaffiltext{5}{Department of Astronomy, University of Washington,
  Box 351580, Seattle, WA 98195, USA}
\altaffiltext{6}{Department of Physics \& Astronomy, Texas A\&M
  University, College Station, TX 77843, USA}
\altaffiltext{7}{Institute of Cosmology and Gravitation, Dennis Sciama
  Building, Burnaby Road, University of Portsmouth, Portsmouth, PO1
  3FX, UK}
\altaffiltext{8}{Institute of Astronomy, University of Cambridge,
  Madingley Road, Cambridge CB3 0HA, UK}
\altaffiltext{9}{Wayne State University, Department of Physics and
  Astronomy, Detroit, MI 48202, USA}
\altaffiltext{10}{Department of Physics and Astronomy, University of
  Utah, Salt Lake City, UT 84112, USA}
\altaffiltext{11}{Center for Particle Astrophysics, Fermi National
  Accelerator Laboratory, P.O. Box 500, Batavia, IL 60510, USA}
\altaffiltext{12}{Las Cumbres Observatory Global Telescope Network,
  6740 Cortona Dr. Suite 102, Goleta, CA 93117, USA}
\altaffiltext{13}{Institute of Astronomy, Graduate School of Science,
  The University of Tokyo, 2-21-1 Osawa, Mitaka, Tokyo 181-0015,
  Japan}
\altaffiltext{14}{Kavli Institute for the Physics and Mathematics of
  the Universe (Kavli IPMU, WPI), Todai Institutes for Advanced Study,
  the University of Tokyo, Kashiwa 277-8583, Japan}
\altaffiltext{15}{Department of Astronomy, University of California,
  Berkeley, CA 94720-3411, USA}
\altaffiltext{16}{Astronomy Department, University of Illinois at
  Urbana-Champaign, 1002 West Green Street, Urbana, IL 61801 USA}
\altaffiltext{17}{Department of Physics, University of Illinois
  Urbana-Champaign, 1110 W. Green Street, Urbana, IL 61801 USA}
\altaffiltext{18}{Department of Astronomy and Astrophysics, The
  University of Chicago, 5640 South Ellis Avenue, Chicago, IL 60637,
  USA}
\altaffiltext{19}{Kavli Institute for Cosmological Physics, The
  University of Chicago, 5640 South Ellis Avenue, Chicago, IL 60637,
  USA}
\altaffiltext{20}{Institut de F\'{\i}sica d'Altes Energies, E-08193
  Bellaterra (Barcelona), Spain}
\altaffiltext{21}{CENTRA Centro Multidisciplinar de Astrof\'isica,
  Instituto Superior T\'ecnico, Av. Rovisco Pais 1, 1049-001 Lisbon,
  Portugal}
\altaffiltext{22}{Department of Physics, University of Notre Dame, 225
  Nieuwland Science Hall, Notre Dame, IN 46556, USA}
\altaffiltext{23}{Oskar Klein Centre, Stockholm University, SE-106 91
  Stockholm, Sweden}
\altaffiltext{24}{Department of Physics, Stockholm University, SE-106
  91 Stockholm, Sweden}
\altaffiltext{25}{Argonne National Laboratory, 9700 South Cass Avenue,
  Lemont, IL 60439, USA}
\altaffiltext{26}{McDonald Observatory, University of Texas at Austin,
  Austin, TX 7871, USA}
\altaffiltext{27}{Lawrence Berkeley National Laboratory, 1 Cyclotron
  Road MS 50B-4206, Berkeley, CA 94720, USA}
\altaffiltext{28}{Department of Astronomy, Princeton University,
  Princeton, New Jersey 08544, USA}
\altaffiltext{29}{Department of Astronomy, MSC 4500, New Mexico State
  University, P.O. Box 30001,Las Cruces, NM 88003, USA}
\altaffiltext{30}{Universitaets-Sternwarte Munich, Scheiner Str 1, D
  81679 Munich, Germany}
\altaffiltext{31}{MPI f.\ Extraterrestrische Physik,
  Giessenbachstrasse, D 85741 Garching, Germany}
\altaffiltext{32}{Department of Physics and Astronomy, Rutgers, the
  State University of New Jersey, 136 Frelinghuysen Road, Piscataway,
  NJ 08854, USA}
\altaffiltext{33}{Institut f\"ur Astrophysik, Universit\"at
  G\"ottingen, Friedrich-Hund Platz 1, 37077 G\"ottingen, Germany}
\altaffiltext{34}{Dark Cosmology Centre, Niels Bohr Institute,
  University of Copenhagen, Juliane Maries Vej 30, 2100 Copenhagen,
  Denmark}
\altaffiltext{35}{Instituci\'o Catalana de Recerca i Estudis
  Avan\c{c}ats, E-08010 Barcelona, Spain}
\altaffiltext{36}{Space Sciences Lab, University of California
  Berkeley, 7 Gauss Way, Berkeley, CA 94720, USA}
\altaffiltext{37}{Department of Astrophysical Sciences, Princeton
  University, Princeton, NJ 08540, USA}
\altaffiltext{38}{School of Physics and Astronomy, Rochester Institute
  of Technology, Rochester, New York 14623, USA}
\altaffiltext{39}{Department of Physics, Stanford University, Palo
  Alto, CA 94305, USA}
\altaffiltext{40}{Department of Astronomy, Stockholm University,
  SE-106 91 Stockholm, Sweden}
\altaffiltext{41}{Department of Physics and Astronomy, Aarhus
  University, Denmark}
\altaffiltext{42}{Department of Astronomy and Astrophysics, The
  Pennsylvania State University, University Park, PA 16802, USA}
\altaffiltext{43}{Institute for Gravitation and the Cosmos,  The
  Pennsylvania State University, University Park, PA 16802, USA}
\altaffiltext{44}{Department of Physics, University of the Western
  Cape, Cape Town, 7535, South Africa}
\altaffiltext{45}{Department of Astronomy, University of Texas at
  Austin, Austin, TX 78712, USA}
\altaffiltext{46}{10497 Anson Avenue, Cupertino, CA 95014, USA}



\slugcomment{Submitted to ApJS}

\shorttitle{SDSS-II SN Data Release}
\shortauthors{Sako et al.}

\begin{abstract}
This paper describes the data release of the Sloan Digital Sky
Survey-II (\sdssii) Supernova Survey conducted between 2005 and 2007.
Light curves, spectra, classifications, and ancillary data are
presented for \ncand\ variable and transient sources discovered
through repeat $ugriz$ imaging of \sdss\ Stripe 82, a 300~deg$^2$ area
along the celestial equator.  This data release is comprised of all
transient sources brighter than $r \simeq 22.5$~mag with no history of
variability prior to 2004.  Dedicated spectroscopic observations were
performed on a subset of \nsnspec\ transients, as well as spectra for
thousands of transient host galaxies using the
\sdssiii\ \boss\ spectrographs.  Photometric classifications are
provided for the candidates with good multi-color light curves that
were not observed spectroscopically. From these observations,
\nsncand\ transients are either spectroscopically confirmed, or likely
to be, supernovae, making this the largest sample of supernova
candidates ever compiled.  We present a new method for
\sn\ host-galaxy identification and derive host-galaxy properties
including stellar masses, star-formation rates, and the average
stellar population ages from our \sdss\ multi-band photometry.  We
derive \saltii\ distance moduli for a total of \nsnzfixed\ \snia\ with
spectroscopic redshifts as well as photometric redshifts for a further
\nsnzfit\ purely-photometric \snia\ candidates.  Using the
spectroscopically confirmed subset of the three-year
\sdssii\ \snia\ sample and assuming a flat {\small $\Lambda$CDM}
cosmology, we determine $\Omega_{M} = 0.315 \pm 0.093$ (statistical
error only) and detect a non-zero cosmological constant at
5.7~$\sigma$.
\end{abstract}

\keywords{cosmology: observations --- supernovae: general --- surveys}


\section{Introduction}\label{section_intro}

In response to the astounding discovery of the late-time acceleration of the
expansion rate of the Universe \citep{riess98,perlmutter99}, a number of
large-scale supernova (\sn) surveys were launched.  These experiments included
programs to observe low redshift \sn\ such as the Nearby Supernova Factory
\citep{aldering02}, the Carnegie Supernova Project \citep{hamuy06}, and the
Center for Astrophysics \sn\ Program \citep{hicken09}.  At higher redshift,
new surveys included \essence\ \citep{miknaitis07}, the Supernova Legacy
Survey \citep[\snls;][]{astier06}, and dedicated \hst\ observations by
\citet{riess07}. At intermediate redshifts, the Sloan Digital Sky Survey
\citep[\sdss;][]{york00} bridged the gap between the local and distant
\sn\ searches by providing repeat observations of a 300~deg$^2$ stripe of sky
at the equator (known as Stripe 82) and discovered thousands of Type Ia
\sn\ (\snia) over the redshift range $0.05 < z < 0.4$ \citep{frieman08}.

This paper presents all data collected over the last decade as part of the
\sdss\ \sn\ Survey.  This search was a dedicated multi-band, magnitude-limited
survey, which provided accurate multi-color photometry for tens of thousands
of transient objects, all with a well-determined detection efficiency.  The
data have lead to precise measurements of the \sn\ rate as a function of
redshift, environment, and \sn\ type \citep{dilday08, dilday10a, dilday10b,
  smith12,taylor14}, and have lead to important new constraints on cosmology
with detailed studies of systematic uncertainties \citep{kessler09a,
  sollerman09,lampeitl10a,betoule14}.  The large survey volume and high
cadence have enabled early discoveries of rare events
\citep{phillips07,mcclelland10,mccully13}, as well as detailed statistical
studies of normal events \citep{hayden10a,hayden10b}.

The extensive, well-calibrated \sdss\ galaxy catalog has also helped
revolutionize the study of \snia\ and the dependence on their host-galaxy
properties.  For example, \citet{lampeitl10b} and \citet{johansson13} showed a
clear correlation between \sn\ Hubble residuals and the stellar mass of the
host.  The origin of this correlation remains unclear, but \citet{gupta11}
found evidence for the correlation being due to the age of the stellar
population \citep[cf.][]{johansson13}, while \citet{dandrea11} found the
correlation was likely related to the gas-phase metallicity using a sub-sample
of star-forming \sdss\ host galaxies. \citet{hayden13} have used the
fundamental metallicity relation \citep{mannucci10} to further reduce the
Hubble residuals, suggesting again that metallicity is the underlying physical
parameter responsible for the correlation.  \citet{galbany12}, however, did
not detect an obvious correlation between Hubble residuals and distance to the
\sn\ from the center of the host galaxy, as might be expected due to
metallicity gradients, but the are not as sensitive as the more direct
metallicity measurements presentend in \citet{dandrea11}.  \citet{galbany12}
also found that extinction and \snia\ color decrease with increasing distance
from the center of the host, and that the average \sn\ light curve shape
differs significantly in elliptical and spiral galaxies as seen in many
previous studies \citep{hamuy96,gallagher05,sullivan06}.  \citet{xavier13}
found that \snia\ properties in rich galaxy clusters are, on average,
different from those in passive field galaxies, possibly due to differences in
age of the stellar populations.  Finally, \citet{smith14} studied the effects
of weak gravitational lensing on the \sdssii\ \snia\ distance measurements.

The \sn\ spectra presented in this data release are a collection of data from
11 different telescopes and includes some spectra taken to determine galaxy
properties long after the \sn\ had faded.  We did not attempt a detailed
spectroscopic analysis of the full sample beyond transient classification and
redshift measurement, but subsets of the data were previously published
\citep{zheng08,konishi11a,ostman11} and analyzed to quantitatively measure
spectral features \citep{konishi11b,nordin11a,nordin11b,foley12}.

Since spectra were not obtained for all discovered transients (as is true for
all \sn\ surveys), \citet{sako11} analyzed the light curves of the full sample
of variable objects and identified $\sim 1100$ purely photometric
\snia\ candidates with quantitative estimates for the classification
efficiency and sample purity.  In the absence of a \sn\ spectrum, the
identification and placement of \snia\ on a Hubble diagram is greatly aided by
a knowledge of the host-galaxy redshift.  Many host galaxy spectroscopic
redshifts were measured by the \sdssi\ and \sdssii\ surveys, but the
\sdssiii\ \citep{eisenstein11} Baryon Oscillation Spectroscopic Survey (\boss;
\citealt{dawson13}) ancillary program \citep{olmstead13} provided redshifts
for most of the observable \sn\ host galaxies.  \citet{hlozek12} and
\citet{campbell13} presented Hubble diagrams using photometric
\sn\ classification and host redshifts, and demonstrated that statistically
competitive cosmological constraints can be obtained with limited
spectroscopic follow up of active \sn\ candidates.  The \sdss\ work on
photometric identification represents an important example analysis for
ongoing and future large surveys, such as
\panstarrs\ \citep{scolnic13a,rest13}, \des\ \citep{bernstein12} and
\lsst\ \citep{tyson02}, where full spectroscopic follow up of all active
\sn\ candidates will be impractical.

This paper presents a catalog of \ncand\ \sdss\ sources that were identified
as part of the \sdss\ \sn\ search.  The images and object catalogs provided
herein were produced by the standard \sdss\ survey pipeline as presented in
\sdss\ Data Release 7 \citep{dr7}.  Our transient catalog is presented as a
machine readable table in the on-line version of this paper, and the format of
the catalog is described in Table \ref{tab:fullCatalog}.  Detailed
descriptions of general properties (\S~\ref{section_catalog}), source
classification (\S~\ref{section_photoid}), \snia\ light curve fits
(\S~\ref{section_lcfits}) for selected sources, and host galaxy
identifications (\S~\ref{section_hostid}) are given along with truncated
tables of catalog data.  The photometric data is described in
\S~\ref{section_photometry}.  Many sources have associated optical spectra,
which are described and cataloged in \S~\ref{section_spec}.

\section{SDSS-II Supernova Survey}\label{section_survey}

The \sdssii\ \sn\ data were obtained during three-month campaigns in the Fall
of 2005, 2006, and 2007 as part of the extension of the original \sdss.  A
small amount of engineering data were collected in 2004 \citep{sako05}, but
are not included in this paper, since the cadence and survey duration were not
adequate for detailed light curve studies.  The \sdss\ telescope
\citep{gunn06} and imaging camera \citep{gunn98} produce photometric
measurements in each of the $ugriz$ \sdss\ filters \citep{fukugita96} spanning
the wavelength range of 350 to 1000~nm.  The most useful filters for observing
\sdss\ \sn, however, are \textit{g}, \textit{r}, and \textit{i} because the
\sn\ are difficult to detect in \textit{u} and \textit{z} except at low
redshifts ($z\lesssim0.1$ for \snia) due to the relatively poor throughput of
those filters.

\begin{figure}[htb]
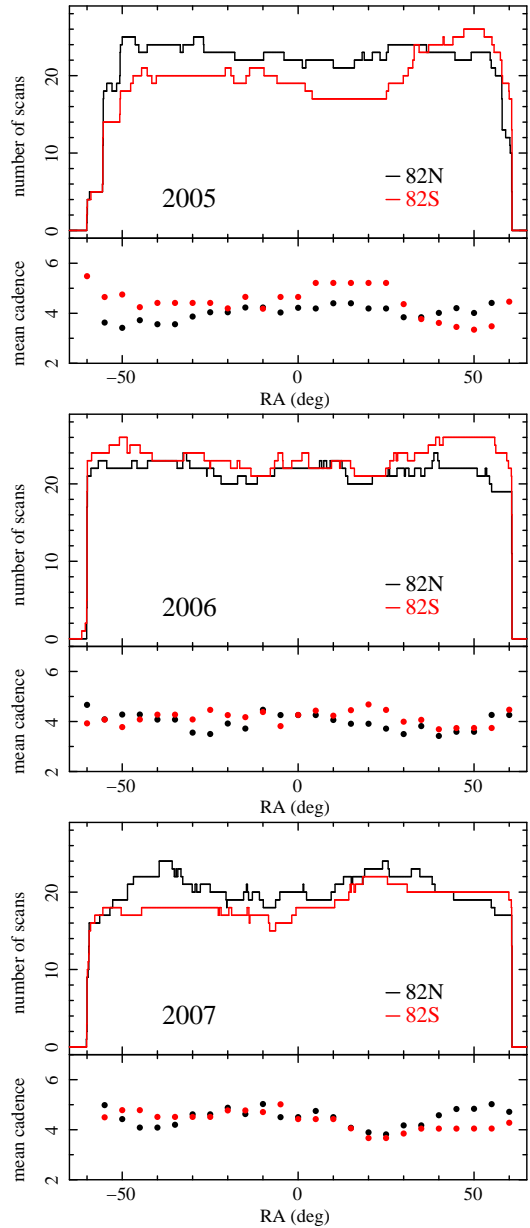

%
%
  \begin{center}
    \includegraphics[angle=-90,scale=0.3]{f1a.eps}
    \includegraphics[angle=-90,scale=0.3]{f1b.eps}
    \includegraphics[angle=-90,scale=0.3]{f1c.eps}
  \end{center}
\caption{Number of scans versus right ascension (shown in degrees) of the
  \sdss\ \sn\ equatorial stripe (Stripe 82) is shown along with the mean
  cadence for each year (2005-2007) of the survey.  The coverage in right
  ascension increased slightly as the template image coverage increased while
  the mean cadence was approximately four days for all three observing
  seasons.}
  \label{fig:run_coverage}
\end{figure}

The \sdss\ \sn\ survey is a ``rolling search'', where a portion of the
sky is repeatedly scanned to discover new \sn\ and to measure the
light curves of the ones previously discovered.  The survey observed
Stripe 82, which is 2.5\arcdeg\ wide in Declination between Right
Ascension of $20^h$ and $04^h$.  The camera is operated in drift scan
mode with all filters being observed nearly simultaneously with a
fixed exposure time of 55 seconds each.  Full coverage of Stripe82 was
obtained in two nights (with offset camera positions), but the average
cadence was approximately four nights because of inclement weather and
interference from moonlight.  The coverage and cadence of the survey
is shown in Figure~\ref{fig:run_coverage}.  The repeated scans were
used by \citet{annis11} to produce and analyze deep coadded images.
The survey is sensitive to \snia\ beyond a redshift of 0.4, but beyond
a redshift of 0.2 the completeness, and the ability to obtain
high-quality photometry, deteriorates.

The \sdss\ camera images were processed by the \sdss\ imaging software
\citep{stoughton02} and \sn\ were identified via a frame subtraction
technique \citep{alard98}.  Objects detected after frame subtraction
in two or more filters were placed in a database of detections.  These
detected objects were scanned visually and were designated candidates
if they were not obvious artifacts.  Spectroscopic measurements were
made for promising candidates depending on the availability and
capabilities of telescopes.  The candidate selection and spectroscopic
identification have been described by \citet{sako08}.  In three
observing seasons, the \sdssii\ \sn\ Survey discovered
\ncand\ \emph{new} variable objects and spectroscopically identified
\nssnia\ \snia\ and \nsccsn\ core-collapse \sn\ (\ccsn).

\section{SN Candidate Catalog}\label{section_catalog}

Table \ref{tab:fullCatalog} describes the format of the
\sdssii\ \sn\ catalog, which includes information on the
\ncand\ sources detected on two or more nights.  The full catalog is
made available online; a small portion is reproduced as an example in
Table~\ref{tbl:cand}.

\begin{figure}
\begin{center}
\epsscale{1.1}
\plotone{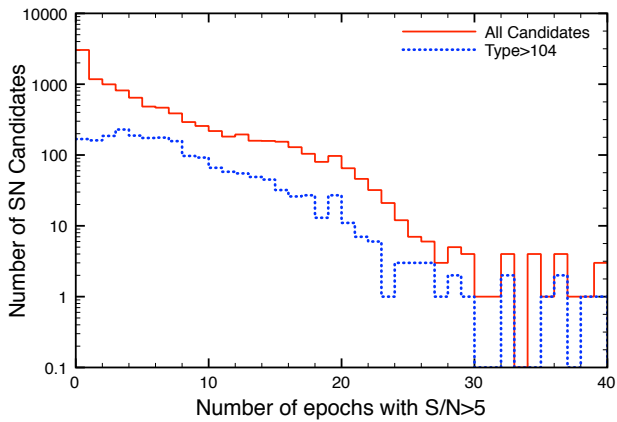}
\caption{The distribution of number of epochs observed per SN is
  shown.  An epoch consists of one night of observation in all 5 SDSS
  filters without any requirement that there was a detectable signal
  in any of the filters.  There are typically about 20 epochs in an
  observing season, but a small fraction of SN lie in the overlap
  region and are observed with twice the cadence or up to 40 times per
  season.\label{fig:nepoch}}
\end{center}
\end{figure}

General photometric properties include the J2000 coordinates of the
\sn\ candidate, the number of epochs detected by the search pipeline
(\verb2Nsearchepoch2) and final photometry pipeline above
\signoi\ $>5$ (\verb2NepochSNR52), and $r$-band magnitude
(\verb2Peakrmag2) and \mjd\ (\verb2MJDatPeakrmag2) of the brightest
measurement.  We show the distribution of \verb2NepochSNR52 for all
candidates in Figure~\ref{fig:nepoch} as an indication of the general
quality of the light curves.

We provide the heliocentric redshift (\verb2zspecHelio2) and
uncertainty (\verb2zspecerrHelio2) when spectroscopic measurements are
available.  The source of the redshift is from the host galaxy
spectrum or, if the host galaxy redshift is not known, from the
\sn\ spectrum.  More details on the spectra are given in
\S~\ref{section_spec}.  The number of spectra available as part of
this Data Release are given as \verb2nSNspec2 (the number of
\sn\ spectra) and \verb2nGALspec2 (the number of host galaxy spectra)
in the catalog.  The galaxy spectra include cases where the galaxy
spectrum is obtained from the \sn\ spectroscopic observation but with
an aperture chosen to enhance the galaxy light and cases where a
spectrum was taken when the \sn\ was no longer visible for the purpose
of measuring the galaxy redshift and possibly other galaxy properties.
Galaxy spectra that were taken with the \sdss\ spectrograph
\citep{smee13} are not included in these totals, but \verb2objIDHost2
gives \sdss\ \dr 8 object index so that the galaxy properties may be
easily extracted from the \sdss\ database.  Spectra as part of the
\sdssiii\ \boss\ program are also not included in these totals.  They
are discussed in \citet{campbell13} and \citet{olmstead13}, but their
redshifts are listed under \verb2zspecHelio2.  Finally, we provide the
{\small CMB}-frame redshifts and uncertainties in \verb2zCMB2 and
\verb2zerrCMB2, respectively.

\begin{figure}
\begin{center}
\epsscale{1.1}
\plotone{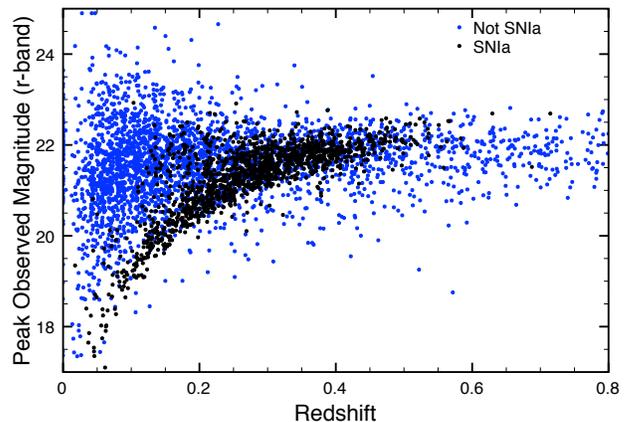}
\caption{The peak $r$-band magnitude observed is shown as a function
  of redshift.  Black points shown are for all candidates classified
  as \snia.  All other SN candidates are shown in
  blue.}\label{fig:magvsz}
\end{center}
\end{figure}

Some sources (most of the spectroscopically identified \sn) were assigned a
standard name by the \iau; the name is listed for those sources that have been
assigned one.  The peak $r$-band magnitude observed is plotted versus redshift
in Figure \ref{fig:magvsz}.

The candidates are classified according to their light curves and spectra
(when available), and the results of the classification are shown in
Table~\ref{tbl:cand}.  Visual scanning removed most of the artifacts, so
almost all of the objects in the catalog are variable astronomical sources,
some of which are only visible for a limited period of time (for example,
supernovae).  The multi-night requirement eliminates rapidly moving objects,
which are primarily main-belt asteroids.  A summary of the number of objects
in each classification is shown in Table~\ref{tbl:typeTotal}.  The
classification ``Unknown'' means that the light curve was too sparse and/or
noisy to make a useful classification, ``Variable'' means that the source was
observed in more than one observing season, and ``AGN'' means that an optical
spectrum was identified as having features associated with an active galaxy,
primarily broad hydrogen emission lines.  The other categories separate the
source light curves into 3 \sn\ types: Type II, Type Ibc (either Ib or Ic),
and Type Ia.  A prefix ``p'' indicates a purely-photometric type where the
redshift is unknown and that the identification has been made with the
photometric data only.  A prefix ``z'' indicates that a redshift is measured
from its candidate host galaxy and the classification uses that redshift as a
prior.  The \sn\ classifications without a prefix are made based on a spectrum
(including a few non-\sdss\ spectra).  The Type Ib and Ic spectra
identifications are shown separately.  The ``\snia ?''  classification is based
on a spectrum that suggests a \snia\ but is inconclusive.  The details and
estimated accuracy of the classification scheme are given in the next section.

Some of the \sn\ candidates in the catalog have associated notes.  Notes
indicate \sn\ where the typing spectrum was obtained by other groups (and is
not included in the \sdss\ data release) and indicate \sn\ candidates that may
have peculiar features.  The bulk of the spectroscopically identified
\snia\ are consistent with normal \snia\ features, but a few were identified
as having some combination of peculiar spectral and light curve features.  We
did not search for these peculiar features in a systematic way, but we have
noted the likely peculiar features that were found.  Some \snia\ have poor
fits to the \snia\ light curve model or unlikely parameters for normal \snia,
but we have not noted these, preferring to just present the fit parameters.
Table \ref{tbl:notes} describes the codes that may appear in the notes column
(item 136) of Table~\ref{tab:fullCatalog}.

\begin{figure*}[htb]
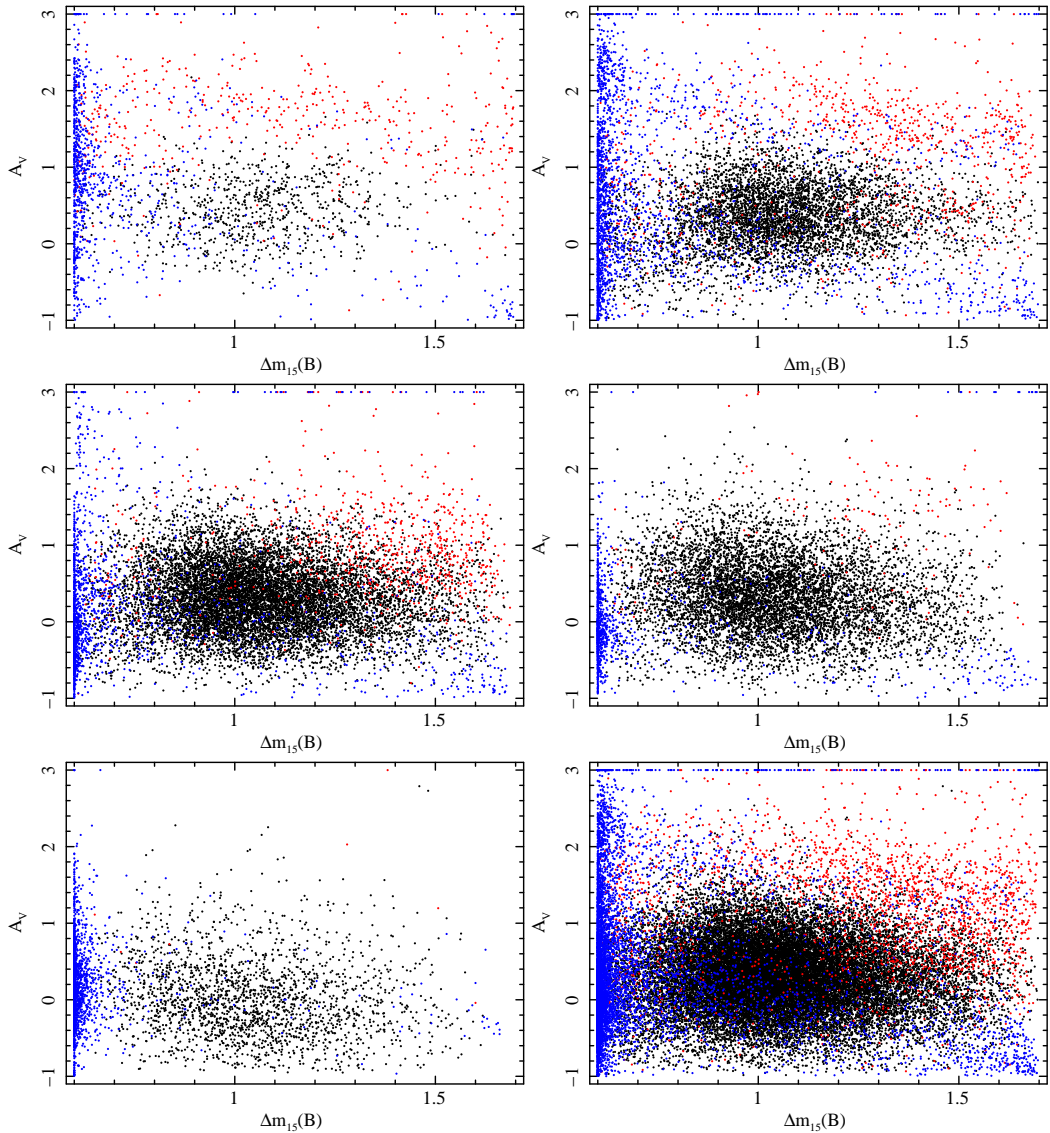

\begin{center}
\includegraphics[angle=-90,scale=0.3]{f4a.eps}
\includegraphics[angle=-90,scale=0.3]{f4b.eps}
\includegraphics[angle=-90,scale=0.3]{f4c.eps}
\includegraphics[angle=-90,scale=0.3]{f4d.eps}
\includegraphics[angle=-90,scale=0.3]{f4e.eps}
\includegraphics[angle=-90,scale=0.3]{f4f.eps}
\end{center}
 \caption{Regions occupied by SN~Ia (black), SN~Ibc (red) and SN~II (blue) in
   \dmB\ -- \av\ space in different redshift slices for a simulated SDSS-II SN
   Survey.  The panels are $z<0.1$ (top left), $0.1 < z < 0.2$ (top right),
   $0.2 < z < 0.3$ (middle left), $0.3 < z < 0.4$ (middle right), $z > 0.4$
   (bottom left), and all $z$ (bottom right).}
    \label{fig:zdm_zav}
\end{figure*}

\section{Photometric Classification}\label{section_photoid}

This section describes our method for photometric classification of the
\sn\ candidates.  First, we reject likely non-\sn\ events as those showing
variability over two or more seasons.  The exact nature of these sources is
not known, but the majority are most likely variable stars and active galactic
nuclei.  A total of \nvariable\ are identified as ``Variable'' in
Table~\ref{tbl:cand}.

\begin{figure*}[htb]
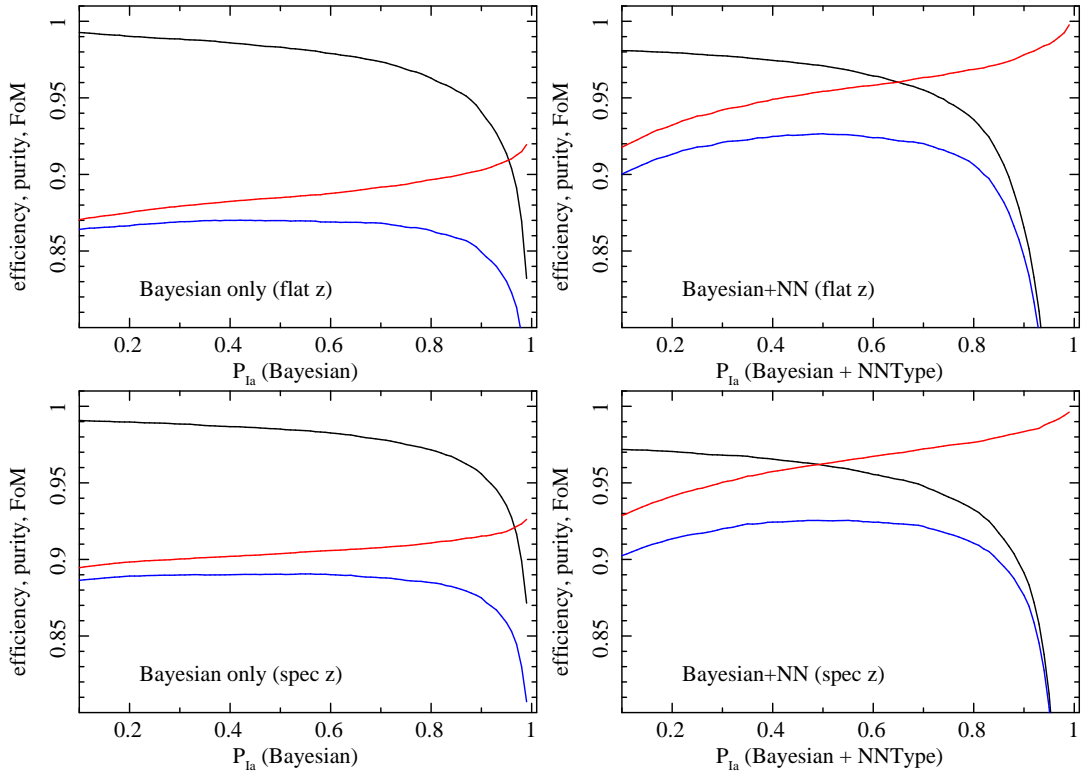

  \begin{center}
  \includegraphics[angle=-90,scale=0.3]{f5a.eps}
  \includegraphics[angle=-90,scale=0.3]{f5b.eps}
  \includegraphics[angle=-90,scale=0.3]{f5c.eps}
  \includegraphics[angle=-90,scale=0.3]{f5d.eps}
  \end{center}
  \caption{The SN~Ia photometric classification efficiency (black),
    purity (red), and figure of merit (product of the efficiency and
    purity; blue) as a function of the \pia\ probability cut for
    simulated SDSS-II SN data.  The top panels show results from
    Bayesian-only (left) and with the nearest-neighbor extension
    (\psnidnn) for a flat redshift prior.  The bottom panels show the
    same for a spectroscopic redshift prior.  We required
    $\log$(\pfit) $>-4.0$.  Note that the purity using the
    Bayesian-only method is never above $\sim 93$\%.}
    \label{fig:type_compare}
\end{figure*}

All remaining candidates showed variability during only a single season and
are therefore viable \sn\ candidates.  Their light curves were then analyzed with
the Photometric \sn\ IDentification (\psnid) software \citep{sako11}, first
developed for spectroscopic targeting and subsequently extended to identify
and analyze photometric \snia\ samples.  In short, the software compares the
observed photometry against a grid of \snia\ light curve models and
core-collapse \sn\ (\ccsn) templates, and computes the Bayesian probabilities
of whether the candidate belongs to a Type Ia, Ib/c, or II \sn.  The technique
is similar to that developed by \citet{poznanski07}, except that we
subclassify the \ccsn\ into Type~Ib/c and {\small II}.  Extensive tests and
tuning were performed using the large (but still limited) sample of
spectroscopic confirmations from \sdssii\ and simulations as described in
\citet{sako11}.  The light curve templates used in the analysis presented here
are the same as those from \citet{sako11}.  \psnid\ and the templates are now
part of the \snana\ package \citep{kessler09b}.

The Bayesian probabilities are useful because they represent the {\it
  relative} likelihood of \sn\ types, whereas the best-fit minimum reduced
$\chi^2$ (\chir), or more precisely the fit probability \pfit, provides an
{\it absolute} measure of the likelihood.  The combination of the Bayesian
probability (\pia) and the goodness-of-fit (\pfit) provides reliable
classification of \snia\ candidates.  The expected level of contamination and
efficiency can be estimated from either large datasets or simulations.
\citet{sako11} used this method to identify \snia\ candidates from \sdssii.
The \snia\ classification purity and efficiency were estimated to be 91\% and
94\%, respectively.  The one major drawback of this techinique, however, was
the general unreliability of classifying \ccsn.

To make further improvements, we developed an extension to \psnid\ that uses
the Bayesian classification described above as an {\it initial} filter, but
subsequently refines the classification using a kd-tree nearest-neighbor
({\small NN}) technique.  We call this method \psnidnn, and it is based on the
fact that different \sn\ types populate a distinct region in extinction,
light-curve shape, and redshift parameter space when fit to an \snia\ model.
This is illustrated in Figure~\ref{fig:zdm_zav}.  \snibc\ are generally redder
(large \av) and they fade more rapidly (large \dmB) compared to \snia.  \snii,
on the other hand, have broad, flat light curves (small \dmB).  As described
below, this method makes substantial improvements to both \snia\ and
\ccsn\ classification.

In this method, every \sn\ in the data sample is compared against a training
set and the most likely type is determined from the statistics of its
neighbors in a multi-dimensional parameter space.  Ideally, the training set
is a large, uniform, and unbiased sample of spectroscopically confirmed \sn,
but such training sets do not exist at the low-flux limit of the
\sdssii\ \sn\ sample.  Our current implementation, therefore, uses simulated
\sn\ from \snana.  The simulation is based on well-measured \ccsn\ template
light curves, which are used to simulate events of different magnitudes and
redshifts.  However, the underlying library is small (only 42 \ccsn\ template
light curves), and adequacy of this sample size has yet to be rigorously
verified.  We simulated 10 seasons worth of \sn\ candidates using a mix of
\snia, \snibc, and \snii\ identical to that used in the \sn\ Classification
Challenge \citep{kessler10b,kessler10c}.  For each \sn\ candidate in the data
sample, we calculate Cartesian distances in 3-dimensional parameter space
(\av, \dmB, $z$) to each simulated \sn\ (labeled $i$) using the following
formula:
\begin{equation}
\begin{split}
  d_\mathrm{SN}^2 = c_z (z_\mathrm{SN}-z_i)^2 + c_{\Delta m_{15}} (\Delta
    m_{15,\mathrm{SN}}-\Delta m_{15,i})^2 + \\ c_{A_V} (A_{V,\mathrm{SN}}-A_{V,i})^2,
  \label{nntype:d}
\end{split}
\end{equation}
where $c_z$, $c_{\Delta m_{15}}$, and $c_{A_V}$ are coefficients determined
and optimized using simulations for both the data and training sets.  The
classification probabilities are determined by counting the numbers of \snia,
\snibc, and \snii\ in the training set that are within a certain distance
$d_{\mathrm{max}}$.  Since this distance is degenerate with the overall
normalization of the other three coefficients, we set $d_{\mathrm{max}}=1.0$.
The optimized set of coefficients are $c_z=160$, $c_{\Delta m_{15}}=60$, and
$c_{A_V}=10$ assuming $d_{\mathrm{max}}=1$.

For each \sn\ candidate in the data sample we count the number of simulated
\sn\ from each type $N_{\mathrm {type}}$ within $d_\mathrm{SN} <
d_{\mathrm{max}}$.  The nearest-neighbor probabilities $P_{\mathrm {NN,type}}$
are then determined using,
\begin{equation}
  P_{\mathrm {NN,type}} = \frac{N_{\mathrm {type}}}{N_{\mathrm
      {Ia}}+N_{\mathrm {Ibc}}+N_{\mathrm {II}}}.
\end{equation}

The final classification is performed using the Bayesian, nearest-neighbor,
and fit probabilities.  For a candidate to be a photometric \snia\ candidate,
we require,
\begin{itemize}
  \item \pia\ $>$ \pibc\ and \pia\ $>$ \pii
  \item \pnnia\ $>$ \pnnibc\ and \pnnia\ $>$ \pnnii
  \item \pfit $\ge 0.01$ for \snia\ model
  \item Detections at $-5 \le$ \trest $\le +5$~days and $+5 <$ \trest $\le
    +15$~days.
\end{itemize}
For the photometric \snibc\ candidates, we require,
\begin{itemize}
  \item (\pibc\ $>$ \pia\ and \pibc\ $>$ \pii) or (\pia\ $>$ \pibc\ and
    \pia\ $>$ \pii)
  \item \pnnibc\ $>$ \pnnia\ and \pnnibc\ $>$ \pnnii.
\end{itemize}
Finally, for the photometric \snii\ candidates, we require,
\begin{itemize}
  \item (\pii\ $>$ \pia\ and \pii\ $>$ \pibc) or (\pia\ $>$ \pibc\ and
    \pia\ $>$ \pii)
  \item \pnnii\ $>$ \pnnia\ and \pnnii\ $>$ \pnnibc.
\end{itemize}
We impose no requirement on detections at any particular \trest\ for the
\ccsn\ selection.  The classification is performed using a spectroscopic
redshift prior if a spectrum of either the \sn\ candidate or its host galaxy
is available.  In these cases, the candidates are classified as \zsnia,
\zsnibc, or \zsnii\ in Table~\ref{tbl:cand}.  Otherwise, we use a flat
redshift prior and the candidates are denoted \psnia, \psnibc, or \psnii.

All candidates that do not meet any of the criteria above are declared
``unknown''.  The statistics of the \sn\ candidate classification are shown in
Table~\ref{tbl:typeTotal}.  Simulation results are shown in
Figure~\ref{fig:type_compare} where we compare classification performance
between the Bayesian-only method and with the nearest-neighbor probabilities.
For the Bayesian-only method, the \snia\ classification figure-of-merit
(defined as the product of the efficiency and purity) has a very broad maximum
when we require \pia\ $>$ 0.5, where the efficiency and purity are 98\% and
90\%, respectively.  For the Bayesian with the nearest-neighbor probabilities,
the figure-of-merit also peaks for \pia\ $>$ 0.5, where the efficiency and
purity are both 96\%.  Note the substantial improvement in the purity at the
expense of some reduction in efficiency.  This level of purity is not
attainable even with the most stringent cut (e.g., \pia\ $>0.99$) with the
Bayesian-only method.  The full summary of efficiencies and purities of
classification of all \sn\ types with flat-$z$ and spec-$z$ priors is listed
in Table~\ref{tbl:nntype_results}.

\section{Photometry}\label{section_photometry}

Light curves are constructed using the Scene Modeling Photometry software
\citep[\smp;][]{holtzman08}.  \smp\ assumes that the pixel data can be
described by the sum of a point source that is fixed in space but varying in
magnitude with time, a galaxy background that is constant in time but has an
arbitrary spatial distribution, and a sky background that is constant over a
wider area but varies in brightness at each observation.  The galaxy
background is parameterized as an arbitrary amplitude on a $15 \times 15$ grid
of pixels of size $0.6 \arcsec$. The fitting process accounts for the
variations in point spread function (\psf) to model the distribution of light
for each night of observation.  The \sn\ magnitudes and \sdss\ reference stars
on the same image are measured simultaneously using the same \psf\ so the
\sn\ magnitudes are measured relative to a calibrated \sdss\ star catalog.

A complete set of light curve photometric data for all \ncand\ \sn\ candidates
is given on the \sdss\ Data Release web page \citep{sdsssndr}.  The format of
the data is described on the web page and is the same as the previously
released first-year data sample \citep{holtzman08}.  The magnitudes quoted in
these data files, and elsewhere in this paper, are the \sdss\ standard inverse
hyperbolic sine magnitudes defined by \citet{lupton99}.  Magnitudes are given
in the \sdss\ native system and differ from the \ab\ system by an additive
constant given in \S\ref{ABoffset}.  The fluxes in those files, however, have
been \ab -corrected and are expressed in $\mu$J.  The magnitudes and fluxes
are reported in a way that is consistent with the first-year data sample
except that the calibration of \sdss\ native magnitudes to $\mu$J has changed
as described below in \S\ref{ABoffset}.  Quality flags defined by
\citet{holtzman08} are provied for each photometric measurement.  Special
attention should be given to the non-zero flags as they are indicators of
subtle problems in \smp\ fitting procedure.

\subsection{Photometric Uncertainties}

A substantial effort has been made to ensure accurate estimates of the
uncertainties in the \sdss\ light curve flux measurements.  An important
feature of \smp\ is that it works on the original images (\textit{i.e.},
without resampling pixels) so that a simple propagation of pixel-by-pixel
uncertainties from photo-electron statistics offers a robust estimate of the
photometric uncertainty. The galaxy model is remapped for each image, but the
galaxy is well measured in the reference images and the error is almost always
subdominant.  In addition to the pixel statistical uncertainty, \smp\ computes
a ``frame error'' that accounts for zero point uncertainty, galaxy model
uncertainty, and systematic sky background uncertainty.

The error model was tested by \citet{holtzman08} using pre-explosion epochs
(known zero flux), artificial supernovae (computer generated), and real stars.
The conclusion was that the error model provides a good description of the
observed photometric errors.

\begin{figure*}
  \begin{center}
  \epsscale{1.0}
  \plottwo{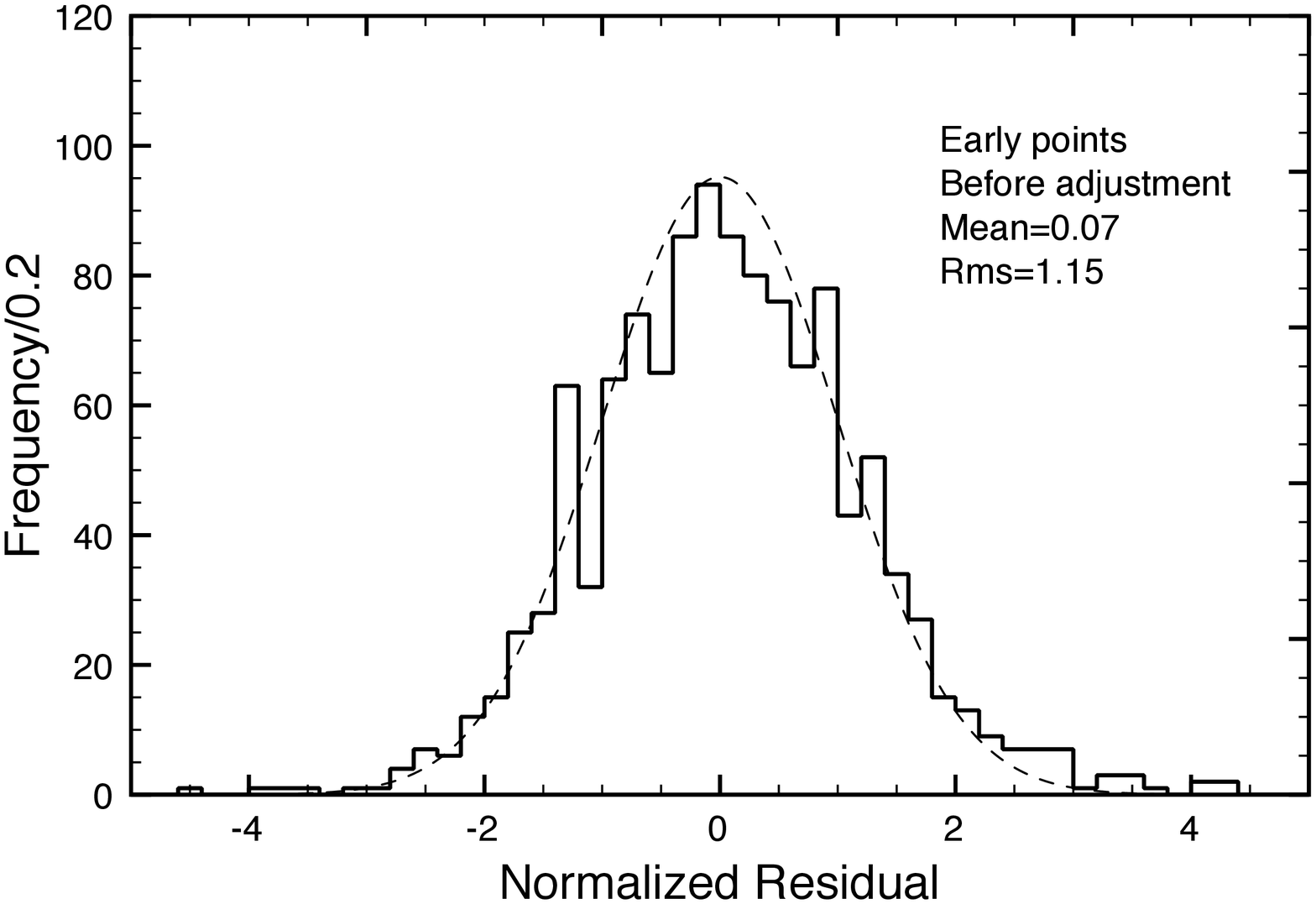}{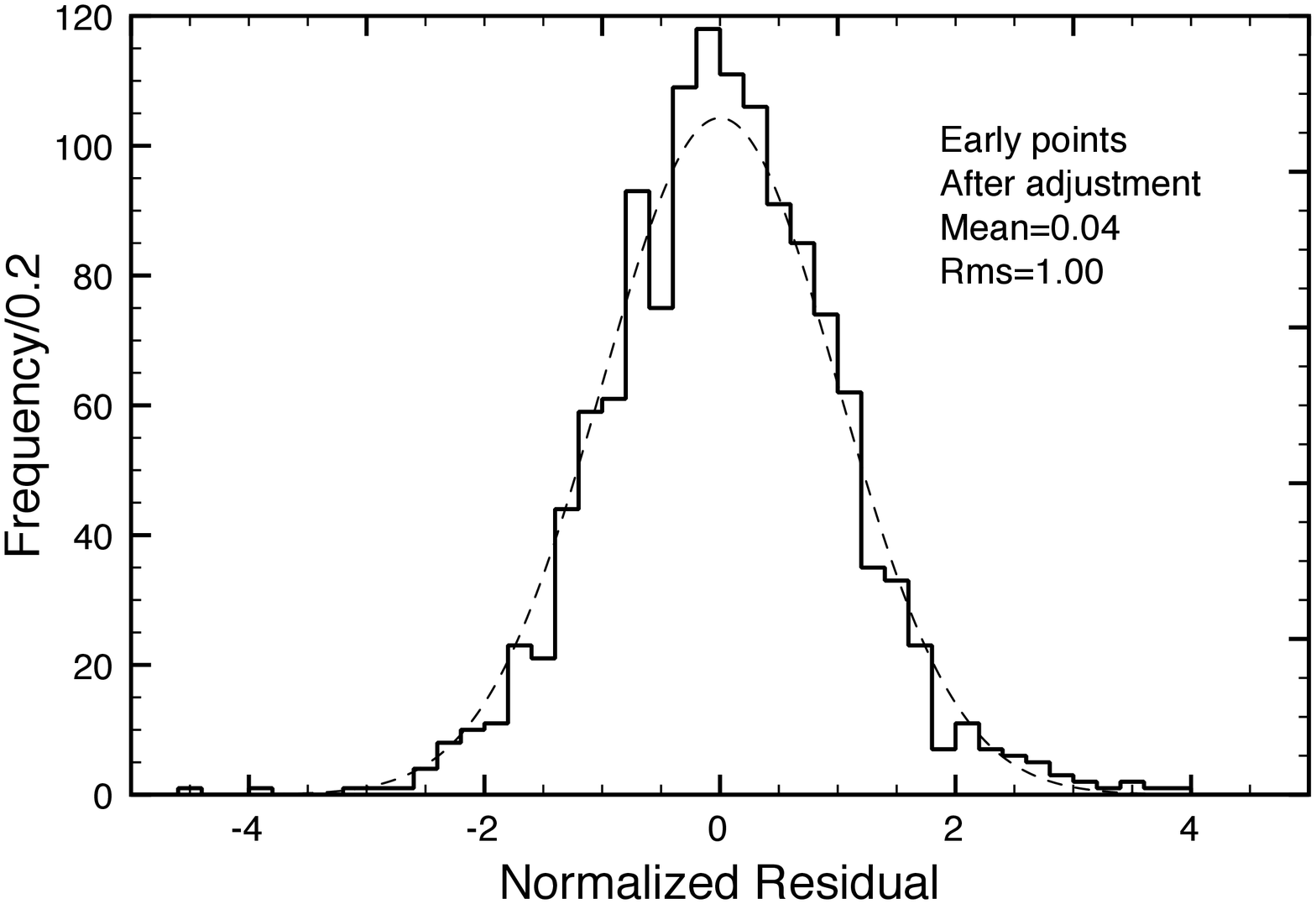}
  \plottwo{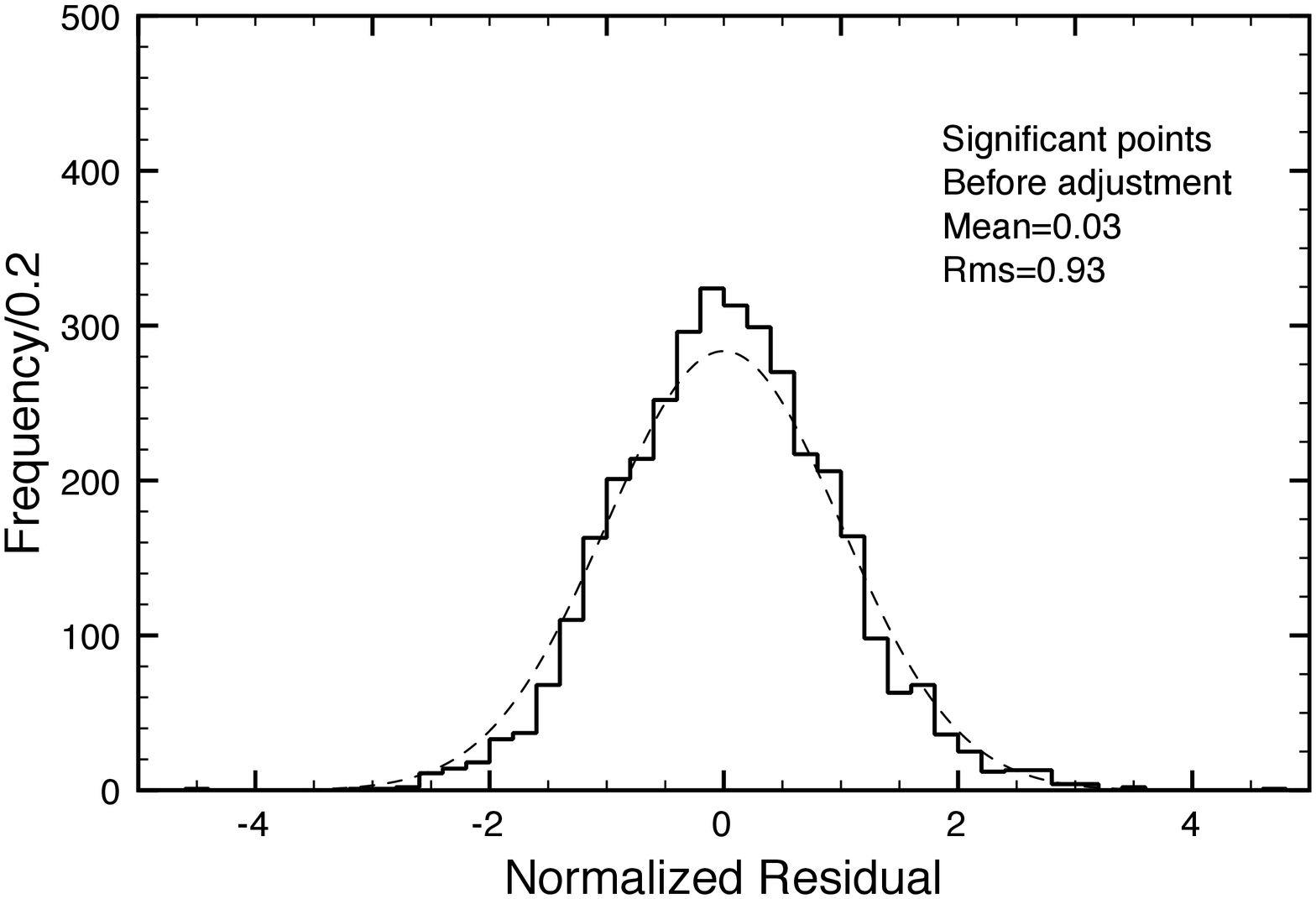}{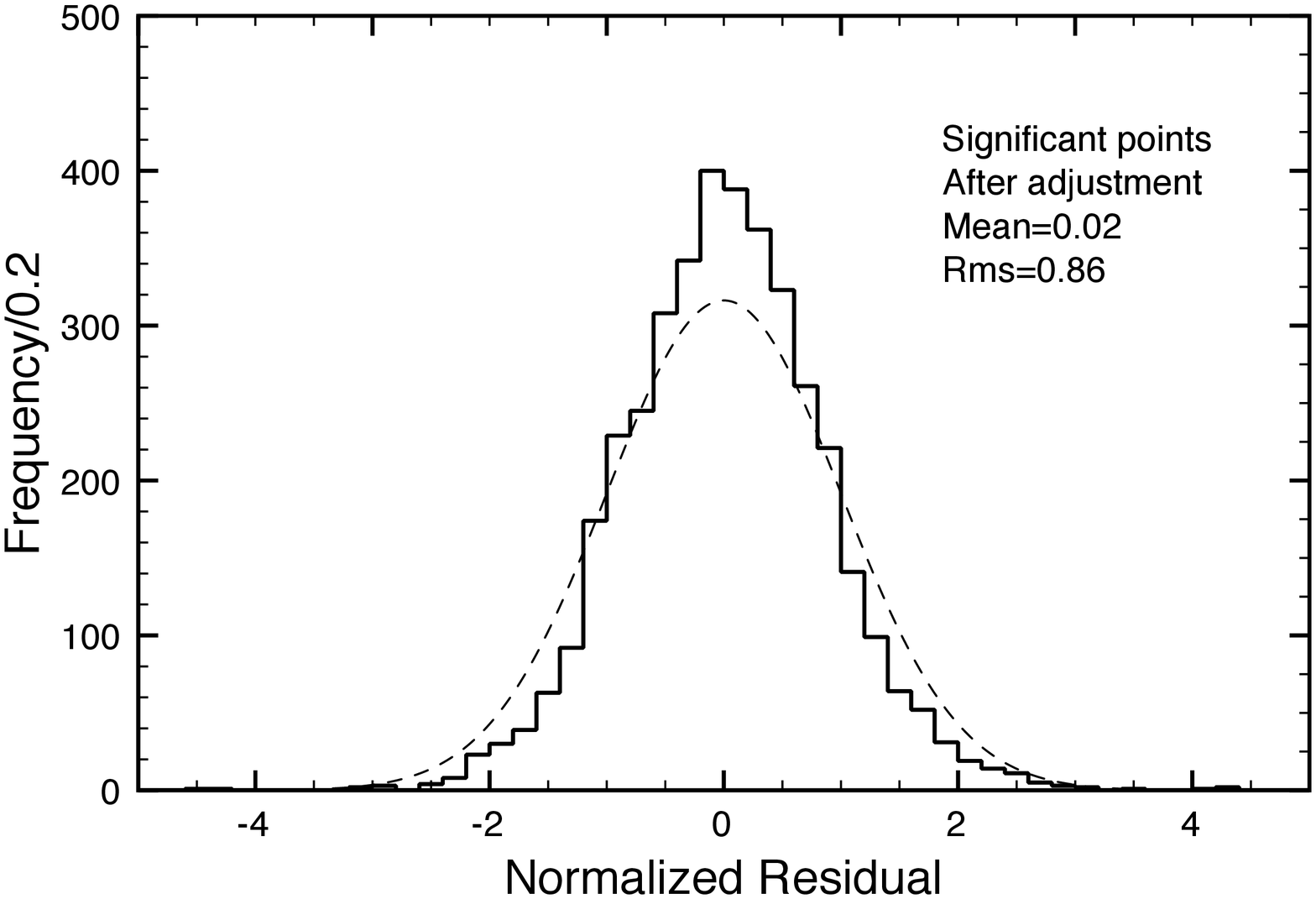}
  \caption{The normalized residuals for SN~Ia light curve fits to the $g$-band
    data before (left-hand panels) and after (right-hand panels) the
    adjustment described in the text.  The top panels use data prior to the SN
    explosion (``Early points''), which is therefore independent of the light
    curve model.  The bottom panels show the residuals for points where the
    detected flux is two or more standard deviations above background
    (``Significant points'').\label{fig:errorFudge}}
  \end{center}
\end{figure*}

After running the \smp\ code, we re-examined the photometric errors by
examining the light curve residuals relative to the \saltii\ \citep{guy10}
model.  We also investigated the distribution of residuals using pre-explosion
epochs, where the residuals do not depend on the \snia\ model.  For these data
the largest errors arise from statistical uncertainties and possible errors in
modeling the galaxy background light.  We also examined the distribution of
residuals relative to the \saltii\ light curve model when there was a
significant signal (more than 2$\sigma$ above the sky background).  In this
latter case, uncertainties in the light curve model and zeropoints contribute
to the width of the distribution of residuals.  For these tests we used
spectroscopically confirmed \snia\ excluding peculiar types and further
limited the sample to those \sn\ whose \saltii\ fit parameters indicated
normal stretch $|x_1|<2$ and low extinction $c<0.2$.  The $g$-band
distributions of the normalized residuals (residual divided by the
uncertainty) are shown in left-hand panels of Figure \ref{fig:errorFudge}.  A
normal Gaussian distribution (not a fit) is shown for comparison.  While both
distributions are quite close to the expected normal Gaussian, the
pre-explosion epoch distribution (upper left) is slightly wider than the curve
and the distribution with significant signal ($\sigma>2$) is narrower.  The
normalized residual distribution for the pre-explosion epochs could be larger
if the photometry underestimates the error in modeling the galaxy background.
When there is significant signal, the distribution of normalized residuals
could be smaller because of an overestimate of the zero-pointing error or the
light curve model uncertainty, which is included in the estimated errors.
Since the zero-point errors are at least partially correlated between epochs,
the fit parameters (especially the \sn\ color parameter) can absorb part of
the zero-point error, and therefore decrease the width of the distribution of
residuals.  While the measurement errors are considerably larger in $u$ and
$z$ bands, the distribution of the normalized residuals are similar for the
other \sdss\ filters, indicating that the error estimates are approximately
correct.

Based on these distributions, we adjusted the errors according to the
prescription
 \begin{equation}
\sigma^\prime = \sqrt{\sigma^2+c_f} \label{eqn:sigmaFudge}
\end{equation}
The constant $c_f$ was adjusted to result in an rms of unity for the
pre-explosion epoch distributions.  These small adjustments are within the
errors quoted by \citet{holtzman08}.  The values used for the error
adjustments for all five filters are shown in Table~\ref{tab:sigmaFudge}.  The
resulting $g$-band distributions of normalized residuals are shown on the
right-hand side of Figure \ref{fig:errorFudge}.  Our choice of the form in
Equation (\ref{eqn:sigmaFudge}) also slightly reduces the width of the
distribution of residuals with $\sigma>2$.  We did not attempt additional
modifications to the errors to bring the $\sigma>2$ distribution closer to a
normal Gaussian because of the additional uncertainties in interpretation.  As
a consequence, our error adjustment has the effect of deweighting low flux
measurements relative to measurements with significant flux.  The adjustment
has the most effect on $u$-band, where it is common to have many points
measured with large errors.  The overall \saltii\ lightcurve fit mean
confidence level (derived from the $\chi^2$/dof) is increased from 0.28 to
0.57 as a result of this change.

We also observe a small, but statistically significant offset in the mean
residual of the pre-explosion epochs.  The largest offset was found for
$r$-band where the offset was $0.12\sigma$, where $\sigma$ is the width of the
normalized distribution.  We did not correct this offset because we were
uncertain whether subtracting a constant flux from all epochs would be an
appropriate correction.  We did determine, however, that adding a constant
flux offset to our data had a negligible effect on the \saltii\ light curve
fit probability.

\subsection{Star catalog calibration}\label{ABoffset}

The star catalog calibration is discussed in detail by \citet{betoule13},
where the \sdss\ stellar photometry calibration is described in detail and the
\sdss\ photometry is compared with the Supernova Legacy Survey (\snls)
photometry.  The starting point for the \sdss\ \sn\ calibration is a
preliminary version of the \citet{ivezic07} star catalog that was used for
\smp\ photometry in \citet{holtzman08}.  This catalog uses the stellar locus
to calibrate the stellar colors but relies on photometry from the
\sdss\ Photometric Telescope (\pt) to establish the relative zeropoint for
$r$-band.  As explained in detail by \citet{betoule13}, there is a significant
flat-fielding error in the \pt\ photometry, leading to a photometry that was
biased as a function of declination.  We determined corrections to the
\citet{ivezic07} star catalog using \sdss\ Data Release~8 \citep{dr8}, whose
calibration is based on the method of \citet{padmanabhan08}.  This method, the
so-called ``Ubercal'' method, re-determines the nightly zeropoints based only
on the internal consistency of the 2.5 m telescope observations.  Our
adjustments to the stellar photometry were typically within a range of 2\%,
but corrections of up to 5\% were made in the $u$-band.  The corrections
improved the agreement with the \snls\ photometry.  Instead of recomputing the
\sn\ magnitudes relative to the new star catalog, we simply applied the
corrections to the \sn\ magnitudes found using the \citet{ivezic07} catalog.

Neither the star catalog of \citet{ivezic07} (based on the stellar locus) nor
\sdss\ Data Release 8 attempts to improve the absolute calibration of
\sdss\ photometry.  The photometry is tied to an absolute scale by
BD+17\arcdeg4708 using the magnitudes determined by \citet{fukugita96}.  We
have followed \citet{holtzman08} and re-determined the absolute scale using
the \sdss\ filter response curves \citep{doi10} and the \hst\ standard spectra
\citep{bohlin07} given in the \hst\ {\small CALSPEC} database \citep{CALSPEC}.
When the synthetic photometry of these standards is compared to the
\sdss\ \pt\ photometry, we obtain an absolute calibration, which is expressed
as ``\ab\ Offsets'' from the nominal \sdss\ calibration (see \citealt{oke83}
for a description of the \ab\ magnitude system).  The differences between our
current results and those of \citet{holtzman08} are that we have: 1) used the
recently published \sdss\ filter response curves, 2) used more recent
\hst\ spectra, and 3) re-derived the \pt\ to 2.5 m telescope photometric
transformation, including corrections for the recently discovered
non-uniformity of the \pt\ flat field.  Details of \ab\ system calibration may
be found in \citet{betoule13}.  Table~\ref{tab:ABoff} lists the \ab\ offsets
to be applied to the \sdss\ \sn\ data.  We use the average of three solar
analogs (P041C, P177D, and P330E) because these stars are similar in color to
the stars used to determine the (assumed) linear color transformation between
the \pt\ and 2.5 m telescope.  The uncertainty is calculated from the
dispersion of the results for the solar analogs.  The value determined for
BD+17\arcdeg4708 is given as a consistency check.  The most significant
numerical difference between the \ab\ offsets presented here and Table 1 of
\citet{holtzman08} is the $u$-band offset with $\Delta$\ab\ $\sim 0.03$, which
differs primarily because of the different filter response curve for $u$-band,
as discussed in detail by \citet{doi10}.

It is important to note that the \sn\ light curve photometry is given in the
\sdss\ natural system -- the same system that is used for all the \sdss\ data
releases.  The \ab\ offsets must be added to the \sn\ light curve magnitudes
in order to place them on a calibrated \ab\ system.

\subsection{$u$-band uncertainties}

\begin{figure}
  \begin{center}
  \epsscale{1.1}
  \plotone{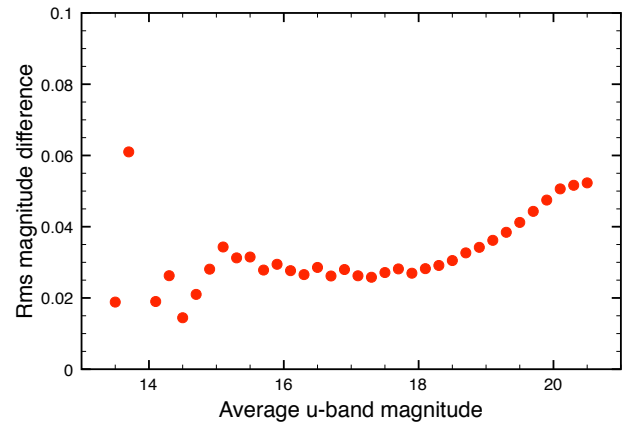}
  \caption{The rms photometric scatter of repeated measurements of SDSS Stripe
    82 stars in $u$-band.\label{fig:uband_rms}}
  \end{center}
\end{figure}

There has been some concern in the literature about the accuracy of the
$u$-band photometry.  The observations reported by \citet{jha06}, for example,
used a diverse set of telescopes and cameras and were not supported by a
large, uniform survey like \sdss .  For these reasons, one might question
whether there are substantial errors in the $u$-band calibration.  For
example, in the \snls 3 cosmology analysis \citep{conley11} measurements in
the $u$ band are de-weighted.  The quality of the \sdss\ $u$-band data
benefits greatly from an extensive, accurate star catalog of \sdss\ Stripe 82.
For example, Figure \ref{fig:uband_rms} shows the variations in stellar
magnitudes in the \citet{ivezic07} catalog, showing a repeatability of 0.03
mag over most of the magnitude range.  These secondary stars, which are the
\smp\ photometric references, are measured several times during photometric
conditions so that the calibration error is typically 0.01 to 0.02 magnitude
per star.  The \smp\ normally uses at least three calibration stars in
$u$-band so that the typical zero point error (which is included in the
\smp\ frame error) is comparable to the overall $u$-band scale error of 0.0089
\citep{betoule13}.

A check of \sdss\ \sn\ photometry is described in \citet{mosher12}, who
compared \sdss\ and Carnegie Supernova Project \citep{contreras10}
measurements on a subset of \snia\ observed by both surveys.  For the 32
$u$-band observations, they find agreement of $0.001 \pm 0.014$~mag, and
comparable agreement in the other bands.

\section{Spectra}\label{section_spec}

\sdss\ \sn\ spectra were obtained with the Hobbey-Eberly Telescope (\het), the
Apache Point Observatory 3.5m Telescope (\apo), the Subaru Telescope, the
2.4-m Hiltner Telescope at the Michigan-Dartmouth-MIT Observatory (\mdm), the
European Southern Observatory (\eso) New Technology Telescope (\ntt), the
Nordic Optical Telescope (\nnot), the Southern African Large Telescope
(\salt), the William Herschel Telescope (\wht), the Telescopio Nazionale
Galileo (\tng), the Keck I Telescope, and the Magellan Telescope.  Table
\ref{tab:spconfig} provides details of the instrumental configurations used at
each telescope.  These observations resulted in confirmation of
\nssnia\ \snia, \nssnibc\ \snibc, and \nssnii\ \snii.  A total of
\nspec\ unique spectra are part of this data release.  In many cases, we
provide extractions of the \sn\ and host galaxy spectra separately.  The
majority of the \sn\ spectra suffer contamination from the host galaxy, and we
did not attempt to remove that contamination.  Contamination of the galaxy
spectrum by \sn\ light may also be an issue in some of the galaxy spectra.

\begin{figure}
  \begin{center}
  \epsscale{1.1}
  \plotone{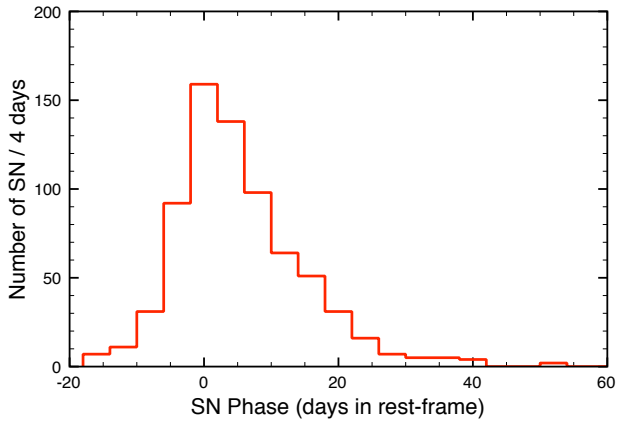}
  \caption{The distribution in time when SN~Ia spectra were observed
    relative to peak brightness in $B$-band.} \label{fig:spec_epochs}
  \end{center}
\end{figure}

Most \sn\ spectra were taken when the \sn\ candidates were near peak
brightness.  The distribution of observation times relative to peak brightness
is shown in Figure~\ref{fig:spec_epochs}.  Of the \nsnspec\ \sn\ candidates
with measured spectra, 177 have two or more spectra, and 16 have five or more
spectra.

The spectra were all observed using long slit spectrographs, but they were
observed under a variety of conditions with the procedures determined by the
individual observers.  Some spectra were observed at the parallactic angle
while other spectra were observed with the slit aligned to pass through both
the \sn\ and the host, or nearest, galaxy.  The different slit sizes and
observing conditions result in slit losses that are not well characterized for
most of the spectra.  The spectra were processed by the observers, or their
collaborators, using procedures developed for each particular telescope.

The spectra are calibrated to standard star observations, but with the
exception of the Keck spectra, the quality of the calibration is not verified.
Telluric lines are generally removed, but residual absorption features or sky
lines may be present.  We provide uncertainties for all the spectra, but the
uncertainties are generally limited to statistical errors.  Because of the
non-uniformities in the sample, and uncontrolled systematic errors, we cannot
make a general statement about the accuracy of all the spectra.  Some
subsamples of spectra have been subjected to detailed analyzes
\citep{ostman11,konishi11a,konishi11b,foley12} and more detailed information
on corrections and systematic errors can be found in these references.

The \sn\ spectral classification and redshift determination methods are
described in \citet{zheng08}.  Briefly, the spectra were compared to template
spectra and the best matching template spectrum was determined.  Each spectrum
was classified as ``None'' (no preferred match, usually because the spectrum
was too noisy), ``Galaxy'' (spectrum of a normal galaxy with no evidence for a
\sn), ``AGN'' (spectrum of an active galaxy) or a \sn\ type: ``Ia'' (Type Ia),
``Ia?'' (possible Type Ia), ``Ia-pec'' (peculiar Type Ia), ``Ib'' (Type Ib),
``Ic'' (Type Ic), or ``II'' (Type II).  The redshifts are generally determined
by cross-correlation with template spectra, but for some of the galaxy
redshifts observed in 2008 were determined by measuring line centroids.  All
redshifts are presented in the heliocentric frame.

The list of spectra is displayed in Table~\ref{tbl:spec}.  Each
observation is uniquely specified by the \sn\ candidate \id\ and
spectrum \id.  The observing telescope is listed and the
classification of the spectrum described above is listed in the column
labeled ``Evaluation''.  Separate redshifts are given for the galaxy
and \sn\ spectra, when available.  The mean value of the
\snia\ redshifts are offset from the host galaxy by
$0.0022\pm0.0004$(galaxy redshift minus \snia\ redshift).  The offset
probably arises from variations in the \sn\ template spectra that were
used to determine the \sn\ redshifts.  A similar offset ($0.003$) was
reported for the first-year sample; see \citet{zheng08} for the result
and a discussion of the offset.

The source of the redshift can generally be discerned from the size of
the uncertainty.  For redshifts measured from broad features of the
\sn\ spectrum, the uncertainty floor is set to $\delta z = 0.005$.
For redshifts measured from narrow galaxy lines, the uncertainty floor
is set to $\delta z = 0.0005$.  Redshifts measured from the \sdss\ and
\boss\ spectrographs have uncertainties set by their respective
pipelines as quoted in their catalogs.


\section{SN~Ia Sample and \saltii\ Analysis}\label{section_lcfits}

\begin{figure}[htb]
  \epsscale{1.1}
  \plotone{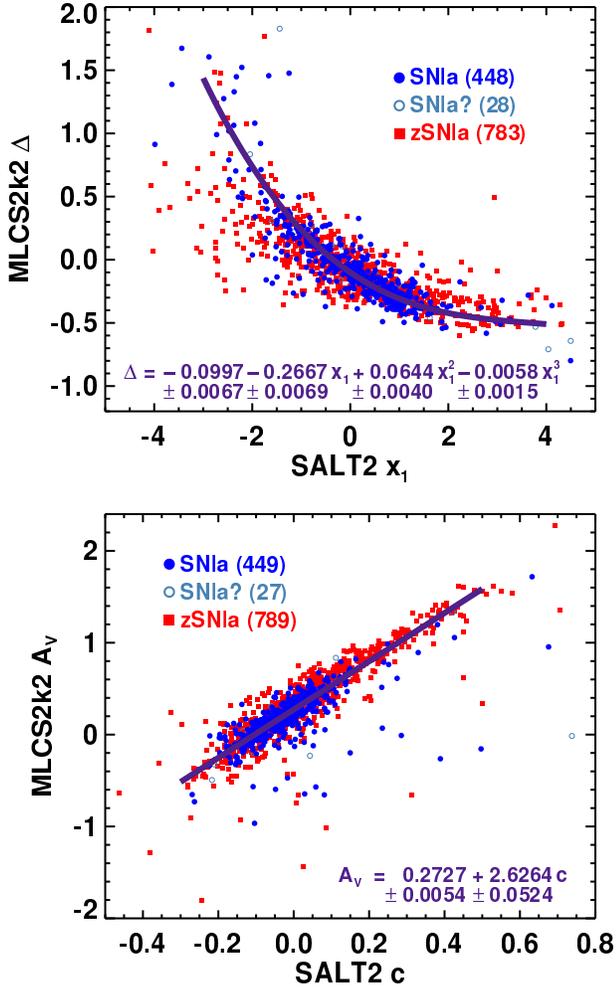}
  \caption{Comparisons of the MLCS2k2 and SALT2 light-curve fit parameters for
    SN~Ia from the spectroscopic (SNIa, SNIa?) or photometric (with host
    redshift; zSNIa) samples. The top panel shows the light curve shape
    parameters, MLCS2k2 $\Delta$ versus SALT2 $x1$, for the 1259 SN~Ia where
    these parameters were well measured. The solid purple curve is a cubic
    polynomial fit to the data, with coefficients as displayed, restricted to
    points with $-3 < x1 < 4$ and $-1 < \Delta < 2$ (1236 objects; 443 SNIa,
    26 SNIa?, 767 zSNIa). The bottom panel shows the reddening parameters,
    MLCS2k2 $A_V$ versus SALT2 $c$, for 1265 SN~Ia. The solid purple line is a
    linear regression fit using the Bayesian Gaussian mixture model of Kelly
    (2007), via the IDL routine {\tt{linmix\_err.pro}}, over the restricted
    data range $-0.3 < c < 0.5$ and $-1 < A_V < 2$ (1248 objects; 447 SNIa, 26
    SNIa?, 775 zSNIa).  For clarity in display, the uncertainties on the data
    points are not shown, but they have been included in deriving the fits.}
\label{fig:mlcscomp}
\end{figure}

We provide results from light curve fits as a reference to serve as a check
for those who wish to make their own fits using different methods or selection
criteria and for those less critical applications that can use our light curve
fits directly.  Using the \snana\ \snanaversion\ package \citep{kessler09b}
implementation of the \saltii\ \snia\ light curve model \citep{guy10}, we
determine light curve parameters for two kinds of fits.  The first uses fixed
spectroscopic redshifts (either from the \sn\ spectrum or the host galaxy),
and fits four parameters: time of peak brightness ($t_0$), color ($c$), the
shape (stretch) parameter ($x_1$), and the luminosity scale ($x_0$).  The
second fit ignores spectroscopic redshift (when known) and includes the
redshift as a fifth fitted parameter as described in \citet{kessler10a}.  For
comparison, we have also used the \mlcs\ light curve fitting method
\citep[JRK07]{jha07}, where the luminosity parameter $\Delta$ and the
extinction parameter $A_V$ play similar roles to the \saltii\ parameters $x_1$
and $c$, respectively.

To ensure reasonable fits, we applied selection criteria as summarized in
Table \ref{tbl:sncuts}.  Note that \snia\ fits are made regardless of the
\sn\ type classification.  The \snana\ input files for these fits are
available on the data release web pages \citep{sdsssndr}.

We also placed some requirements on the photometric measurements that were
used in the fit. We exclude epochs where \smp\ was judged to be unreliable (a
photometric flag\footnote{The meaning of the photometric flags is detailed in
  \citet{holtzman08}.} of 1024 or larger) and epochs earlier than 15 days or
later than 45 days (in the rest frame).  In addition, 151 epochs in 105
different \sn\ were designated outliers based on the inspection of the light
curve fits and were not used to determine the light curve parameters.  These
outlier epochs are included in both the {\small ASCII} and \snana\ data
releases, and a list of these epochs is included in the \snana\ release.
 
Some representative 4-parameter fit results are shown in Table
\ref{tbl:salt2_4par} (\saltii\ 4-parameter fits).  Similar data for
the \mlcs\ fits and \saltii\ 5-parameter fits may be found in the full
machine readable table (see Table 1).  We show a comparison of the
4-parameter \saltii\ and \mlcs\ fits in Figure \ref{fig:mlcscomp},
where \saltii\ $c$ is compared with \mlcs\ $A_V$ and \saltii\ $x_1$ is
compared with \mlcs\ $\Delta$.  There is generally a strong
correlation between the \saltii\ and \mlcs\ parameters (indicated by
the lines shown), with modest scatter and some outliers.  In
particular, the \mlcs\ $\Delta$ parameter spans a large range in the
vicinity of $x_1=-2$ (fast-declining light curves). The correlation
between the reddening parameters $A_V$ and $c$ is tighter, with just a
handful of outliers. There is also a clear color zeropoint offset
between the fitters, $c\approx-0.1$ when $A_V=0$.

\begin{figure}
 \begin{center}
  \epsscale{1.1}
  \plotone{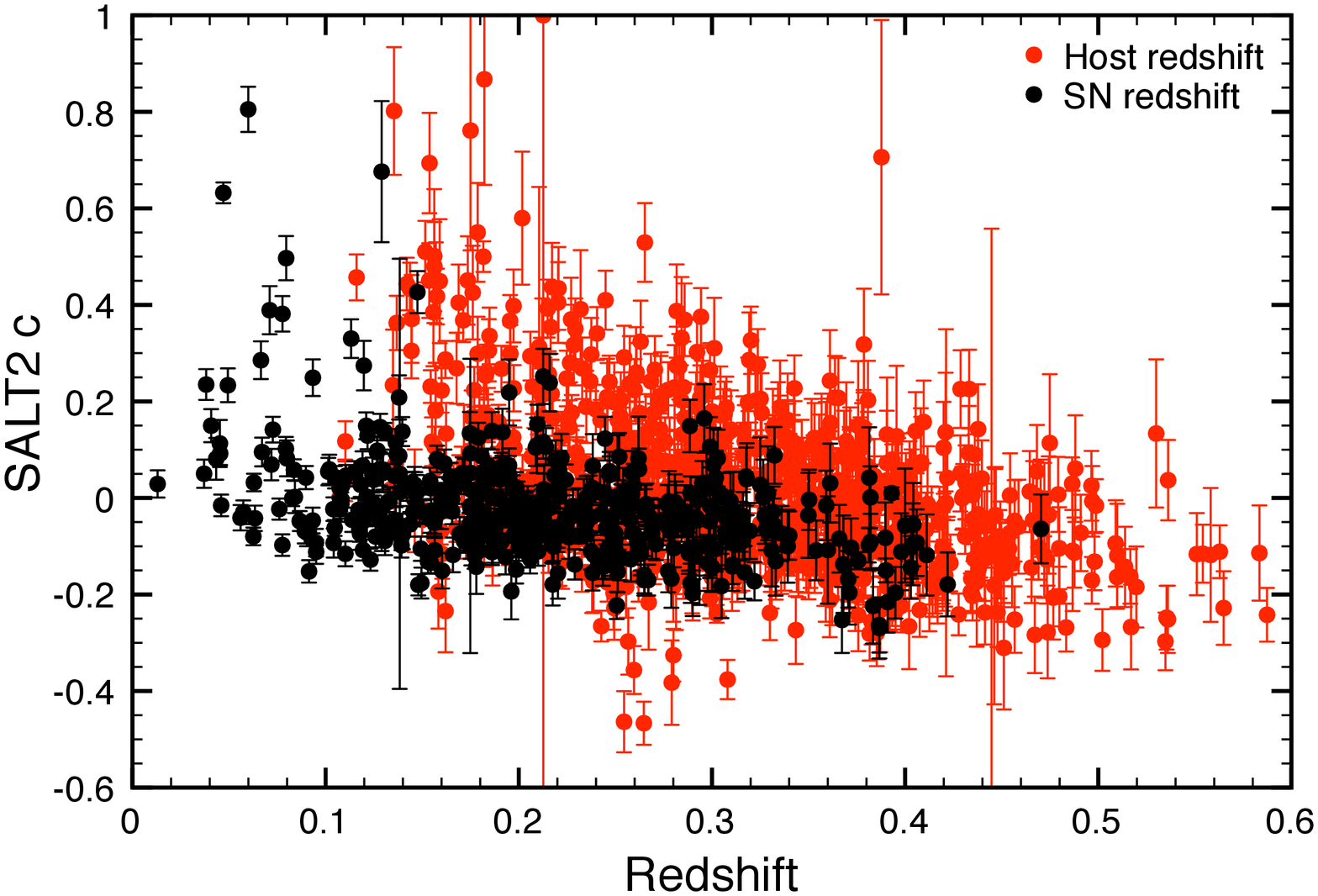}
  \plotone{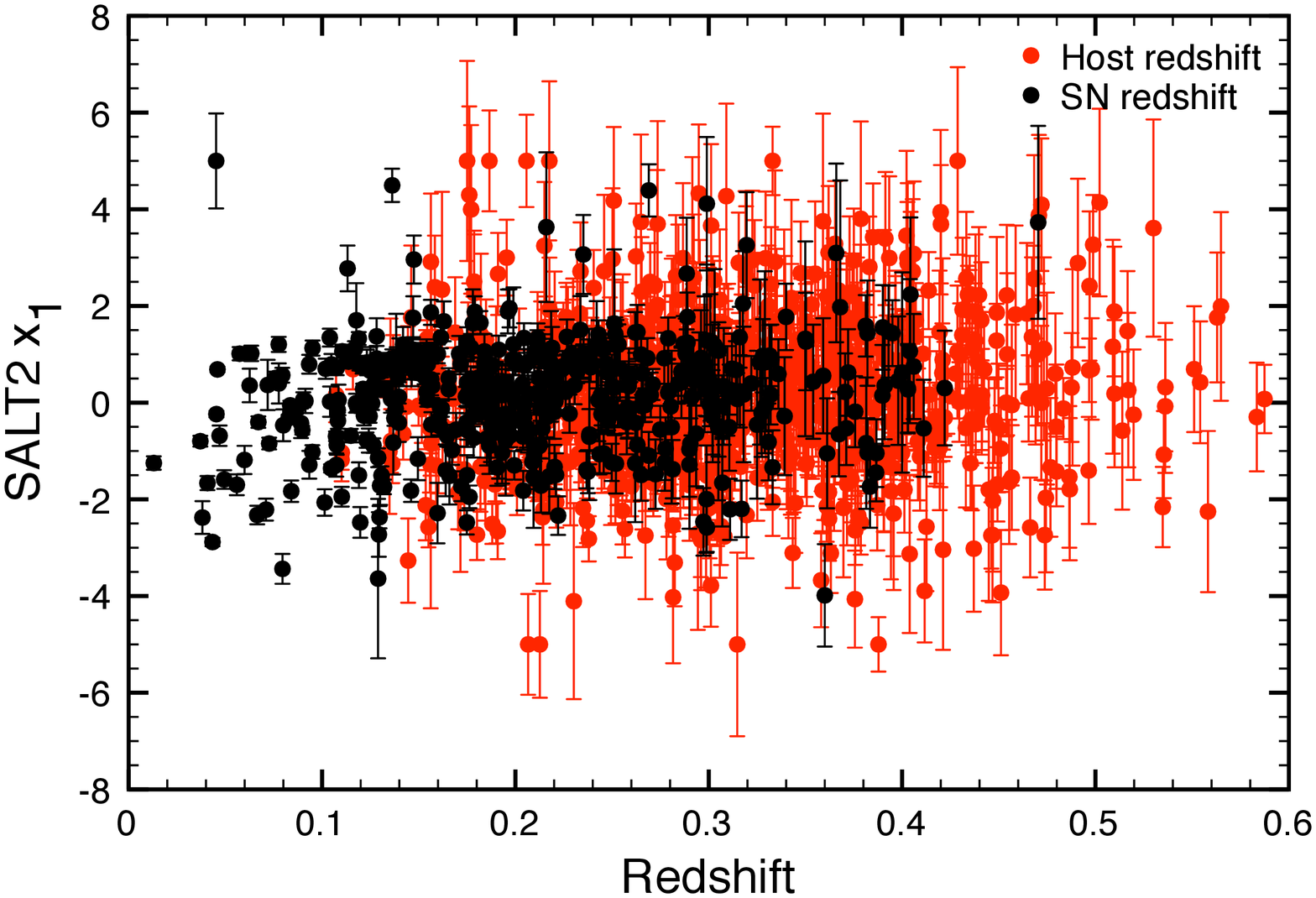}
  \end{center}
  \caption{SALT2 color (top) and $x_1$ as a function of redshift for
    the spectroscopic SN~Ia (black) and \zhost -Ia (red) samples.  The
    \zhost -Ia sample is noticeably redder (more positive values of
    $c$) than the spec-Ia sample, but the $x_1$ distributions are
    indistinguishable.}\label{fig:salt2params}
\end{figure}

We have also compared the \saltii\ parameters $x_1$ and $c$ in Figure
\ref{fig:salt2params} for the 4-parameter fit sample, showing separately the
sample where the redshift is obtained from the \sn\ spectrum as opposed to the
host galaxy spectrum.  We expect the sample with \sn\ spectra to be biased
because of the spectroscopic target selection.  Figure \ref{fig:salt2params}
shows no evidence of a bias in $x_1$ but a clear difference in $c$, which is
consistent with the findings of \citet{campbell13} where the weighted mean
\saltii\ colors of the spectroscopically-confirmed \snia\ where slightly bluer
than for the whole sample (including many photometrically--classified \snia).
This effect is presumably because reddened Type Ia \sn\ were less likely to be
selected for spectroscopy.

\subsection{Distance moduli}

We have used the results of our 4-parameter \saltii\ fits to compute the
distance modulus to the \sdssii\ \sn, excluding those events where the fit
parameter uncertainty was large ($\delta t_0>1$ or $\delta x_1>1$).  The
distance moduli are presented in the on-line version of Table
\ref{tab:fullCatalog}; a subset is displayed in Table \ref{tbl:salt2_4par}.
We used \saltii mu \citep{marriner11}, which is also part of \snana, to
compute the \saltii\ $\alpha$ and $\beta$ parameters and computed the distance
modulus according to the relationship,
\begin{equation}
\mu=-2.5\log_{10}{x_0}-M_0+\alpha x_1 - \beta c,\label{eqn:stBmag}
\end{equation}
where $\mu$ is the distance modulus, and $M_0=-29.967$ based on the average of
the input data.  We do not include a correction for host galaxy stellar masses
as discussed in \citet{lampeitl10b} and \citet{johansson13}. The results of
the fit are $\alpha=0.155\pm0.010$ and $\beta=3.17\pm0.13$.  Only the
spectroscopically confirmed \snia\ were used to determine these parameters and
the intrinsic scatter was assumed to be entirely due to variations in peak
$B$-band magnitude with no color variations \citep{marriner11}.  Including the
photometric \snia\ sample, we get $\alpha=0.187\pm0.009$ and
$\beta=2.89\pm0.09$.

\begin{figure}
  \begin{center}
  \epsscale{1.1}
  \plotone{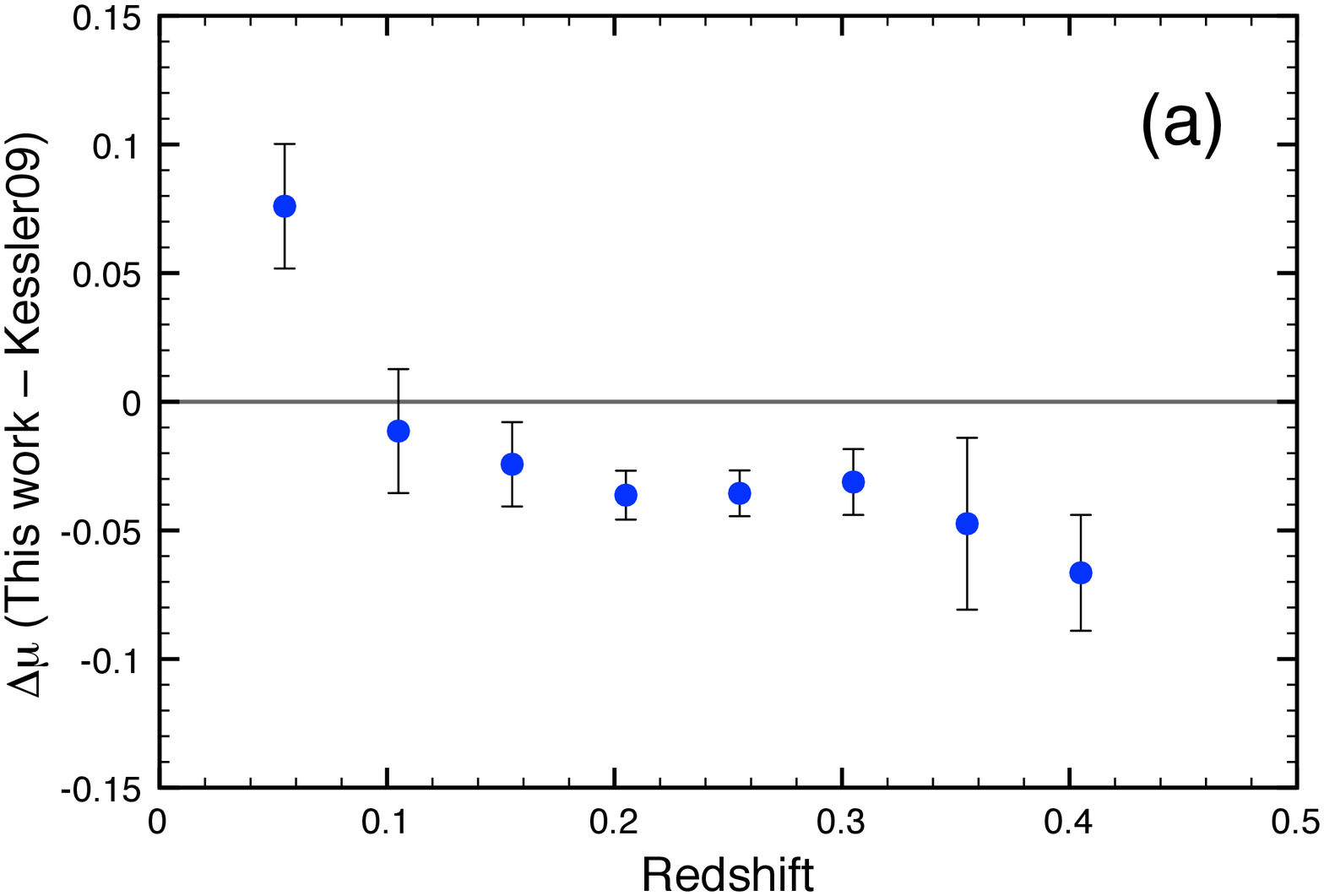}
  \plotone{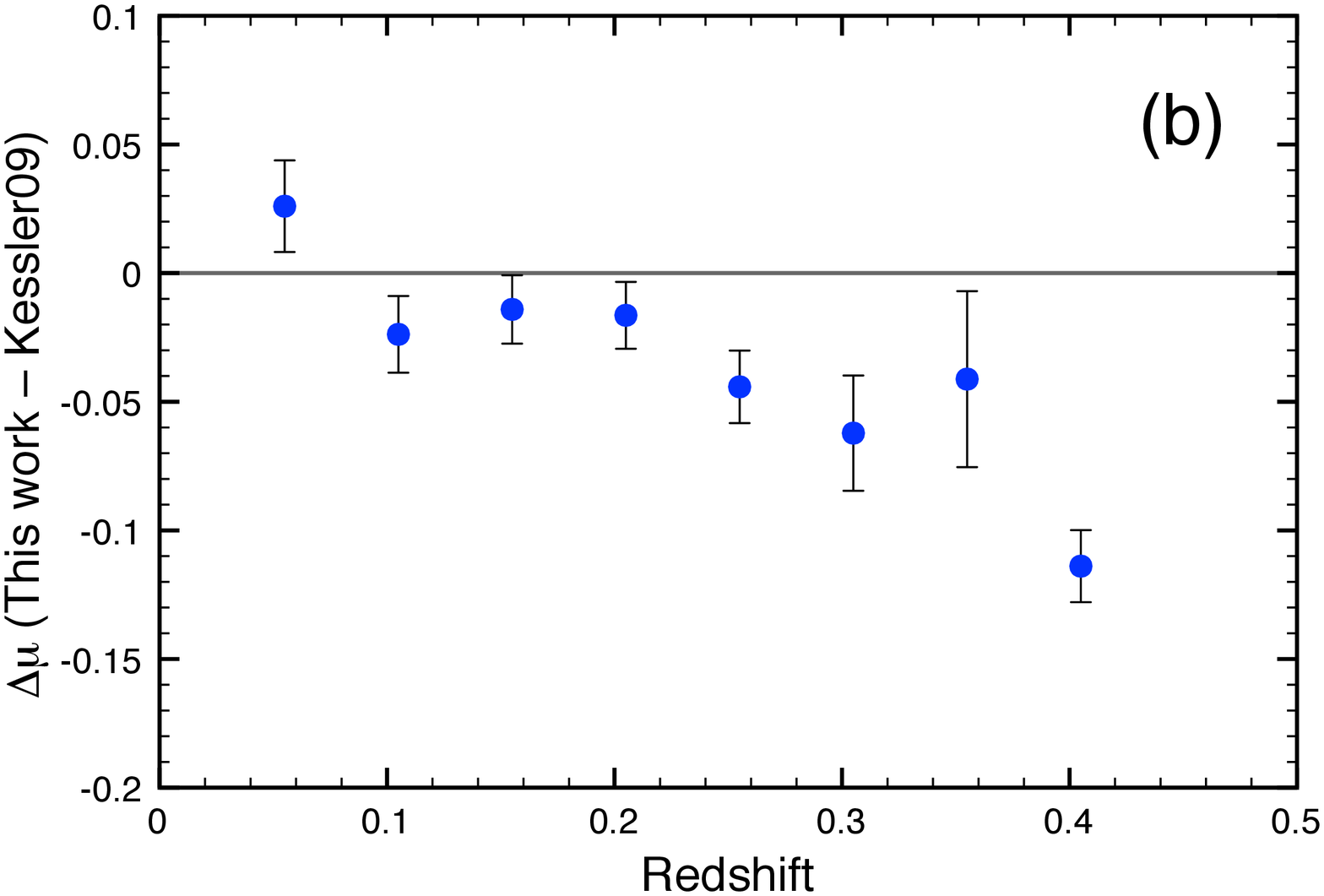}
  \caption{The differences in distance modulus between the results in
    this paper and the results published in \citet{kessler09a} for (a)
    \mlcs\ and (b) SALT2 are shown as a function of
    redshift. \label{fig:mudiff}}
  \end{center}
\end{figure}

There are subtle differences in these distance moduli compared to the light
curve fits reported for the \snia\ reported in \citet{kessler09a} and
elsewhere \citep{lampeitl10a}.  The differences arise from the following
changes:
\begin{itemize}
\item Re-calibration and updated \ab\ offsets \citep{betoule13}
\item Fitting $ugriz$ instead of $gri$
\item For \mlcs\ an approximation of host-galaxy extinction from JRK07 was
  replaced with an exact calculation (effect is negligible)
\item Updated \citep{guy10} \saltii\ model (see Section
  \S~\ref{salt_versions})
\end{itemize}
For the 103 \sn\ previously published in \citet{kessler09a}, the difference in
$\mu$ versus redshift is shown in Fig. \ref{fig:mudiff} for \mlcs\ and
\saltii.  For \mlcs\ the difference is mostly constant with redshift, except
in the lowest-redshift bin where the $u$ band has an important effect.  For
\saltii\ the difference increases with redshift.

\subsection{\saltii\ versions}\label{salt_versions}

The results of the \saltii\ fits depend on the version of the code used, the
spectral templates, and the color law.  Our fits use the \saltii\ model as
implemented in \snana\ \snanaversion\ and the spectral templates and color law
reported in \citet[G10]{guy10}.  Most of the prior work with the \sdss\ sample
used the earlier versions of the spectral templates and color law given in
\citet[G07]{guy07} with the notable exception of \citet{campbell13}, which
used G10.  For the \sdss\ data, the largest differences in the fitted
parameters arises from the difference in the color law between G07 and G10.
The \sdssii - \snls\ joint light curve analysis paper on cosmology
\citep{betoule14} releases a new version of the \saltii\ model that is based
on adding the full \sdssii\ spectroscopically confirmed \sn\ sample to the
\saltii\ training set.

\begin{figure}[bht]
  \begin{center}
    \epsscale{1.1}
    \plotone{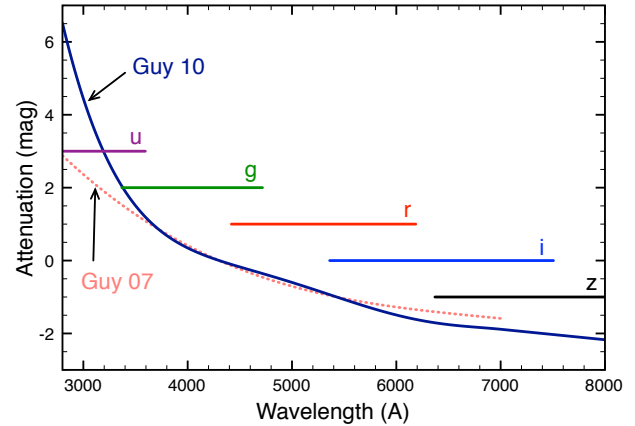}
  \end{center}
  \caption{The figure displays the color laws from \citet[][dotted
      line]{guy07} and \citet[][solid line]{guy10}. The horizontal
    lines ($ugriz$) indicate the range for the mean wavelength
    response of each filter, respectively, over the redshift range of
    $0.0 < z < 0.4$.  There is a significant difference for $i$-band
    and, at higher redshift, in $g$-band.}
    \label{fig:colorlaw}
\end{figure}

\begin{figure}
  \begin{center}
    \epsscale{1.1}
    \plotone{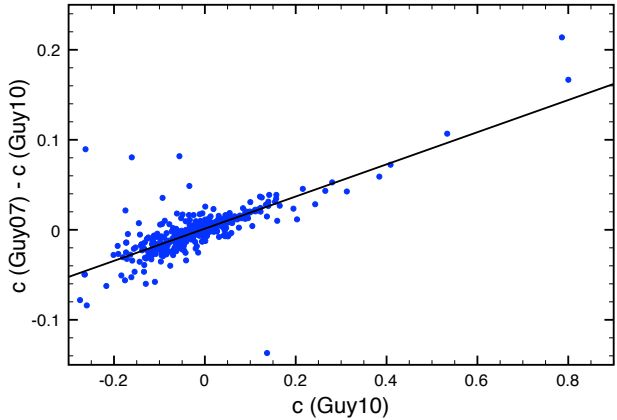}
  \end{center}
  \caption{A comparison of derived SALT2 $c$ using color laws from
    \cite{guy07} and \cite{guy10} for the spectroscopically confirmed
    SN~Ia sample.  
    straight line fit, where the errors along the horizontal axis are
    ignored and all data is given equal weight.  The result is
    $\delta_c = 0.18*c + 0.00$. }
    \label{fig:colorversion}
\end{figure}

Figure \ref{fig:colorlaw} shows the different versions of the color law and
the range of wavelengths sampled for each photometric band assuming an
\sdss\ redshift range of $0<z<0.4$.  The color laws are significantly
different, particularly at bluer wavelengths.  Figure \ref{fig:colorversion}
shows a comparison of the \sn\ fits for the \saltii\ color parameter ($c$),
where each point is a particular \sn\ with both fits using the spectral
templates from G10 but different color laws.

Although there is some scatter, the relationship between the two fits can be
described approximately by a line.
\begin{equation}
\delta c = 0.18c + 0.00
\end{equation}
We conclude that the G07 color law results in a value of the c parameter that
is ~20\% higher than G10 on average.  The effects of the differences in the
spectral templates and changes to the \snana\ code are much smaller.

\subsection{Comparison of \sdss\ u-band with model}

\begin{figure}
  \begin{center}
  \epsscale{1.1}
  \plotone{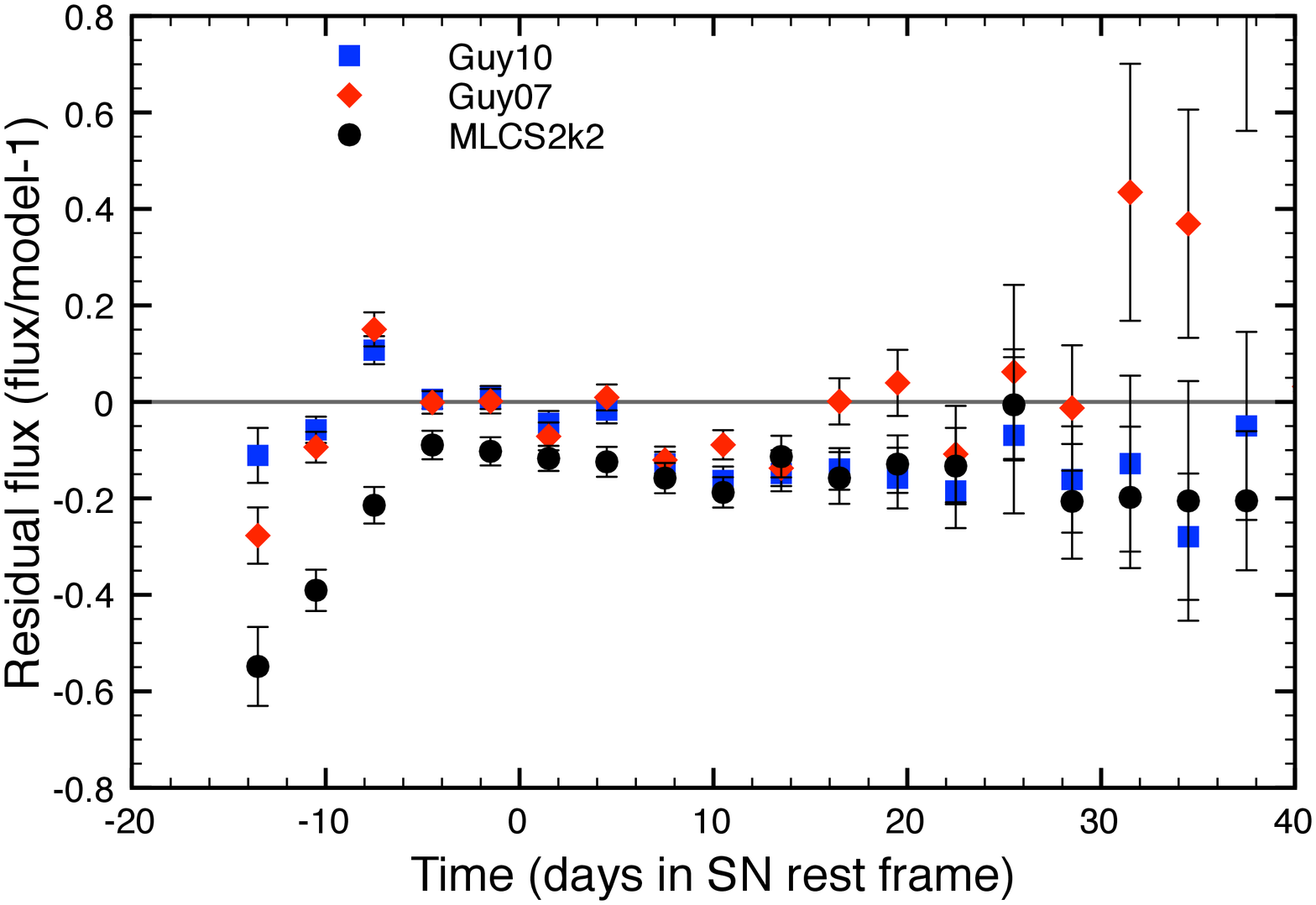}
  \plotone{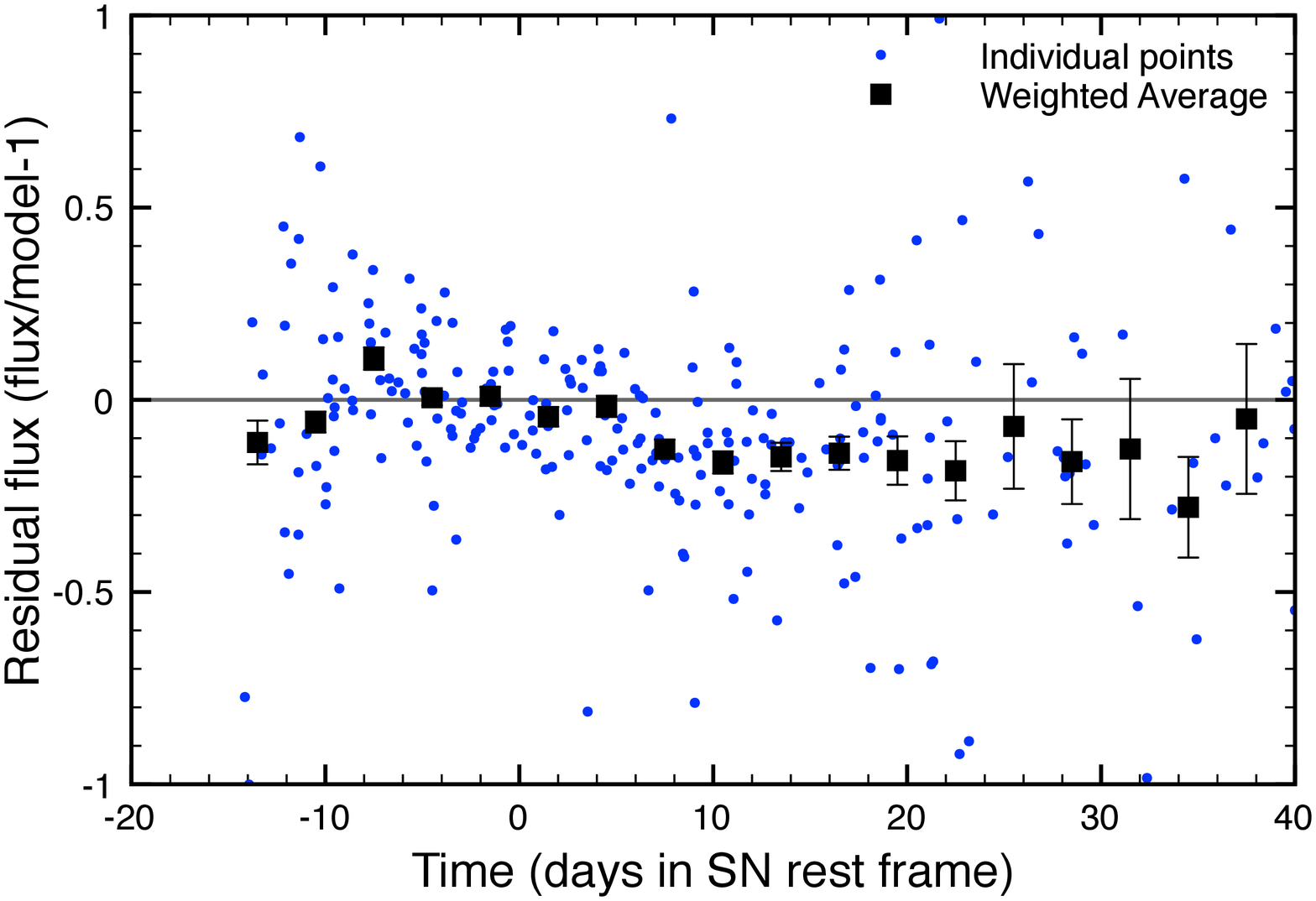}
  \end{center}
  \caption{The average $u$ band residual for the SALT2 (G07 and G10) models
    and the MLCS2k2 model are shown (top).  The measured $u$-band flux is
    compared with the prediction from the fit using $g$, $r$, and $i$ band
    data only.  The points are a weighted average of the residuals shown in
    3-day intervals measured in the SN~Ia rest frame.  The bottom panel shows
    the same weighted average for the G10 model, but also shows the individual
    points that comprise the average.\label{fig:salt_uband}}
\end{figure}

To address concerns about ultraviolet measurements, we compared our $u$-band
data with the predictions of the \saltii\ and \mlcs\ models by fitting the
$gri$ band data and comparing the measured $u$-band flux with that predicted
by the model.  The results are shown in Figure \ref{fig:salt_uband} for the
G07 model, the G10 model, and \mlcs.  All the models predict too much $u$-band
flux compared to our data at early times with the exception of the earliest
point for the G10 model.  Both the G07 and G10 models lie above \mlcs\ in
Figure \ref{fig:salt_uband}, indicating that these models predict lower flux.
We determine that the G10 model is on average $0.050\pm0.008$ magnitudes
higher than our data, G07 is $0.038\pm0.009$ and mlcs2k2 is $0.156\pm0.010$
higher.  These conclusions confirm the observations of \citet{kessler09a}
based on the first year of \sdss\ data concerning the differences between
\mlcs\ and \saltii.  The \sdss\ light curve fits are relatively insensitive to
this difference because of the poor instrumental sensitivity in the $u$-band;
it is more important for the high redshift data where an accurate rest-frame
$u$-band measurement is necessary to obtain an accurate measurement of the
color.

\subsection{Hubble Diagram and Cosmological Constraints}

\begin{figure}[bht]
  \begin{center}
    \includegraphics[angle=-90,scale=0.4]{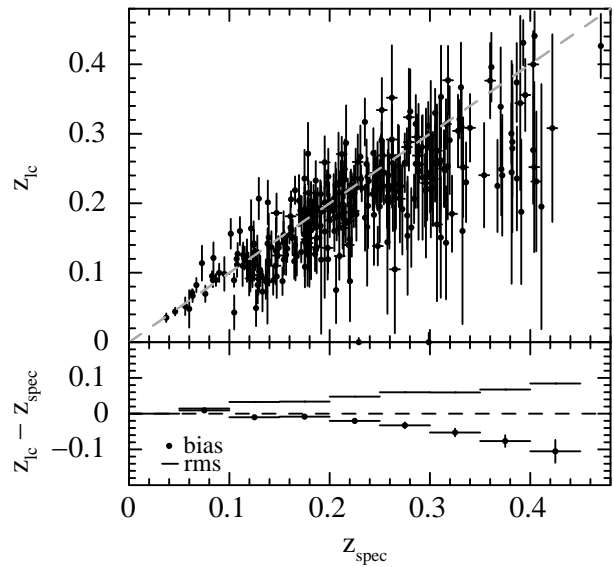}
  \end{center}
  \caption{Comparison of spectroscopic and light curve photometric
    redshifts for the confirmed SN~Ia sample.}
    \label{fig:photoz}
\end{figure}

We compare the redshift determination from the 5-parameter \saltii\ fit and
spectroscopic redshift in Figure \ref{fig:photoz}.  Good agreement between the
two redshifts is seen although the photometric error is often large.
Averaging \sn\ in bins of redshift reveals a net redshift bias of the
photometric redshifts relative to those measured spectroscopically.  The bias
has been seen previously \citep{kessler10a, campbell13}, and was shown to
agree well with the bias observed with simulated \snia\ light curves.

\begin{figure}[bht]
  \begin{center}
    \epsscale{1.1}
    \plotone{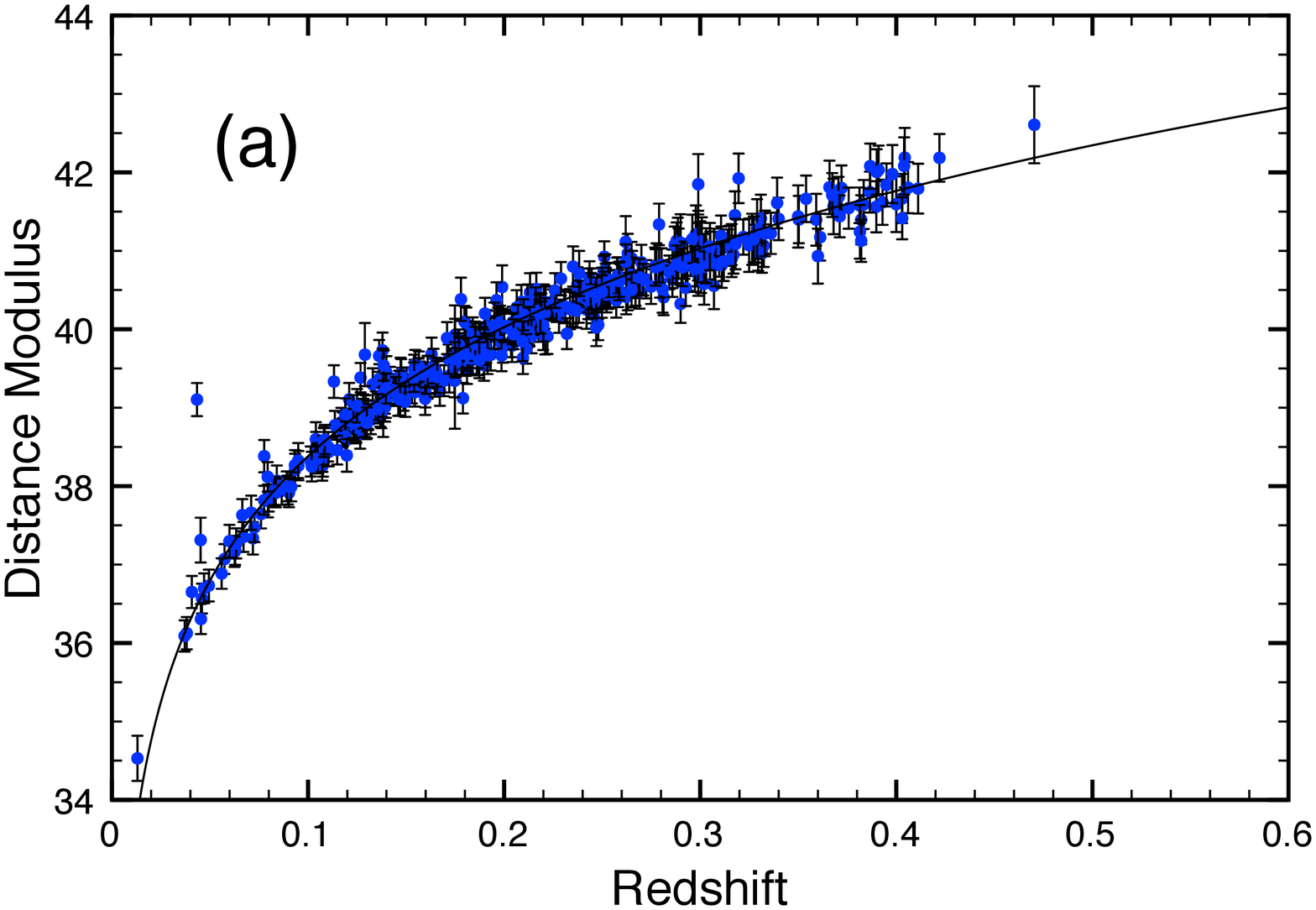}
    \plotone{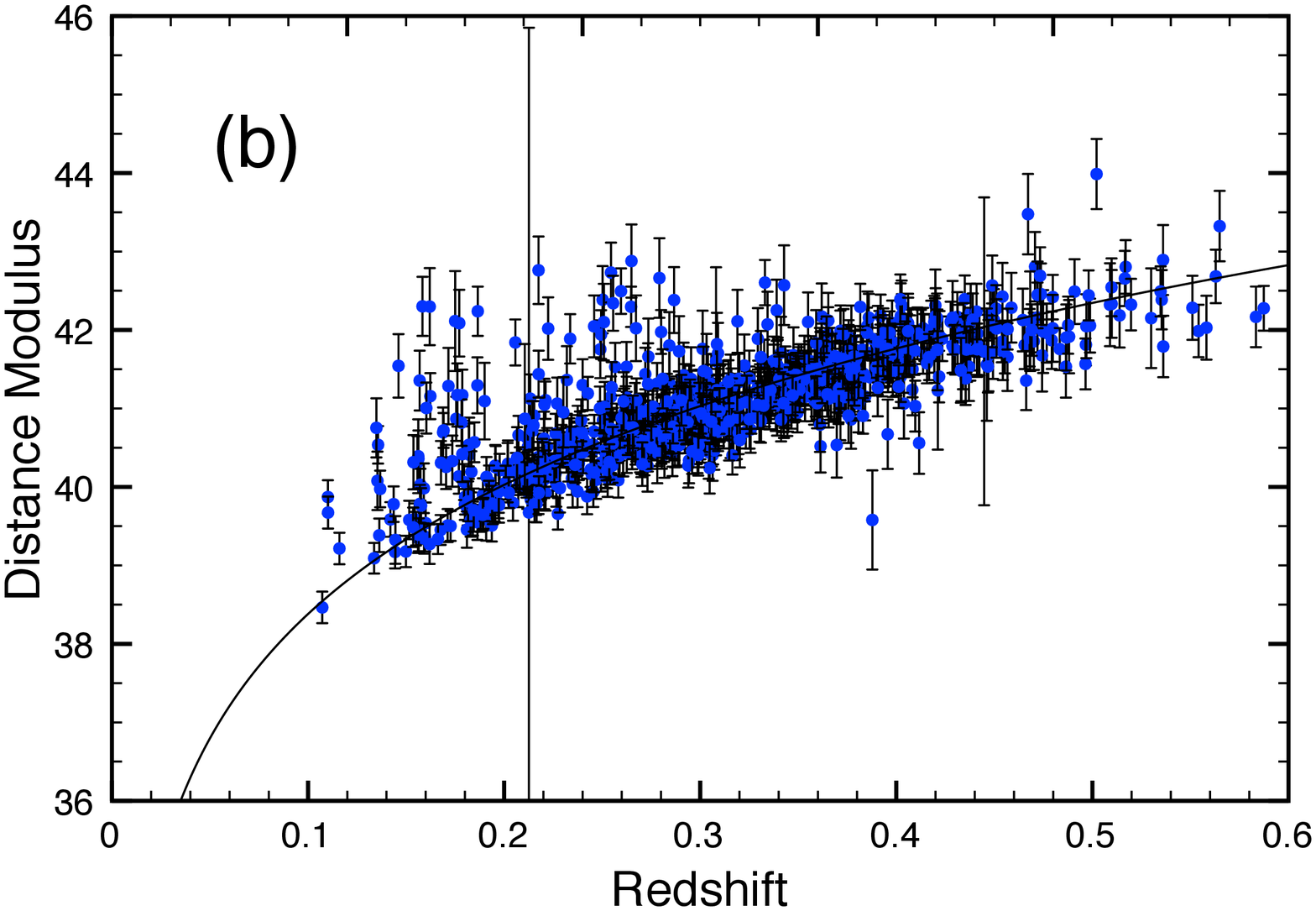}
  \end{center}
  \caption{Hubble diagram of the spectroscopic SN~Ia (top) and \zhost
    -Ia (bottom).  The large scatter in the \zhost -Ia sample
    especially at low redshift ($z < 0.2$) is most likely due to
    contamination from CC~SN.}
    \label{fig:hubble}
\end{figure}

Figure \ref{fig:hubble} shows the Hubble diagram for the \sn\ that meet our
fit selection criteria and have spectroscopic redshifts ($\delta z <0.01$):
the top panel (a) shows the 457 \sn\ that have been typed with spectra and the
bottom panel (b) shows the 827 \sn\ where the redshift is determined from the
host galaxy.  We present these plots to show the full sample and have not
attempted to make selections or to optimize the determination of cosmological
parameters.  The obvious outlier at $z=0.043$ is the under-luminous \sn
2007qd, which was discussed by \citet{mcclelland10} as a possible explosion by
pure deflagration \citep[see also,][]{foley13}.  The photometrically
identified sample (b) in Figure \ref{fig:hubble} shows a considerably larger
scatter as seen before in \citet{campbell13} and is primarily due to lower
signal-to-noise light curves being included.  Selection criteria to obtain a
sample of photometrically identified \sn\ for determination of cosmological
parameters were presented previously \citep{campbell13}.

\begin{figure}
  \begin{center}
    \epsscale{1.1}
    \plotone{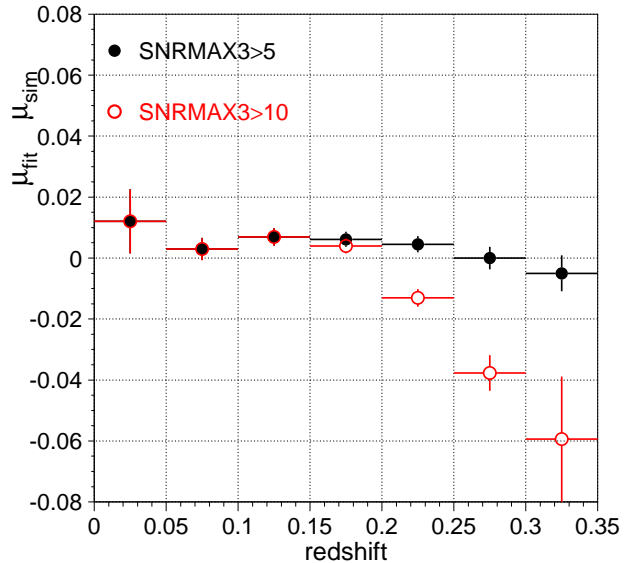}
  \end{center}
  \caption{The bias in distance modulus as a function of redshift for the SDSS
    sample for two different example selection criteria.}
   \label{fig:mubias}
\end{figure}

The Hubble diagrams shown in Figure \ref{fig:hubble} are not corrected for
biases due to selection effects.  Since the \sdss\ \sn\ survey is a magnitude
limited survey a bias towards brighter \sn\ is expected, particularly at the
higher redshifts.  Correction for bias was a particularly important effect in
the analysis of \citet{campbell13} and \citet{betoule14}, who used
photometrically identified \sn\ in addition to the spectroscopically confirmed
sample.  Figure \ref{fig:mubias} shows the bias expected from a simulation of
the \sdss\ \sn\ survey for two sample detection thresholds: requiring at least
one light curve point to be observed in each of 3 filters above background by
5$\sigma$ ({\small SNRMAX}3) and 10$\sigma$.  The expected bias for a
5$\sigma$ threshold, which is typical for \sdssii, is small but still
significant for a precise determination of cosmological parameters.  These two
different bias corrections illustrate that the correction is important and
that it depends on the selection criteria for each particular analysis.  The
\sn\ detection efficiency is discussed in more detail elsewhere
\citep{dilday10a}.

\begin{figure}
  \epsscale{1.15}
  \plotone{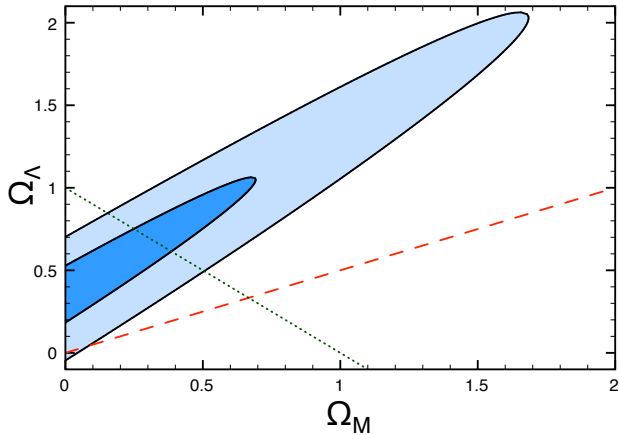}
  \caption{The 68\% and 95\% contours (statistical errors only) for the joint
    fit to $\Omega_M$ and $\Omega_{\Lambda}$ for the full three-year SDSS-II
    spectroscopic sample.  The dashed line represents $\Omega_{\Lambda} =
    \Omega_M/2$.  Assuming a flat {\small $\Lambda$CDM} cosmology, we
    determine $\Omega_M=0.315\pm0.093$.}
   \label{fig:cosmo}
\end{figure}

\begin{figure}
  \epsscale{1.1}
  \plotone{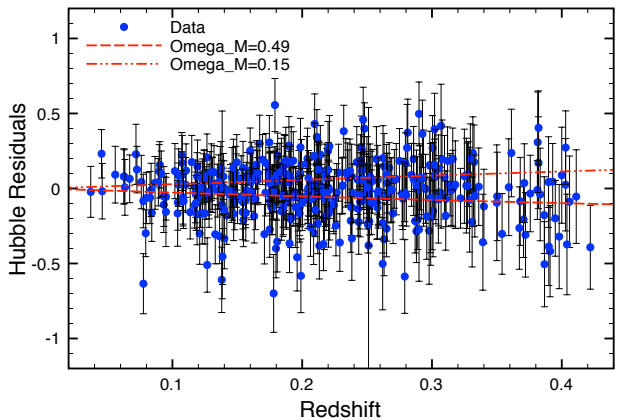}
  \caption{Residuals of Hubble diagram for the full three-year SDSS-II
    spectroscopic sample relative to a best-fit cosmology assuming a flat
    $\Lambda$CDM cosmology with $\Omega_M=0.315$.  Reference lines showing the
    expected trend for $\Omega_M$ two standard deviations higher and lower are
    also shown.}
   \label{fig:Hubresid}
\end{figure}

We present a brief cosmological analysis of our full three-year
spectroscopically--confirmed \snia\ sample in Table \ref{tbl:typeTotal} and
with \saltii\ fit parameters in the range of normal \snia\ ($-0.3<c<0.5$ and
$-2.0<x_1<2.0$).  These selections result in a sample of 413 \snia.  Assuming
a $\Lambda$CDM cosmology, we simultaneously fit $\Omega_M$ and
$\Omega_{\Lambda}$ using the {\tt sncosmo\_mcmc} module within \snana, and
show their joint constraints in Figure~\ref{fig:cosmo}.  In this analysis, we
have corrected for the expected selection biases (including Malmquist bias)
using the 5$\sigma$ threshhold curve (Figure~\ref{fig:mubias}), and have
marginalised over $H_0$ and the peak absolute magnitude of \snia, but only
show statistical errors in Figure~\ref{fig:cosmo}.  Acceleration
($\Omega_{\Lambda} > \Omega_M/2$) is detected at a confidence of 3.1$\sigma$.
If we further assume a flat geometry, then we determine
$\Omega_M=0.315\pm0.093$ and $Omega_{\Lambda} > 0$ is required at 5.7$\sigma$
confidence (statistical error only).  In Figure \ref{fig:Hubresid}, we show
the residuals of the distance moduli with respect to this best fit cosmology,
including varying $\Omega_M$ by $\pm2\sigma$ from this best fit.  Overall, our
cosmological constraints are not as competitive as higher redshift samples of
\snia\ because of the limited redshift range of our \sdssii\ \sn\ sample.
Therefore, we refer the reader to \citet{betoule14} for a more extensive
analysis of the full \sdssii\ spectroscopically--confirmed \sn\ sample
combined with other \sn\ datasets (low redshift samples, \snls, \hst) and
other cosmological measurements.

\section{Host Galaxies}\label{section_hostid}

A wealth of data on the \sn\ host galaxies is available from the \sdss\ Data
Release 8 \citep[\dr 8;][]{dr8}.  In Section \ref{HGID} we describe the
host-galaxy identification method used in this paper, which we suggest for
future analyses.  In Section \ref{HGProp} we describe the host-galaxy
properties computed from \sdss\ data and presented in Table
\ref{tab:fullCatalog}, and explain differences with values reported in
previous analyses \citep{lampeitl10b,smith12,gupta11}.

\subsection{Host Galaxy Identification\label{HGID}}

We use a more sophisticated methodology for selecting the correct \sn\ host
galaxy than trivially selecting the nearest galaxy with the smallest angular
separation to the supernova.  We instead use a technique that accounts for the
probability based on the local surface brightness similar to that used in
\citet{sullivan06}.

We begin by searching \dr 8 for primary objects within a $30\arcsec$
radius of each \sn\ candidate position and consider all the objects as
possible host galaxy candidates.  We characterize each host galaxy by
an elliptical shape.  We chose the elliptical approximation, because
the model-independent isophotal parameters were determined to be less
reliable\footnote{\url{http://www.sdss3.org/dr8/algorithms/classify.php}}
and were therefore not included in \dr 8.  The shape of the ellipse
was determined from second moments of the distribution of light in the
{\it r}-band.  The second moments are given in \dr 8 in the form of
the Stokes parameters {\tt Q} and {\tt U}, from which one can compute
the ellipticity and orientation of the ellipse.  The major axis of the
ellipse is set equal to the Petrosian half-light radius
(\sdss\ parameter {\tt PetroR50}) in the {\it r}-band; this radius
encompasses 50\% of the observed galaxy light. We found this parameter
to be a more robust representation of the galaxy size than the {\tt
  deVRad} and {\tt expRad} profile fit radii, which too often had
values that indicated a failure of the profile fit.

For each potential host galaxy, we calculate the elliptical light radius in
the direction of the \sn\ and call this the directional light radius (\dlr).
Next, we compute the ratio of the \sn -host separation to the \dlr\ and denote
this normalized distance as \ddlr.  We then order the nearby host galaxy
candidates by increasing \ddlr\ and designate the first-ranked object as the
host galaxy.  For particular objects where this fails 
(the mechanism for determining this is described later), 
due to values of {\tt Q,  U} or {\tt PetroR50} 
that are missing or poorly measured, we select the next
nearest object in \ddlr\ as the host.  In addition, we impose a cut on the
maximum allowed \ddlr\ for a nearby object to be a host.  This cut is chosen
to maximize the fraction of correct host matches while minimizing the fraction
of incorrect ones as explained below.  If there is no host galaxy candidate
meeting these criteria, we consider the candidate to be \hostless.

\begin{figure}[htb]
  \begin{center}
    \epsscale{1.0}
    \plotone{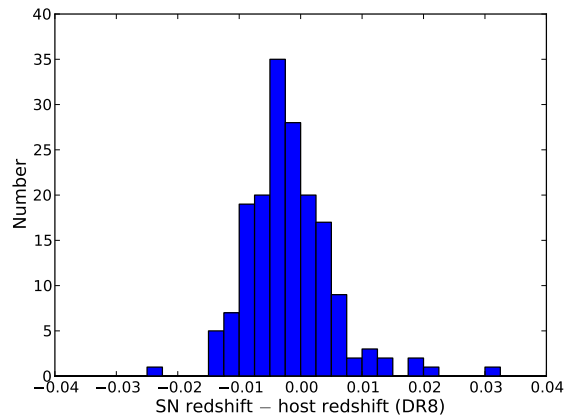}
  \end{center}
  \caption{The distribution of the difference in redshift (SN spectrum
    redshift minus host galaxy spectrum redshift) for the sample of
    spectroscopically-confirmed SNe whose hosts have redshifts in
    DR8.}
\label{fig:zDifference}
\end{figure}

\begin{figure}[tbp]
  \begin{center}
    \epsscale{1.1}
    \includegraphics[angle=-90,scale=0.32]{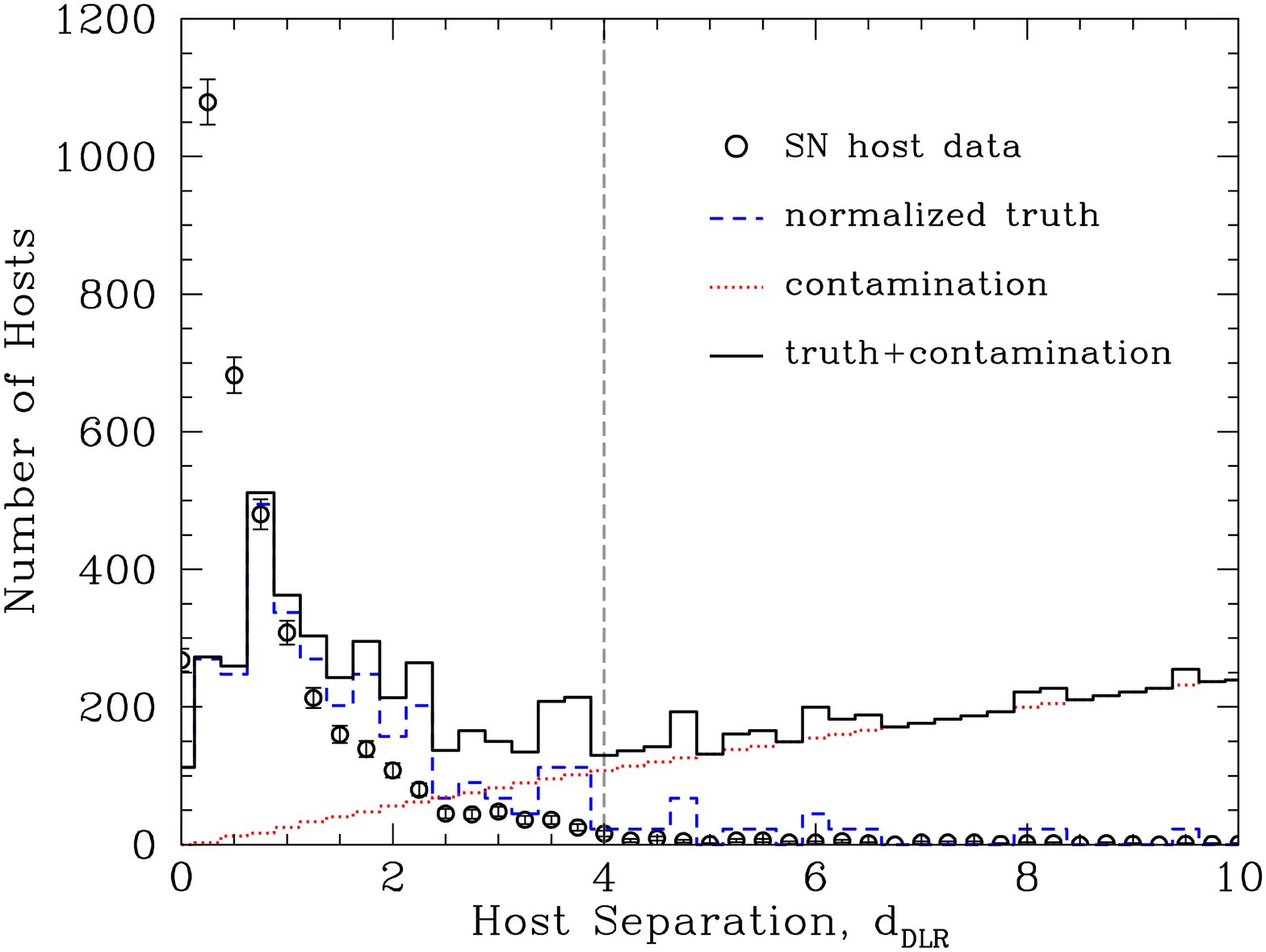}
    \includegraphics[angle=-90,scale=0.32]{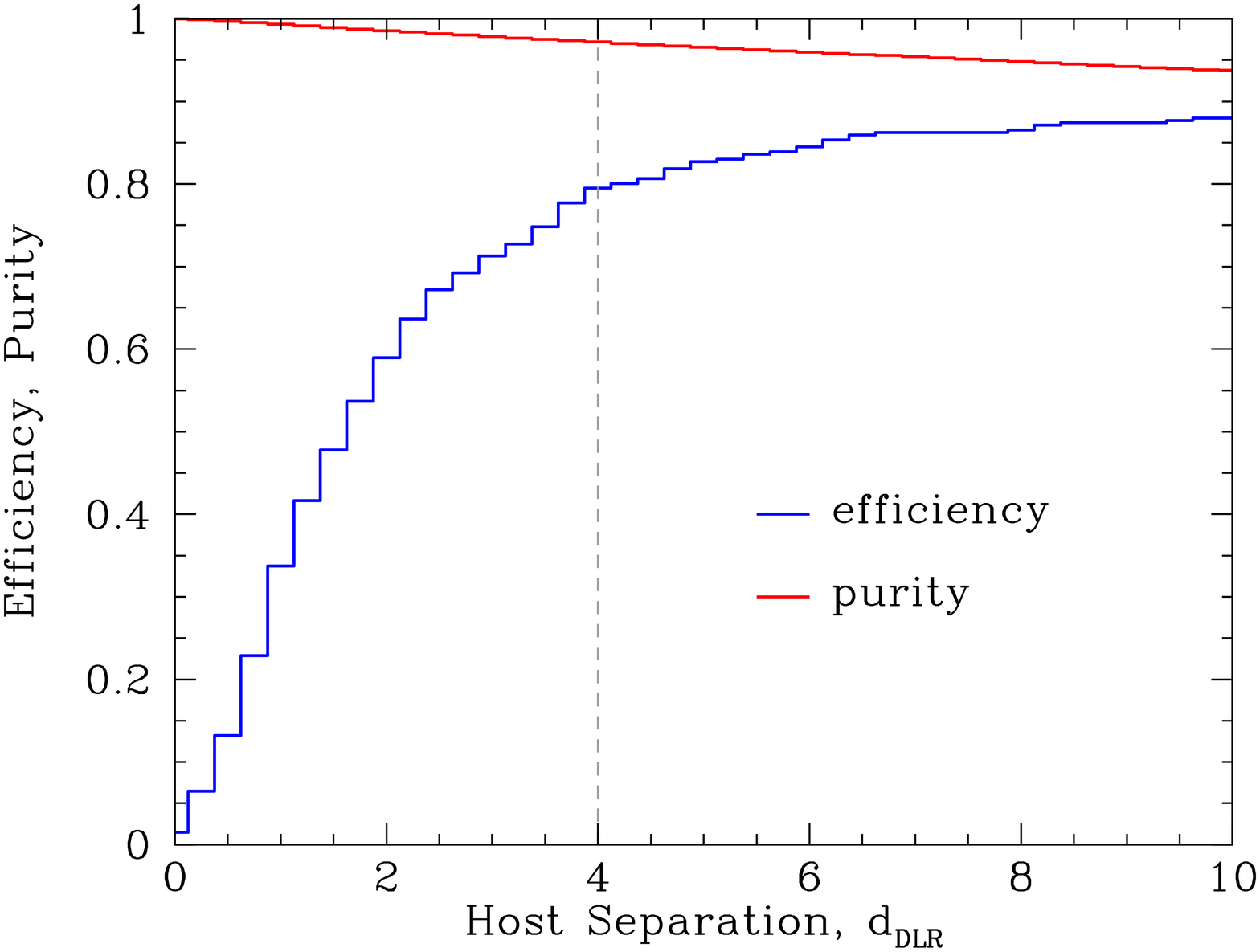}
  \end{center}
  \caption{{\it Top}: The distribution in \ddlr\ is shown for the
    truth sample (dashed blue line), the contamination of false host
    galaxy matches (dotted red line) and the sum of the two
    distributions (solid blackline).  The full sample is shown as the
    open circles with Poisson error bars and should be compared to the
    solid line.  {\it Bottom}: The efficiency and purity of the host
    galaxy selection is shown as a function of \ddlr\ and the matching
    criterion at \ddlr$=4$ indicated.}
\label{HostMatching}
\end{figure}

Determining an appropriate \ddlr\ cutoff requires that we first estimate the
efficiency of our matching algorithm.  We estimate our efficiency by selecting
a sample of positively identified host galaxies based on the agreement between
the \sn\ redshift and the redshift of the host galaxy from \sdss\ \dr 8
spectra.  We select host galaxies from our sample of several hundred
spectroscopically-confirmed \sn\ of all types via visual inspection of images.
We then consider the 172 host galaxies that have redshifts in \dr 8.  The
distribution of differences in the \sn\ redshift and host galaxy redshift for
this sample is shown in Figure \ref{fig:zDifference}.  The prominent peak at
zero and lack of outliers is proof that these \sn\ are correctly matched with
the host galaxy. The small offset between the host galaxy redshift and the
redshift obtained from the \sn\ spectrum was discussed in
\S\ref{section_spec}. Of the 172 host galaxies, 150 have a redshift agreement
of $\pm0.01$ or better, and we designate this sample of \sn -host galaxy pairs
as the ``truth sample".  We plot the distribution as a function of \ddlr\,
normalized to the data, as the dashed blue curve in Fig.~\ref{HostMatching}
(top panel).

The efficiency for the identification of the full \sdss\ sample needs to
include the \sn\ which are \hostless.  Using the sample of spectroscopic
\snia\ with $z<0.15$, the redshift below which the \sdssii\ \sn\ survey is
estimated to be 100\% efficient \citep{dilday10a} for spectroscopic
measurement, we estimated the rate of \hostless\ \sn\ under the assumption
that this low-$z$ host sample is representative of the true \snia\ host
distribution.  We obtained \sdss\ {\it ugriz} model magnitudes and errors for
the low-$z$ host sample from the \dr 8 Catalog Archive Server
(CAS)\footnote{\url{http://skyservice.pha.jhu.edu/casjobs/}} and used them and
the measured redshifts to compute the best-fit model spectral energy
distributions (\sed) using the code {\tt kcorrect v4\_2} \citep{kcorrect}. The
spectra were shifted to redshift bins of 0.05 up to $z=0.45$, and we computed
the expected apparent magnitudes of the hosts at those redshifts.  We then
weighted these magnitudes in the various $z$-bins by the redshift distribution
of the entire spectroscopic \snia\ sample to mimic the observed $r$-band
distribution for the whole redshift range.  We identified those hosts that
fell outside the \dr 8 $r$-band magnitude limit of 22.2 as \hostless.  From this
analysis, we predict a \hostless\ rate of 12\% for the \sdss\ sample.
Normalizing the truth distribution to 88\% and taking the cumulative sum gives
us an estimate of the efficiency of our matching method as a function of
\ddlr, which is shown as the blue curve in Fig.~\ref{HostMatching} (bottom
panel).

Unfortunately, we do not have spectroscopic redshifts for all candidates nor
all potential host galaxies, so we can not rely on agreement between redshifts
for the purity of the sample.  In order to estimate the rate of
misidentification, we chose a set of 10,000 random coordinates in the
\sn\ survey footprint and applied our matching algorithm using the \dr 8
catalog.  We use these random points to determine the distribution in
\ddlr\ of \sn\ candidates with unrelated galaxies.  
We realize that in reality, SNe will occur galaxies rather than randomly on the 
sky but a more sophisticated background estimate involving random galaxies 
and an assumed \ddlr\ distribution is left for future work.  
The top panel of
Fig. \ref{HostMatching} summarizes the situation: the distribution in
\ddlr\ is shown for the truth galaxies (dashed blue line), the expected
distribution of background galaxies is shown as the dotted red line, and the
solid black line is the sum of the two.  The data sample is shown as the open
circles.  While the data is similar to expectations, it is notably more peaked
at low values of \ddlr\ than the truth sample would lead us to expect.  
The difference in the distributions is partly due to the fact that the truth sample 
(being constructed from the sample of spectroscopically-confirmed SNe) is 
biased against SNe that occur near the core of their host galaxy where a 
spectroscopic confirmation is very difficult or impossible.  We therefore expect 
that many more SNe will reside at low \ddlr\ than the truth sample predicts.  
In addition, there may be difficulty in determining accurate galaxy shape 
parameters for the fainter galaxies that comprise our full sample.  
Normalizing the host distribution for the random
points and taking the cumulative sum yields the contamination rate as a
function of \ddlr.  In the bottom panel we plot the estimated sample purity
($1-$ contamination) as the red curve on the bottom panel of
Figure~\ref{HostMatching}.

We choose \ddlr$=4$ as our matching criterion in order to obtain high purity
(97\%) while still obtaining a good efficiency (80\%).  For that criterion we
find that 16\% of our \sn\ candidates are \hostless.  We expect the observed
rate of \hostless\ \sn\ to be higher than the predicted rate because of the
inefficiency of our \ddlr$<4$ selection, partly offset by candidates added by
visual scanning and an estimated contamination of incorrect matches of 2\% at
\ddlr$=4$.  While the measured rate of \hostless\ galaxies agrees fairly well
with expectations, we suspect that our efficiency is underestimated
because of the difference in \ddlr\ distributions between the truth sample and
the full sample and also the corrections made by visual scanning, which are
described below.

There are many ways in which a host galaxy can be misidentified.  If the true
host is not found (which can happen when it is too faint or near a bright star
or satellite track), it will not be selected.  Even having a matching
\sn\ redshift and host galaxy redshift does not guarantee a correct match in
the presence of galaxy groups, clusters, or mergers.  For nearby ($z \lesssim
0.05$) candidates, the \sn\ can be offset by more than $30 \arcsec$ from the
center of the host galaxy or the \sdss\ galaxy reconstruction may erroneously
detect multiple objects in a large, extended galaxy.  More distant candidates
suffer from a higher density of plausible host galaxies, which also tend to be
fainter and more point-like.

We attempted to mitigate these issues by examining a subset of all candidates
and manually correcting any obvious mistakes made by the host-matching
algorithm.  In total, only 116 host galaxies were corrected and the details
regarding their selection are given below.  First we examined the images of
several hundred of the lowest redshift candidates, since there are relatively
few of them and they can exhibit some of the issues with host matching listed
in the previous paragraph.  In addition, of the 3000 host galaxies targeted by
\boss\ \citep{campbell13,olmstead13} we found that for $\approx 350$
candidates either the \dr 8 host was not found or the host coordinates
differed from the \boss\ target coordinates by more than $1.5 \arcsec$.  We
visually inspected these $\approx 350$ cases as well.  Based on the inspection
of the images from the lowest redshift candidates and the discrepant
\boss\ targets, the \ddlr\ algorithm choice was changed for 116 host galaxies.
The majority of these (69) had no host identified because of our selection
criterion of \ddlr$<4$, but 4 had no host identified because of highly
inaccurate galaxy shape parameters that gave incorrect estimates for \ddlr.
However, we did not assign a host based on visual inspection if there was no
corresponding object in the \dr 8 catalog.  We found 36 cases where the host
selected by choosing the smallest \ddlr\ disagreed with the visual scanning
result.  Most of these were caused by improperly deblended galaxies or regions
where there were multiple candidates and visual pattern recognition proved
superior; poor estimate of the galaxy size parameters was likely a factor in
these corrections.  There are 7 candidates that were changed to be
\hostless\ because the \ddlr$<4$ candidate was a foreground star, a spurious
source, or a host galaxy with an incompatible spectroscopically measured
redshift.


\subsection{Host Galaxy Properties\label{HGProp}}

Much can be learned about \sn\ through the properties of their host galaxies.
We can derive several such properties by fitting host galaxy photometry to
galaxy spectral energy distribution (\sed) models.  We begin by retrieving the
\sdss\ {\it ugriz} model magnitudes (which yield the most accurate galaxy
colors) and their errors from \dr 8 for the \sn\ host galaxy sample.  For all
\sn\ host galaxies with a spectroscopic redshift from either the host or the
\sn\ we use the redshifts, host magnitudes, and magnitude errors in
conjunction with stellar population synthesis (SPS) codes to estimate physical
properties of our hosts such as stellar mass, star-formation rate, and average
age.  In this work, we obtain these properties using two different methods
used by \citet{gupta11} and \citet{smith12}, respectively.  The former method
utilizes \sed\ models from the code Flexible Stellar Population Synthesis
\citep[\fsps;][]{conroy09, conroy10} while the latter utilizes \sed\ models
from the code \PEGASE\ \citep{fioc97, leborgne04}.  The current results,
however, are not identical to the previously published results in that
\sdss\ \dr 8 photometry is now used while magnitudes from the \sdss\ co-add
catalog \citep{annis11} were used previously, and \citet{gupta11} augmented
\sdss\ photometry with UV and near-IR data.  While the co-add catalog is
certainly deeper, it is more prone to problems like artifacts or galactic
substructures being detected as objects.  The previous works cited above used
relatively small \sn\ samples and identified host galaxies by visual
inspection, while in this paper we must rely on our automated algorithm which
would fail on such problematic cases in the co-add catalogs.

In Table \ref{tbl:fsps}, we display the host properties calculated
using \fsps\ for a few \sn\ candidates.  Table \ref{tab:fullCatalog}
contains the full sample and also calculations using \PEGASE\ of the
analogous quantities using the same photometric data.  The galaxy
stellar mass ($\log(M)$, where $M$ is expressed in units of $M_\odot$)
is shown in Table \ref{tbl:fsps} with its uncertainty while the same
information is presented as a range ($\log(M_{\mathrm{lo}}$) and
$\log(M_{\mathrm{hi}}$) in Table \ref{tab:fullCatalog}).  All the
calculated parameters are presented in the same way: Table
\ref{tbl:fsps} shows the uncertainties while Table
\ref{tab:fullCatalog} gives the range.  Table \ref{tbl:fsps} also
shows the logarithm of the specific star-forming rate
$\log(\textrm{sSFR}$), where \ssfr\ is the mass of stars formed in
$M_\odot$ per year per galaxy stellar mass) averaged over the most
recent 250 Myr.  The mass-weighted average age of the galaxy is also
given in units of Gyr.  We give analogous quantities for \PEGASE\ in
Table \ref{tab:fullCatalog} except that we give the logarithm of the
star-forming rate (\textit{i.e.}, not normalized to the galaxy stellar
mass) and age is the age of the best fitting template (in Gyr) since
these are the natural, more fundamental outputs of the \PEGASE\ code.

\begin{figure}[tbp]
  \begin{center}
    \epsscale{1.1}
    \plotone{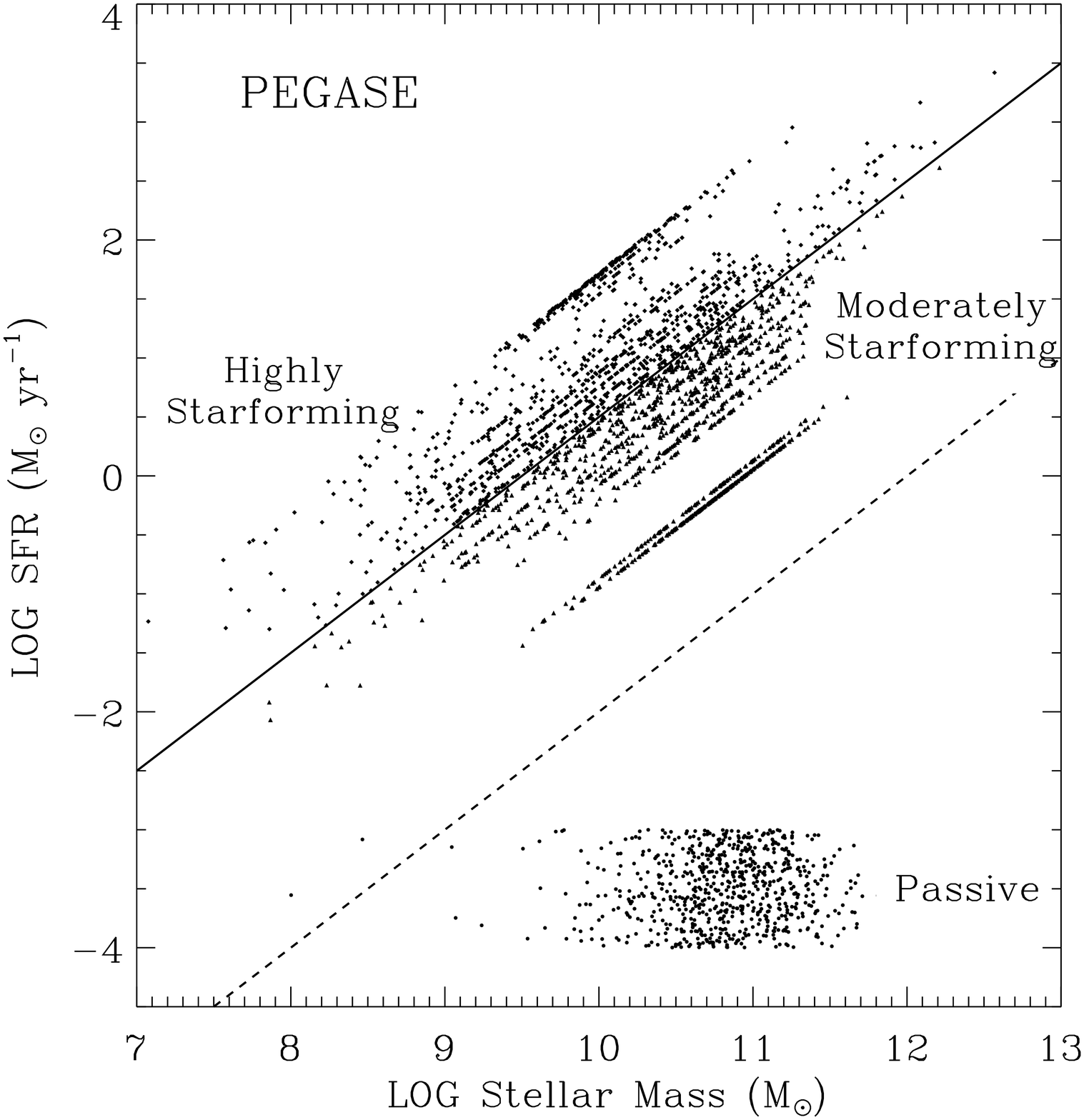}
    \plotone{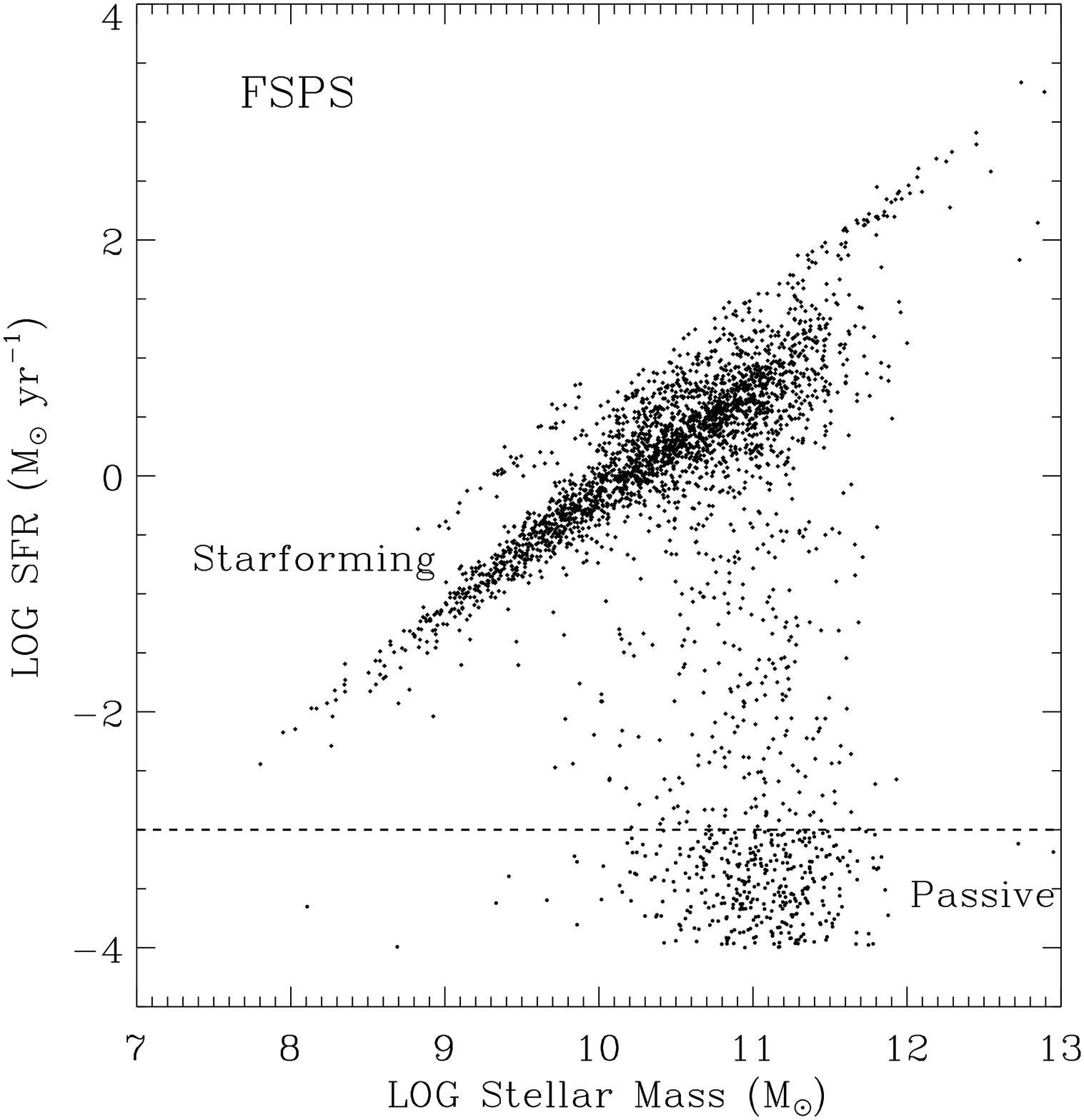}
  \end{center}
  \caption{ The distribution stellar mass and star-formation rate for
    the SN candidate host galaxies with a spectroscopic redshift for
    the P\'EGASE.2 analysis \citep[][top panel]{smith12} and the FSPS
    analysis \citep[][bottom panel]{gupta11}.  Lines of constant
    specific star formation rate separate the regions of high and
    moderate star formation (top panel) and the separation between
    star-forming and passive galaxies is shown with dashed lines in
    each panel.  For P\'EGASE.2, galaxies with a $\log(\textrm{SFR}) <
    -3$ are considered to be underflow and are displayed with an
    artificial $\log(\textrm{SFR})$, which is number randomly chosen
    between $-4$ and $-3$.}
\label{HostProps}
\end{figure}

Figure~\ref{HostProps} shows the distribution of galaxies as a
function of logarithm of galaxy stellar mass versus the logarithm of
star-forming rate for \PEGASE\ (top) and \fsps\ (bottom).  The two
distributions are similar overall but there are significant
differences as well.  In the analysis of \cite{smith12}, galaxies are
split into groups based on their \ssfr: highly star-forming galaxies
have $\log(\textrm{sSFR}) \ge -9.5$, moderately star-forming galaxies
have $-12.0 < \log(\textrm{sSFR}) < -9.5$, and passive galaxies have
$\log(\textrm{sSFR}) \le -12.0$.  Galaxies classified as passive by
\PEGASE\ were assigned random $\log(\textrm{SFR})$ values between $-4$
and $-3$ for plotting purposes.  Additionally, as noted in
\cite{smith12}, a population of galaxies with
$\log(\textrm{sSFR})\sim-10.6$ is present; these galaxies lie on a
boundary of the \PEGASE\ templates between star-forming and completely
passive galaxies.  The \fsps\ calculations do not provide such a clear
distinction between passive and star-forming, so we somewhat
arbitrarily define passive galaxies as those with $\log(\textrm{SFR})
< -3.0$.  For both analyses we see that, with a few exceptions, the
most massive galaxies are classified as passive compared to the less
massive galaxies which are classified as star-forming.

\begin{figure}[tbp]
  \begin{center}
    \epsscale{1.1}
    \plotone{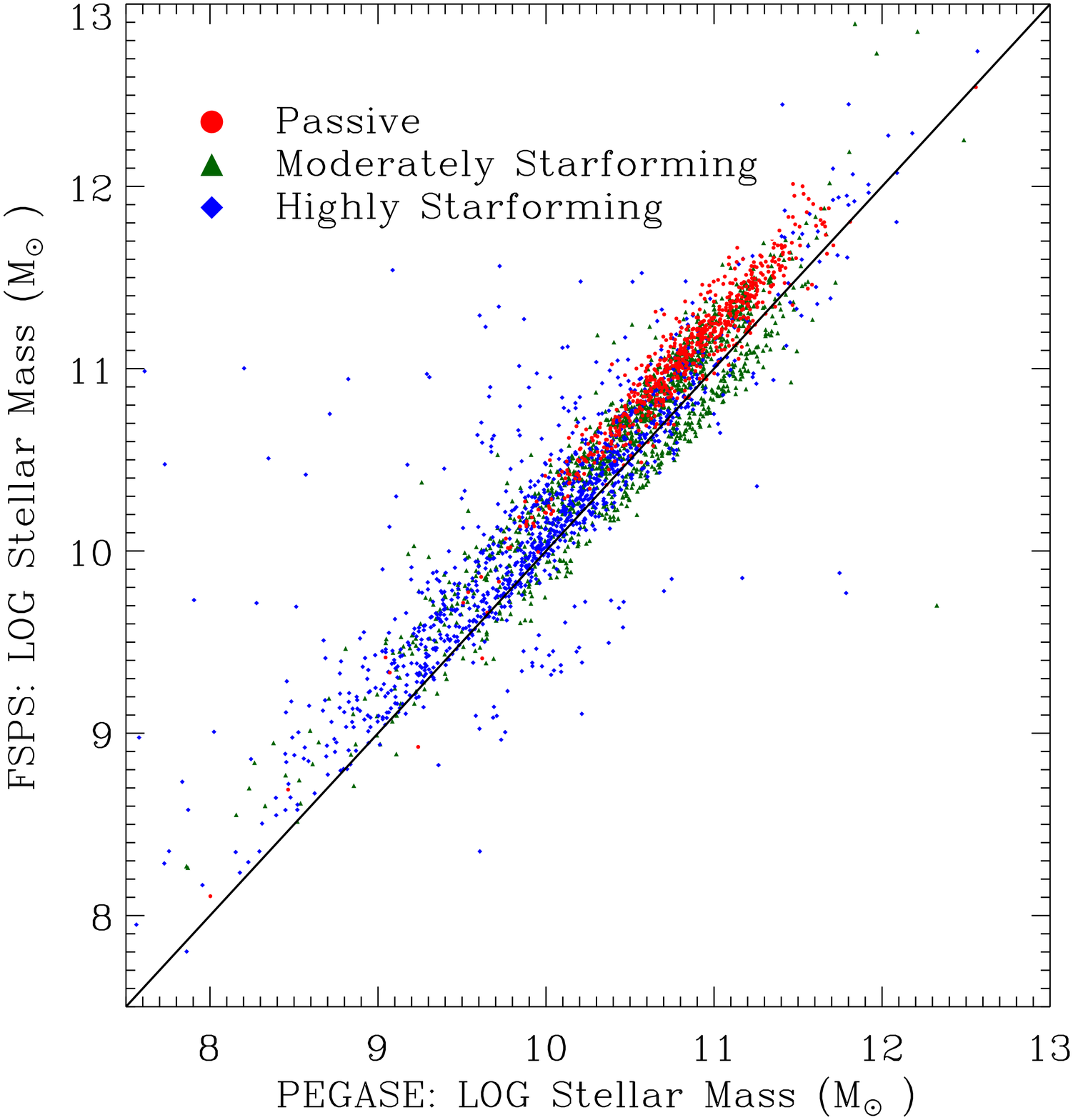}
    \plotone{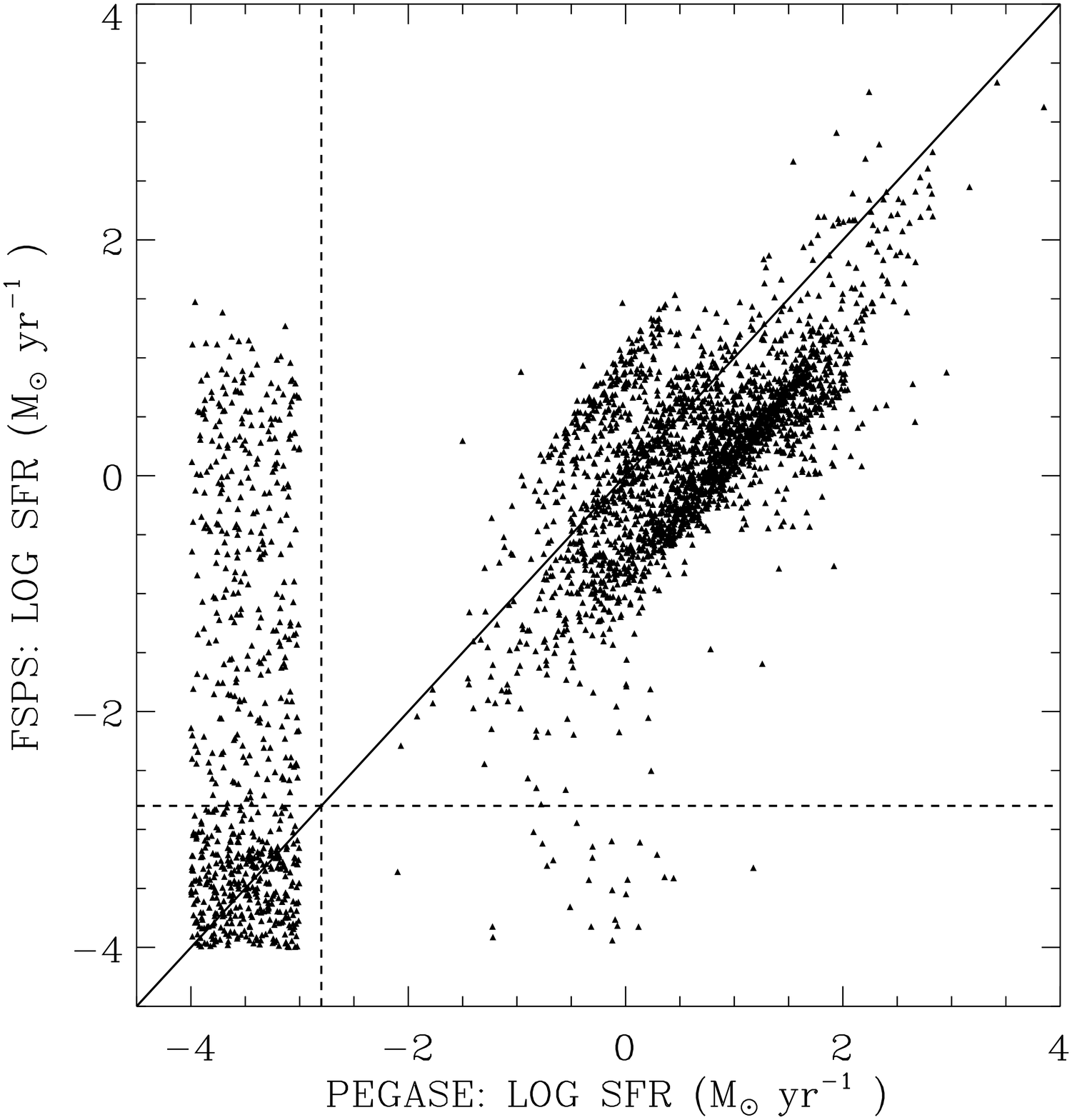}
  \end{center}
\caption{The comparison between P\'EGASE.2 and FSPS galaxy stellar masses and
  SFRs for the the SN candidate host galaxies with a spectroscopic redshift.
  As in Figure~\ref{HostProps} for P\'EGASE.2, galaxies with
  $\log(\textrm{SFR}) < -3$ are considered to be underflow and are displayed
  with an artificial $\log(\textrm{SFR})$ value randomly chosen between $-4$
  and $-3$.}
\label{HostComp}
\end{figure}

Figures~\ref{HostComp} (top) compares the stellar mass calculated with
\PEGASE\ and \fsps.  Galaxies are split according to the \ssfr\ scheme
described above, with red circles indicating passive, green triangles
indicating moderately star-forming, and blue diamonds indicating highly
star-forming.  The mass estimates show good agreement, with the stellar mass
estimated from \fsps\ being marginally higher than that estimated from the
\PEGASE\ templates.  Figure~\ref{HostComp} (bottom) compares the
\sfr\ estimated by both methods.  We find that 68\% (24\%) of galaxies are
found to be star-forming (passive), respectively, by both analyses, and 6\%
are found to be passive by \PEGASE\ and star-forming by \fsps\ and 2\% vice
versa.  In general, the \sfr\ show good agreement between the two methods,
with a larger scatter than that observed for the mass estimates.  For galaxies
classified as star-forming, the \sfr\ estimated by \PEGASE\ are systematically
higher than those estimated by \fsps.  The differences in derived galaxy
properties are likely due to the differences in the available \sed\ templates
and how they are parametrized in \fsps\ compared with \PEGASE.

\section{Summary}\label{section_summary}

This paper represents the final Data Release of the \sdssii\ \sn\ Survey of
\ncand\ candidates.  A new method of classification based on the light curve
data has been presented and applied to the candidates.  Reference light curve
fits are provided for \saltii\ and \mlcs.  A new method to associate
\sn\ observations with their host galaxies was presented, including a
quantitative estimate of efficiency and false-positive association.  Host
galaxy properties were computed from the photometric data using two computer
programs: \PEGASE\ and \fsps.  A table listing the \nspec\ spectra that were
obtained in conjunction with the \sdss\ \sn\ search was presented.  A web page
reference to the complete light curve data and reduced spectra was given.  A
complete set of photometric data for all the \sdss\ \sn\ candidates has been
presented and is released on the \sdss\ \sn\ data release web page.  All the
spectra taken in conjunction with the \sdssii\ \sn\ survey are also released.
In addition, we have provided light curve fits and host galaxy identifications
and estimated host galaxy parameters.

\clearpage

  \acknowledgements Funding for the \sdss\ and \sdssii\ has been provided by
  the Alfred P. Sloan Foundation, the Participating Institutions, the National
  Science Foundation, the U.S. Department of Energy, the National Aeronautics
  and Space Administration, the Japanese Monbukagakusho, the Max Planck
  Society, and the Higher Education Funding Council for England. The
  \sdss\ Web Site is \verb9http://www.sdss.org/9.

  The \sdss\ is managed by the Astrophysical Research Consortium for the
  Participating Institutions. The Participating Institutions are the American
  Museum of Natural History, Astrophysical Institute Potsdam, University of
  Basel, Cambridge University, Case Western Reserve University, University of
  Chicago, Drexel University, Fermilab, the Institute for Advanced Study, the
  Japan Participation Group, Johns Hopkins University, the Joint Institute for
  Nuclear Astrophysics, the Kavli Institute for Particle Astrophysics and
  Cosmology, the Korean Scientist Group, the Chinese Academy of Sciences
  ({\small LAMOST}), Los Alamos National Laboratory, the Max-Planck-Institute
  for Astronomy ({\small MPIA}), the Max-Planck-Institute for Astrophysics
  ({\small MPA}), New Mexico State University, Ohio State University,
  University of Pittsburgh, University of Portsmouth, Princeton University,
  the United States Naval Observatory, and the University of Washington.

  The Hobby-Eberly Telescope (\het) is a joint project of the University of
  Texas at Austin, the Pennsylvania State University, Stanford University,
  Ludwig-Maximillians-Universit\"at M\"unchen, and Georg-August-Universit\"at
  G\"ottingen.  The \het\ is named in honor of its principal benefactors,
  William P. Hobby and Robert E. Eberly.  The Marcario Low-Resolution
  Spectrograph is named for Mike Marcario of High Lonesome Optics, who
  fabricated several optics for the instrument but died before its completion;
  it is a joint project of the Hobby-Eberly Telescope partnership and the
  Instituto de Astronom\'{\i}a de la Universidad Nacional Aut\'onoma de
  M\'exico.  The Apache Point Observatory 3.5-meter telescope is owned and
  operated by the Astrophysical Research Consortium.  We thank the observatory
  director, Suzanne Hawley, and site manager, Bruce Gillespie, for their
  support of this project.  The Subaru Telescope is operated by the National
  Astronomical Observatory of Japan.  The William Herschel Telescope is
  operated by the Isaac Newton Group, and the Nordic Optical Telescope is
  operated jointly by Denmark, Finland, Iceland, Norway, and Sweden, both on
  the island of La Palma in the Spanish Observatorio del Roque de los
  Muchachos of the Instituto de Astrofisica de Canarias.  Observations at the
  {\small ESO} New Technology Telescope at La Silla Observatory were made
  under programme {\small ID}s 77.A-0437, 78.A-0325, and 79.A-0715.  Kitt Peak
  National Observatory, National Optical Astronomy Observatory, is operated by
  the Association of Universities for Research in Astronomy, Inc. ({\small
    AURA}) under cooperative agreement with the National Science Foundation.
  The {\small WIYN} Observatory is a joint facility of the University of
  Wisconsin-Madison, Indiana University, Yale University, and the National
  Optical Astronomy Observatories.  The W.M.\ Keck Observatory is operated as
  a scientific partnership among the California Institute of Technology, the
  University of California, and the National Aeronautics and Space
  Administration.  The Observatory was made possible by the generous financial
  support of the W.M.\ Keck Foundation.  The South African Large Telescope of
  the South African Astronomical Observatory is operated by a partnership
  between the National Research Foundation of South Africa, Nicolaus
  Copernicus Astronomical Center of the Polish Academy of Sciences, the
  Hobby-Eberly Telescope Board, Rutgers University, Georg-August-Universit\"at
  G\"ottingen, University of Wisconsin-Madison, University of Canterbury,
  University of North Carolina-Chapel Hill, Dartmough College, Carnegie Mellon
  University, and the United Kingdom \salt\ consortium.  The Telescopio
  Nazionale Galileo ({\small TNG}) is operated by the Fundaci\'on Galileo
  Galilei of the Italian {\small INAF} (Istituo Nazionale di Astrofisica) on
  the island of La Palma in the Spanish Observatorio del Roque de los
  Muchachos of the Instituto de Astrof\'{\i}sica de Canarias.

  A.~V.~Filippenko has received generous financial assistance from the
  Christopher R. Redlich Fund, the {\small TABASGO} Foundation, and
  {\small NSF} grant {\small AST}-1211916.  Supernova research at
  Rutgers University is supported in part by {\small NSF CAREER} award
  {\small AST}-0847157 to S.~W.~Jha.  G.~Leloudas is supported by the
  Swedish Research Council through grant No.\ 623-2011-7117.
  M.~D.~Stritzinger gratefully acknowledges generous support provided
  by the Danish Agency for Science and Technology and Innovation
  realized through a Sapere Aude Level 2 grant.



\LongTables
\begin{deluxetable}{rllp{3.5 in}}

  \tabletypesize{\footnotesize}
  \tablewidth{0pt}
  \tablecaption{SDSS SN Catalog\tablenotemark{a}\label{tab:fullCatalog}}

  \tablehead{
  \colhead{Item} &
    \colhead{Format} &
    \colhead{Symbol} &
    \colhead{Description (units)} 
  }

\startdata
  1 & I5     &     CID               &        \sdss\ Candidate Identification Number \\
  2 & F12.6  &     RA                &        SN Right ascension (J2000, degrees) \\
  3 & F11.6  &     DEC               &        SN Declination (J2000, degrees)  \\
  4 & I5     &     Nsearchepoch      &        Number of search detection epochs \\
  5 & A13    &     IAUName           &        Name assigned by the International Astronomical Union   \\
  6 & A11    &     Classification    &        Candidate PSNID type (see Table \ref{tbl:typeTotal})      \\
  7 & F6.1   &     Peakrmag          &        Measured peak asinh magnitude ($r$-band)         \\
  8 & F10.1  &     MJDatPeakrmag     &        Modified Julian Date (MJD) of observed peak brightness ($r$-band) \\
  9 & I5     &     NepochSNR5        &        Number of epochs with $S/N>5$          \\
 10 & I5     &     nSNspec           &        Number of SN spectra             \\
 11 & I5     &     nGALspec          &        Number of host galaxy spectra           \\
 12 & F10.6  &     zspecHelio        &        Heliocentric redshift         \\
 13 & F10.6  &     zspecerrHelio     &        Heliocentric redshift uncertainty      \\
 14 & F10.6  &     zCMB              &        CMB-frame redshift      \\
 15 & F10.6  &     zerrCMB           &        CMB-frame redshift uncertainty    \\
\sidehead{\saltii\ 4-parameter fits}
\hline
 16 & E10.2  &     x0SALT2zspec          &    \saltii\ $x_0$ (normalization) parameter \\
 17 & E10.2  &     x0errSALT2zspec       &    \saltii\ $x_0$ (normalization) parameter uncertainty \\
 18 & F6.2   &     x1SALT2zspec          &    \saltii\ $x_1$ (shape) parameter   \\
 19 & F6.2   &     x1errSALT2zspec       &    \saltii\ $x_1$ (shape) parameter uncertainty  \\
 20 & F6.2   &     cSALT2zspec           &    \saltii\ $c$ (color) parameter  \\
 21 & F6.2   &     cerrSALT2zspec        &    \saltii\ $c$ (color) uncertainty    \\
 22 & F10.2  &     PeakMJDSALT2zspec     &    \saltii\ MJD at peak in $B$-band \\
 23 & F7.2   &     PeakMJDderrSALT2zspec &    \saltii\ MJD at peak in $B$-band uncertainty \\
 24 & F7.2   &     muSALT2zspec          &    SALT2mu distance modulus  \\
 25 & F6.2   &     muerrSALT2zspec       &    SALT2mu distance modulus uncertainty    \\
 26 & F8.3   &     fitprobSALT2zspec     &    \saltii\ fit chi-squared probability     \\
 27 & F8.2   &     chi2SALT2zspec        &    \saltii\ fit chi-squared    \\
 28 & I5     &     ndofSALT2zspec        &    \saltii\ number of light curve points used   \\
 \sidehead{\mlcs\ 4-parameter fits}
\hline
 29 & F6.2   &     deltaMLCS2k2zspec      &   \mlcs\ shape parameter ($\Delta$) \\
 30 & F6.2   &     deltaerrMLCS2k2zspec   &   \mlcs\ shape parameter ($\Delta$) uncertainty  \\
 31 & F6.2   &     avMLCS2k2zspec         &   \mlcs\ $V$-band extinction ($A_V$)     \\
 32 & F6.2   &     averrMLCS2k2zspec      &   \mlcs\ $V$-band extinction ($A_V$) uncertainty  \\
 33 & F10.2  &     PeakMJDMLCS2k2zspec    &   \mlcs\ MJD of peak brightness in $B$-band \\
 34 & F7.2   &     PeakMJDerrMLCS2k2zspec &   \mlcs\ MJD of peak brightness in $B$-band uncertainty  \\
 35 & F7.2   &     muMLCS2k2zspec         &   \mlcs\ distance modulus   \\
 36 & F6.2   &     muerrMLCS2k2zspec      &   \mlcs\ distance modulus uncertainty      \\
 37 & F8.3   &     fitprobMLCS2k2zspec    &   \mlcs\ chi-squared fit probability     \\
 38 & F8.2   &     chi2MLCS2k2zspec       &   \mlcs\ chi-squared     \\
 39 & I5     &     ndofMLCS2k2zspec       &   \mlcs\ number of light curve points used    \\
 \sidehead{PSNID parameters using spectroscopically observed redshift}
\hline
 40 & F7.3   &     PIaPSNIDzspec          &   \snia\ Bayesian probability (\zspec\ prior)   \\
 41 & E10.2  &     logprobIaPSNIDzspec    &   \snia\ log(\pfit) (\zspec\ prior)  \\
 42 & I5     &     lcqualityIaPSNIDzspec  &   \snia\ light curve quality (\zspec\ prior)    \\
 43 & F7.3   &     PIbcPSNIDzspec         &   \snibc\ Bayesian probability (\zspec\ prior)  \\
 44 & E10.2  &     logprobIbcPSNIDzspec   &   \snibc\ log(\pfit) (\zspec\ prior)  \\
 45 & I5     &     lcqualityIbcPSNIDzspec &   \snibc\ light curve quality (\zspec\ prior)   \\
 46 & F7.3   &     PIIPSNIDzspec          &   \snii\ Bayesian probability (\zspec\ prior)   \\
 47 & E10.2  &     logprobIIPSNIDzspec    &   \snii\ log(\pfit) (\zspec\ prior)  \\
 48 & I5     &     lcqualityIIPSNIDzspec  &   \snii\ light curve quality (\zspec\ prior)    \\
 49 & I5     &     NnnPSNIDzspec          &   Number of nearest neighbors (\zspec\ prior)   \\
 50 & F7.3   &     PnnIaPSNIDzspec        &   \snia\ nearest-neighbor probability (\zspec\ prior)   \\
 51 & F7.3   &     PnnIbcPSNIDzspec       &   \snibc\ nearest-neighbor probability (\zspec\ prior)   \\
 52 & F7.3   &     PnnIIPSNIDzspec        &   \snii\ nearest-neighbor probability (\zspec\ prior)   \\
 53 & F8.4   &     zPSNIDzspec            &   \psnid\ redshift (\zspec\ prior)   \\
 54 & F8.4   &     zerrPSNIDzspec         &   \psnid\ redshift uncertainty (\zspec\ prior)   \\
 55 & F6.2   &     dm15PSNIDzspec         &   \psnid\ \dmB\ (\zspec\ prior)   \\
 56 & F6.2   &     dm15errPSNIDzspec      &   \psnid\ \dmB\ uncertainty (\zspec\ prior)   \\
 57 & F6.2   &     avPSNIDzspec           &   \psnid\ \av\ (\zspec\ prior)   \\
 58 & F6.2   &     averrPSNIDzspec        &   \psnid\ \av\ uncertainty (\zspec\ prior)   \\
 59 & F10.2  &     PeakMJDPSNIDzspec      &   \psnid\ \tmax\ (\zspec\ prior)   \\
 60 & F7.2   &     PeakMJDerrPSNIDzspec   &   \psnid\ \tmax\ uncertainty (\zspec\ prior)   \\
 61 & I5     &     SNIbctypePSNIDzspec    &   Best-fit \snibc\ template (\zspec\ prior)   \\
 62 & I5     &     SNIItypePSNIDzspec     &   Best-fit \snii\ template (\zspec\ prior)   \\
 \sidehead{SALT2 5-parameter fits (ignoring spectroscopic redshift information)}
\hline
 63 & E10.2  &     x0SALT2flat            &   \saltii\ $x_0$ (normalization) parameter (flat-$z$ prior)    \\
 64 & E10.2  &     x0errSALT2flat         &   \saltii\ $x_0$ (normalization) parameter uncertainty (flat-$z$ prior) \\
 65 & F6.2   &     x1SALT2flat            &   \saltii\ $x_1$ (shape) parameter (flat-$z$ prior) \\
 66 & F6.2   &     x1errSALT2flat         &   \saltii\ $x_1$ (shape) parameter uncertainty (flat-$z$ prior)      \\
 67 & F6.2   &     cSALT2flat             &   \saltii\ $c$ (color) parameter(flat-$z$ prior)            \\
 68 & F6.2   &     cerrSALT2flat          &   \saltii\ $c$ (color) parameter uncertainty (flat-$z$ prior)  \\
 69 & F10.2  &     PeakMJDSALT2flat       &   \saltii\ \tmax\  (flat-$z$ prior)    \\
 70 & F7.2   &     PeakMJDerrSALT2flat    &   \saltii\ \tmax\ uncertainty (flat-$z$ prior)   \\
 71 & F7.2   &     zphotSALT2flat         &   \saltii\  fitted redshift (heliocentric frame)   \\
 72 & F6.2   &     zphoterrSALT2flat      &   \saltii\  fitted redshift uncertainty (heliocentric frame)    \\
 73 & F8.3   &     fitprobSALT2flat       &   \saltii\ fit chi-squared probability                   \\
 74 & F8.2   &     chi2SALT2flat          &   \saltii\ fit chi-squared                   \\
 75 & I5     &     ndofSALT2flat          &   \saltii\ number of light curve points used  \\
  \sidehead{PSNID parameters ignoring spectroscopic redshift information}
\hline
 76 & F8.3   &     PIaPSNIDflat           &   \snia\ Bayesian probability (flat-$z$ prior)   \\
 77 & F8.2   &     logprobIaPSNIDflat     &   \snia\ log(\pfit) (flat-$z$ prior)  \\
 78 & I5     &     lcqualityIaPSNIDflat   &   \snia\ light curve quality (flat-$z$ prior)    \\
 79 & F7.3   &     PIbcPSNIDflat          &   \snibc\ Bayesian probability (flat-$z$ prior)  \\
 80 & E10.2  &     logprobIbcPSNIDflat    &   \snibc\ log(\pfit) (flat-$z$ prior)  \\
 81 & I5     &     lcqualityIbcPSNIDflat  &   \snibc\ light curve quality (flat-$z$ prior)   \\
 82 & F7.3   &     PIIPSNIDflat           &   \snii\ Bayesian probability (flat-$z$ prior)   \\
 83 & E10.2  &     logprobIIPSNIDflat     &   \snii\ log(\pfit) (flat-$z$ prior)  \\
 84 & I5     &     lcqualityIIPSNIDflat   &   \snii\ light curve quality (flat-$z$ prior)    \\
 85 & I5     &     NnnPSNIDflat           &   Number of nearest neighbors (flat-$z$ prior)   \\
 86 & F7.3   &     PnnIaPSNIDflat         &   \snia\ nearest-neighbor probability (flat-$z$ prior)   \\
 87 & F7.3   &     PnnIbcPSNIDflat        &   \snibc\ nearest-neighbor probability (flat-$z$ prior)   \\
 88 & F7.3   &     PnnIIPSNIDflat         &   \snii\ nearest-neighbor probability (flat-$z$ prior)   \\
 89 & F8.4   &     zPSNIDflat             &   \psnid\ redshift (flat-$z$ prior)   \\
 90 & F8.4   &     zerrPSNIDflat          &   \psnid\ redshift uncertainty (flat-$z$ prior)   \\
 91 & F6.2   &     dm15PSNIDflat          &   \psnid\ \dmB\ (flat-$z$ prior)   \\
 92 & F6.2   &     dm15errPSNIDflat       &   \psnid\ \dmB\ uncertainty (flat-$z$ prior)   \\
 93 & F6.2   &     avPSNIDflat            &   \psnid\ \av\ (flat-$z$ prior)   \\
 94 & F6.2   &     averrPSNIDflat         &   \psnid\ \av\ uncertainty (flat-$z$ prior)   \\
 95 & F10.2  &     PeakMJDPSNIDflat       &   \psnid\ \tmax\ (flat-$z$ prior)   \\
 96 & F7.2   &     PeakMJDerrPSNIDflat    &   \psnid\ \tmax\ uncertainty (flat-$z$ prior)   \\
 97 & I5     &     SNIbctypePSNIDflat     &   Best-fit \snibc\ template (flat-$z$ prior)   \\
 98 & I5     &     SNIItypePSNIDflat      &   Best-fit \snii\ template (flat-$z$ prior)   \\
  \sidehead{Host galaxy information}
\hline
  99 & I21    &     objIDHost             &   Host galaxy object ID in \sdss\ DR8 Database  \\
 100 & F13.6  &     RAhost                &   Right ascension of galaxy host (degrees)     \\
 101 & F11.6  &     DEChost               &   Declination of galaxy host (degrees)         \\
 102 & F6.2   &     separationhost        &   Distance from SN to host (arc-sec)            \\
 103 & F6.2   &     DLRhost               &   Normalized distance from SN to host ($d_{DLR}$)         \\
 104 & F7.2   &     zphothost             &   Host photometric redshift (KF algorithm)       \\
 105 & F6.2   &     zphoterrhost          &   zphothost uncertainty               \\
 106 & F7.2   &     zphotRFhost           &   Host photometric redshift (RF algorithm)  \\
 107 & F6.2   &     zphotRFerrhost        &   zphotRFhost uncertainty                   \\
 108 & F8.3   &     dereduhost            &   Host galaxy $u$-band magnitude (dereddened)     \\
 109 & F7.3   &     erruhost              &   Host galaxy $u$-band magnitude uncertainty      \\
 110 & F8.3   &     deredghost            &   Host galaxy $g$-band magnitude (dereddened)     \\ 
 111 & F7.3   &     errghost              &   Host galaxy $g$-band magnitude uncertainty      \\
 112 & F8.3   &     deredrhost            &   Host galaxy $r$-band magnitude (dereddened)     \\
 113 & F7.3   &     errrhost              &   Host galaxy $r$-band magnitude uncertainty      \\
 114 & F8.3   &     deredihost            &   Host galaxy $i$-band magnitude (dereddened)     \\
 115 & F7.3   &     errihost              &   Host galaxy $i$-band magnitude uncertainty      \\
 116 & F8.3   &     deredzhost            &   Host galaxy $z$-band magnitude (dereddened)     \\  
 117 & F7.3   &     errzhots              &   Host galaxy $z$-band magnitude (dereddened)     \\  
 \sidehead{Galaxy Parameters Calculated with FPPS}
\hline
 118 & F7.2   &     logMassFSPS           &   \fsps\ $\log(M)$,$M$=Galaxy Mass (M in units of $M_\odot$)             \\     
 119 & F7.2   &     logMassloFSPS         &   \fsps\ Lower limit of uncertainty in $\log(M)$               \\    
 120 & F7.2   &     logMasshiFSPS         &   \fsps\ Upper limit of uncertainty in $log(M)$                 \\  
 121 & F8.2   &     logSSFRFSPS           &   \fsps\ $\log(sSFR)$ $sSFR$=Galaxy Specific Star Forming Rate ($SFR$ in $M_\odot$/yr)                    \\
 122 & F8.2   &     logSSFRloFSPS         &   \fsps\ Lower limit of uncertainty in $\log(sSFR)$                   \\
 123 & F8.2   &     logSSFRhiFSPS         &   \fsps\ Upper limit of uncertainty in $\log(sSFR)$                   \\
 124 & F7.2   &     ageFSPS               &   \fsps\ galaxy age (Gyr)                   \\
 125 & F7.2   &     ageloFSPS             &   \fsps\ Lower limit of uncertainty in age          \\
 126 & F7.2   &     agehiFSPS             &   \fsps\ Upper limit of uncertainty in age          \\
 127 & F8.2   &     minredchi2FSPS        &   Reduced chi-squared of best \fsps\ template fit   \\
\sidehead{Galaxy Parameters Calculated with \PEGASE\ }
\hline
 128 & F8.2   &     logMassPEGASE         &  \PEGASE\ $\log(M)$, M=Galaxy Mass (M in units of $M_\odot$)     \\  
 129 & F8.2   &     logMassloPEGASE       &  \PEGASE\ Lower limit of uncertainty in $\log(M)$                  \\ 
 130 & F8.2   &     logMasshiPEGASE       &  \PEGASE\ Upper limit of uncertainty in $\log(SFR)$                   \\
 131 & F9.2   &     logSFRPEGASE          &  \PEGASE\ $\log(SFR)$ SFR=Galaxy star forming rate ($M_\odot$/yr)                   \\
 132 & F9.2   &     logSFRloPEGASE        &  \PEGASE\ Lower limit of uncertainty in $\log(SFR)$                   \\
 133 & F9.2   &     logSFRhiPEGASE        &  \PEGASE\ Upper limit of uncertainty in $\log(SFR)$                   \\
 134 & F8.2   &     agePEGASE             &  \PEGASE\ galaxy age (Gyr)                   \\
 135 & F8.2   &     minchi2PEGASE         &  Reduced chi-squared of best \PEGASE\ fit                   \\
 136 & I3     &     notes                 &  See list of notes in Table \ref{tbl:notes}
 \enddata
 \tablenotetext{a}{The full table is published in its entirety in the electronic
    edition of The Astrophysical Journal.  Only the column names and table format is shown here.}
\end{deluxetable}


\setlength{\tabcolsep}{0.015in} 
\begin{deluxetable*}{rrrrrl@{}r@{}rrrrrrr}
  \tabletypesize{\tiny}
  \tablewidth{0pt}
  \tablecaption{SDSS-II SN Candidates\tablenotemark{a}\label{tbl:cand}}

  \tablehead{
    \colhead{\texttt{CID}} &
    \colhead{\texttt{RA}} &
    \colhead{\texttt{DEC}} &
    \colhead{$n_e$\tablenotemark{b}} &
    \colhead{\texttt{IAUName}} &
    \colhead{Type} &
    \colhead{\texttt{Peakrmag}} &
    \colhead{\texttt{MJDatPeakrmag}} &
    \colhead{$n_5$\tablenotemark{b}} &
    \colhead{$n_s$\tablenotemark{b}} &
    \colhead{$n_g$\tablenotemark{b}} &
    \colhead{$z_{\mathrm{Helio}}$} &
    \colhead{$\delta z_{\mathrm{Helio}}$} &
    \colhead{\texttt{objIDHost}}
  }
  \startdata
   679 &   327.434978 &    0.657569 &     3 & 2005eh  & Unknown   &   21.8 &   53699.2 &     1 &   0    &   0    &   0.124957 &   0.000017 &  1237656238472888902 \\
   680 &   327.555405 &    0.842584 &    21 & \nodata & Variable  &   21.6 &   53685.1 &     1 &   0 &   0    &     \nodata       &     \nodata       &  1237678617403654778 \\
   682 &   331.239470 &    0.845158 &     2 & \nodata & Unknown   &   21.8 &   53656.2 &     0 &   0   &    0   &   0.048551 &   0.000022 &  1237678617405227407 \\
   685 &   337.823273 &   -0.882037 &    14 & \nodata & pSNII     &   21.7 &   53656.2 &    10 &   0   &    0  &    \nodata       &     \nodata       &  1237656906345349717 \\
   688 &   343.171604 &   -0.962902 &     4 & \nodata & Unknown   &   21.4 &   53616.3 &     5 &   0   &   0    &   0.067866 &   0.000010 &  1237656906347708594 \\
   689 &   345.314592 &   -0.866253 &    15 & \nodata & Variable  &   21.3 &   53680.2 &    15 &   0   &    0   &     \nodata      &    \nodata        &  1237656906348626518 \\
   691 &   329.729408 &   -0.498538 &     9 & \nodata & Unknown   &   20.3 &   53616.2 &     9 &   0   &   0   &   0.130903 &   0.000021 &  1237663542608986381 \\
   692 &   351.071097 &   -0.945665 &    18 & \nodata & Variable  &   21.2 &   53663.2 &    15 &   0   &    0  &   0.197275 &   0.000030 &  1237656906351182046 \\
   694 &   330.154633 &   -0.623472 &    22 & \nodata & Unknown   &   19.7 &   53627.2 &    28 &   0    &   0   &   0.127493 &   0.000018 &  1237663542609183155 \\
   695 &   352.963374 &   -0.963772 &     3 & \nodata & Variable  &   22.7 &   53637.3 &     0 &   0   &    0   &   0.058267 &   0.000009 &  1237656906351968456 \\
   696 &   354.180048 &   -1.020436 &     6 & \nodata & psNIa     &   21.3 &   53623.3 &     5 &   0    &   0   &   \nodata        & \nodata       &  \nodata                    \\
   697 &   335.002430 &   -0.626145 &     8 & \nodata & Unknown   &   21.6 &   53627.2 &     5 &   0   &   0   &   0.156675 &   0.000032 &  1237663542611280181 \\
   698 &   335.302586 &   -0.554336 &    17 & \nodata & Variable  &   21.6 &   53663.2 &    14 &   0   &    0  &    \nodata      & \nodata     &  1237663542611411980 \\
   699 &   332.585653 &    0.625899 &    21 & \nodata & Variable  &   21.7 &   53656.2 &    15 &   0    &    0  &    \nodata      &  \nodata     &  1237663479795352223 \\
   700 &   335.579430 &   -0.518987 &    16 & \nodata & AGN       &   21.5 &   53616.2 &    14 &   0   &   0   &   0.595712 &   0.000242 &  1237663542611542434 \\
 \nodata
 \enddata
\tablenotetext{a}{This table is a portion of the full SN catalog, which is published in its entirety as Table \ref{tab:fullCatalog} in the electronic edition of The Astrophysical Journal.  Selected columns relating to general properties of the entries are shown here for guidance regarding the form and content of these columns.}
\tablenotetext{b}{In the electronic edition $n_e$, $n_5$, $n_s$, $n_g$ are called  \texttt{Nsearchepoch}, \texttt{NepochSNR5}, \texttt{nSNspec} and \texttt{nGALspec}, respectively.}
    
\end{deluxetable*}

\begin{deluxetable}{lrr}
\tabletypesize{\scriptsize}
\tablecaption{Number of SN Candidates by type category\label{tbl:typeTotal}}
\tablewidth{0pt}
\tablehead{
\colhead{Type} & \colhead{Type Code} & \colhead{Number} }

\startdata
Unknown         &   0   &  \nunknown \\
Variable        &   5   &  \nvariable \\
pSNII           & 101   &  \npsnii  \\
pSNIbc          & 102   &  \npsnibc \\
pSNIa           & 103   &  \npsnia  \\
zSNII           & 104   &  \nzsnii  \\
zSNIbc          & 105   &  \nzsnibc \\
zSNIa           & 106   &  \nzsnia  \\
AGN             & 110   &  \nagn \\
SLSN            & 114   &  \nslsn \\
SNIb            & 111,115\tablenotemark{a}  & \nssnib \\
SNIc            & 112   &      \nssnic \\
SNII            & 113,117\tablenotemark{a}  & \nssnii \\
SNIa?           & 119   &    \nssniap \\
SNIa            & 118\tablenotemark{a},120 &  \nssnia \\ 
\hline
Total                                 &          & \ncand
\enddata

\tablenotetext{a}{The indicated types were confirmed with spectra obtained by
  observers who were unaffiliated with SDSS.}

\end{deluxetable}

\clearpage

\begin{deluxetable}{rl}
\tabletypesize{\scriptsize}
\tablecaption{Explanation of SN Notes column (Item 136) \label{tbl:notes}}
\tablewidth{0pt}
\tablehead{
\colhead{Note} & \colhead{Explanation} 
}
\startdata
1            &    SN typing based on spectra obtained by groups outside SDSS.    \\
             &    The spectra used for typing are not included in the data release.   \\ 
2            &    Peculiar type Ia SN possibly similar to sn91bg   \\ 
3            &    Peculiar type Ia SN possibly similar to sn00cx   \\ 
4            &    Peculiar type Ia SN possibly similar to sn02ci   \\ 
5            &    Peculiar type Ia SN possibly similar to sn02cx
\enddata
\end{deluxetable}


\begin{deluxetable}{cccc}

  \tabletypesize{\footnotesize}
  \tablewidth{0pt}
  \tablecaption{PSNID/NN Typing Efficiency and Purity\label{tbl:nntype_results}}

  \tablehead{
    \colhead{SN Type} &
    \colhead{$z$-prior} &
    \colhead{Efficiency} &
    \colhead{Purity}
  }

  \startdata
  Ia        &   flat     &   97.5\%   &  94.8\% \\
  \nodata   &   \zspec   &   96.5\%   &  95.8\% \\
  Ibc       &   flat     &   34.3\%   &  85.5\% \\
  \nodata   &   \zspec   &   38.0\%   &  82.4\% \\
  II        &   flat     &   68.4\%   &  96.6\% \\
  \nodata   &   \zspec   &   54.9\%   &  95.8\%
  \enddata

\end{deluxetable}


\begin{deluxetable}{lrrrrrrrr}
\tabletypesize{\scriptsize}
\tablecaption{Normalized residuals.  \label{tab:sigmaFudge}}
\tablewidth{0pt}
\tablehead{
\colhead{Band} & \colhead{Nominal} & \colhead{Adjustment} &  \colhead{Corrected}
& \colhead{Corrected $s<1$} & \colhead{Corrected} $s>2$
}
\startdata
$u$   & 1.182 &    630  & 1.003 & 1.057  & 0.932  \\
$g $  & 1.159 &     85  & 1.003 & 0.994  & 0.842  \\
$r$   & 1.196 &    200  & 1.000 & 0.947  & 0.892 \\
$i$   & 1.222 &    550  & 1.003 & 0.891  & 0.876  \\
$z$   & 1.181 &   2600  & 1.002 & 0.956  & 0.721
\enddata
\end{deluxetable}

\begin{deluxetable}{cc}
\tabletypesize{\scriptsize}
\tablecaption{SDSS AB Offsets.  \label{tab:ABoff}}
\tablewidth{0pt}
\tablehead{
\colhead{Band} & \colhead{AB Offset}
}
\startdata
$u$   & $-0.0679$  \\
$g$   & $+0.0203$  \\
$r$   & $+0.0049$  \\
$i$   & $+0.0178$  \\
$z$   & $+0.0102$
\enddata

\tablecomments{All magnitudes in this paper are SDSS asinh magnitudes
  \citep{lupton99} in the native system used by SDSS.  The AB offsets
  should be added to the native magnitudes to obtain magnitudes
  calibrated to the AB system.  Fluxes are expressed in $mu$J and have
  the AB offsets already applied.  The derivation of the AB offsets is
  described in the text and in more detail in \citet{betoule13}.}

\end{deluxetable}

\clearpage


\begin{deluxetable}{l l c c l}
\tablecaption{Instrument Configurations \label{tab:spconfig}}
\tablehead{
\colhead{Telescope}	
& \colhead{Instrument}	
& \colhead{Wavelength Range }
& \colhead{Resolution}
& \colhead{Reference or Link}   
}
\startdata
 & & $\AA$ & $\AA$ & \\
\hline

HET	& LRS	& 4070 -- 10700	& 20 & Hill et al., 1998 \\
ARC	& DIS	&  3100 -- 9800	& 8-9	& Link\tablenotemark{a} \\
Subaru	& FOCAS	& 3650 -- 6000 	& 8	&  Kashikawa et al., 2000 \\
	&	& 4900 -- 9000	& 12	&	\\
WHT	& ISIS	& 3900 -- 8900	& 4.3 \& 7.5	&  Link\tablenotemark{b} \\
MDM	& CCDS	& 3800 -- 7300 	& 15	& Link\tablenotemark{c} \\
Keck	& LRIS	& 3200 -- 9400	& 4.5 \& 8.9	& Oke et al., 1995 \\
TNG	& DOLORES	& 3800 -- 7300	& 10	& Link\tablenotemark{d} \\
NTT	& EMMI	& 3800 -- 9200		& 17	& Dekker et al., 1986 \\
NOT	& ALFOSC-FASU	& 3200 -- 9100	& 21	& Link\tablenotemark{e}  \\
Magellan	& LDSS3	& 3800 -- 9200	& 9.5	& Link\tablenotemark{f}  \\
SALT	& RSS	& 3800 -- 8000	& 5.7	& Burgh et al., 2003

\enddata
\tablenotetext{a}{www.apo.nmsu.edu/arc35m/Instruments/DIS/\#B}
\tablenotetext{b}{www.ing.iac.es/PR/wht\_info/whtisis.html}
\tablenotetext{c}{www.astronomy.ohio-state.edu/MDM/CCDS/}
\tablenotetext{d}{www.tng.iac.es/instruments/lrs/}
\tablenotetext{e}{www.not.iac.es/instruments/alfosc/}
\tablenotetext{f}{www.lco.cl/telescopes-information/magellan/instruments/ldss-3}
\end{deluxetable}


\begin{deluxetable}{rrrrrrrr}

  \tabletypesize{\footnotesize}
  \tablewidth{0pt}
  \tablecaption{Spectroscopic Data\tablenotemark{a}\label{tbl:spec}}

  \tablehead{
    \colhead{SDSS ID\tablenotemark{b}} &
    \colhead{Spec ID\tablenotemark{c}} &
    \colhead{Telescope} &
    \colhead{Type(s)} &
    \colhead{Observation Date} &
    \colhead{Evaluation} &
    \colhead{SN redshift} &
    \colhead{Galaxy redshift}
  }

  \startdata
    701  & 2795 &        APO  &        Gal  &   2008-09-02 &        Gal &    \nodata &     0.2060 \\ 
    703  & 1963 &        NTT  &        Gal  &   2007-09-21 &        Gal &    \nodata &     0.2987 \\ 
    722  &   58 &        APO  &     SN,Gal  &   2005-09-09 &         Ia &      0.087 &     0.0859 \\ 
    739  &   59 &        APO  &     SN,Gal  &   2005-09-09 &         Ia &      0.105 &     0.1071 \\ 
    744  &   60 &        APO  &     SN,Gal  &   2005-09-08 &         Ia &      0.123 &     0.1278 \\ 
    762  &   61 &        APO  &     SN,Gal  &   2005-09-09 &         Ia &      0.189 &     0.1908 \\ 
    774  &   62 &        APO  &     SN,Gal  &   2005-09-09 &         Ia &      0.090 &     0.0937 \\ 
    774  &  577 &        MDM  &         SN  &   2005-09-17 &        Gal &    \nodata &     0.0933 \\ 
    779  &  592 &        HET  &        Gal  &   2005-12-29 &        Gal &    \nodata &     0.2377 \\ 
    841  & 2757 &        HET  &        Gal  &   2008-01-06 &        Gal &    \nodata &     0.2991 \\ 
    911  & 1894 &        APO  &         SN  &   2007-11-14 &        Gal &    \nodata &     0.2080 \\ 
   1000  & 2827 &        APO  &        Gal  &   2008-09-27 &        Gal &    \nodata &     0.1296 \\ 
   1008  &  436 &        APO  &     SN,Gal  &   2005-11-26 &        Gal &    \nodata &    \nodata \\ 
   1008  & 2752 &        APO  &        Gal  &   2008-09-27 &        Gal &    \nodata &     0.2260 \\ 
   1032  &  149 &        APO  &     SN,Gal  &   2005-09-25 &         Ia &      0.133 &     0.1296 \\ 
   1112  &   87 &        HET  &     SN,Gal  &   2005-09-26 &         Ia &      0.258 &     0.2577 \\ 
   1114  &  270 &        APO  &     SN,Gal  &   2005-11-04 &         II &      0.031 &     0.0245 \\ 
   1119  &  189 &     Subaru  &     SN,Gal  &   2005-09-27 &         Ia &      0.298 &     0.2974 \\
  \nodata
  \enddata

  \tablenotetext{a}{The full table is published in its entirety in the electronic
    edition of The Astrophysical Journal.  A portion is shown here for
    guidance regarding its form and content.}
  \tablenotetext{b}{Internal SN candidate designation.}
  \tablenotetext{c}{Internal spectrum identification number.}

\end{deluxetable}


\begin{deluxetable}{lrrr}

  \tabletypesize{\footnotesize}
  \tablewidth{0pt}
  \tablecaption{Selection criteria for \saltii\ light curve fits\label{tbl:sncuts}}

\tablehead{
\colhead{SNANA variable} &
\colhead{MLCS2k2} &
\colhead{\saltii\ (4-par)} &
\colhead{\saltii\ (5-par)} 
 }

\startdata
redshift\tablenotemark{a}          &   \nodata (0.45)          &    \nodata (0.7)     &  \nodata (\nodata)   \\
redshift\_err\tablenotemark{b}    &   \nodata(0.011)      &     \nodata(0.011)  &  \nodata  (\nodata) \\
SNRMAX\tablenotemark{c}         &   \nodata(3.0)         &  \nodata(3.0)         &  \nodata(3.0)           \\
Trestmin\tablenotemark{e}        &   \nodata(10.0)  & \nodata(10.0) & \nodata(0.0) \\
Trestmax\tablenotemark{f}       &   \nodata(0.0)    &   \nodata(0.0) & \nodata(10.0)

 \enddata
\tablenotetext{a}{Maximum redshift selected.}
\tablenotetext{b}{Maximum redshift uncertainty for 4-parameter fits.}
\tablenotetext{c}{Maximum signal-to-noise ratio among epochs in one band.}
\tablenotetext{d}{The number of filters that must have at least one epoch meeting the SNRMAX requirement.}
\tablenotetext{e}{The earliest epoch measured in days in the rest frame relative to maximum light in $B$-band must occur before this time.}
\tablenotetext{f}{The latest epoch measured in days in the rest frame relative to maximum light in $B$-band must occur after this time.}

\end{deluxetable}

\begin{deluxetable}{rrrrrrrrrrrrrrrr}

  \tabletypesize{\footnotesize}
\tabletypesize{\scriptsize}
  \tablewidth{0pt}
  \tablecaption{SALT2 4-parameter Fit Results\tablenotemark{a}\label{tbl:salt2_4par}}
  \tablehead{
    \colhead{\texttt{CID}} &
    \colhead{\texttt{Classification}} &
    \colhead{$z$\tablenotemark{b}} &
    \colhead{$x_0$\tablenotemark{b}} &
    \colhead{$\delta x_0$\tablenotemark{b}} &
    \colhead{$x_1$\tablenotemark{b}} &
    \colhead{$\delta x_1$\tablenotemark{b}} &
    \colhead{$c$\tablenotemark{b}} &
    \colhead{$\delta c$\tablenotemark{b}} &
    \colhead{$t_{max}$\tablenotemark{b}} &
    \colhead{$\delta t_{max}$\tablenotemark{b}} &
    \colhead{$\mu$\tablenotemark{c}} &
    \colhead{$\delta \mu$\tablenotemark{c}} &
    \colhead{$P$\tablenotemark{c}} &
    \colhead{$\chi^2$\tablenotemark{c}} &
    \colhead{dof\tablenotemark{c}}
  }

  \startdata
703 & zSNIa & 0.298042 & 5.43e-05 & 3.47e-06 & 0.73 & 0.63 & -0.01 & 0.05 & 53626.5 & 0.7 & 40.80 & 0.25 & 0.966 & 40.80 & 59 \\
735 & zSNIa & 0.190858 & 8.82e-05 & 1.23e-05 & -2.66 & 0.58 & 0.01 & 0.09 & 53610.7 & 1.8 & 39.59 & 0.27 & 0.955 & 20.60 & 33 \\
739 & SNIa & 0.107638 & 4.05e-04 & 3.34e-05 & -0.88 & 0.20 & -0.00 & 0.04 & 53609.5 & 1.1 & 38.31 & 0.19 & 0.001 & 58.80 & 29 \\
744 & SNIa & 0.128251 & 2.74e-04 & 1.57e-05 & 1.37 & 0.37 & 0.06 & 0.03 & 53612.9 & 0.9 & 38.97 & 0.20 & 0.983 & 17.40 & 32 \\
762 & SNIa & 0.191381 & 1.29e-04 & 4.84e-06 & 1.09 & 0.29 & -0.05 & 0.03 & 53625.2 & 0.3 & 40.04 & 0.20 & 0.802 & 46.90 & 56 \\
774 & SNIa & 0.093331 & 6.30e-04 & 2.63e-05 & 0.79 & 0.19 & -0.05 & 0.03 & 53608.5 & 0.0 & 38.27 & 0.19 & 0.806 & 25.00 & 32 \\
779 & zSNIa & 0.238121 & 7.72e-05 & 3.68e-06 & 0.46 & 0.38 & 0.02 & 0.04 & 53626.9 & 0.4 & 40.30 & 0.21 & 0.991 & 42.80 & 67 \\
822 & zSNIa & 0.237556 & 6.82e-05 & 3.48e-06 & -0.38 & 0.54 & -0.09 & 0.04 & 53621.3 & 0.5 & 40.60 & 0.24 & 0.454 & 53.50 & 53 \\
841 & zSNIa & 0.299100 & 5.59e-05 & 3.88e-06 & 0.33 & 0.64 & -0.14 & 0.05 & 53624.9 & 0.6 & 41.07 & 0.26 & 0.994 & 39.30 & 64 \\
859 & zSNIa & 0.278296 & 6.57e-05 & 3.33e-06 & 0.68 & 0.51 & 0.03 & 0.04 & 53624.2 & 0.7 & 40.49 & 0.23 & 0.710 & 69.70 & 77 \\
893 & zSNIa & 0.110133 & 8.20e-05 & 4.06e-06 & -1.18 & 0.45 & 0.04 & 0.04 & 53620.2 & 0.5 & 39.87 & 0.21 & 0.006 & 69.90 & 43 \\
904 & zSNIa & 0.385316 & 3.79e-05 & 3.07e-06 & 1.13 & 2.42 & -0.28 & 0.07 & 53620.6 & 4.0 & 42.06 & 0.43 & 0.992 & 40.00 & 64 \\
911 & zSNIa & 0.207264 & 4.97e-05 & 3.59e-06 & -0.39 & 0.74 & 0.23 & 0.06 & 53621.7 & 0.8 & 40.00 & 0.26 & 0.800 & 45.10 & 54 \\
932 & zSNIa & 0.391335 & 3.13e-05 & 3.35e-06 & 3.39 & 1.38 & 0.01 & 0.07 & 53619.0 & 0.8 & 41.84 & 0.40 & 0.796 & 58.20 & 68 \\
986 & zSNIa & 0.280578 & 4.22e-05 & 2.74e-06 & -0.23 & 1.03 & 0.01 & 0.06 & 53619.8 & 1.6 & 40.86 & 0.30 & 0.991 & 43.40 & 68 \\
   \nodata
  \enddata
 \tablenotetext{a}{This table is a portion of the full SN catalog, which is
   published in its entirety as Table \ref{tab:fullCatalog} in the electronic
   edition of The Astrophysical Journal.  Selected columns relating to
   4-parameter SALT2 light curve fits are shown here for guidance regarding
   the form and content of these columns.}
\tablenotetext{b}{In the electronic edition $z$, $x_0$, $\delta x_0$, $x_1$,
  $\delta x_1$, $c$, $\delta c$, $t_{max}$, and $\delta t_{max}$ are called
  \texttt{zspecHelio},  \texttt{x0SALT2zspec},  \texttt{x1SALT2zspec},
  \texttt{x1errSALT2zspec}, \texttt{cSALT2zspec}, \texttt{cerrSALT2zspec},
  \texttt{peakMJDSALT2zspec}, \texttt{peakMJDerrSALT2zspec}, respectively.}
\tablenotetext{c}{In the electronic edition $\mu$, $\delta \mu$, $P$,
  $\chi^2$, and $dof$ are called \texttt{muSALT2zspec},
  \texttt{muerrSALT2zspec}, \texttt{fitprobSALT2zspec},
  \texttt{chi2SALT2zspec}, and \texttt{ndofSALT2zspec}, respectively.}
 \end{deluxetable}

\begin{deluxetable}{cccccrccrr}

  \tabletypesize{\scriptsize}
  \tablewidth{0pt}
  \tablecaption{Derived Host Galaxy Parameters from FSPS and P\'EGASE.2\tablenotemark{a}\label{tbl:fsps}}
  \tablehead{
    & &
    \multicolumn{4}{c}{FSPS} &
    \multicolumn{4}{c}{P\'EGASE.2} \\
    \colhead{\texttt{CID}} &
    \colhead{\texttt{objIDHost}} &
    \colhead{$\log(M)$\tablenotemark{b}} &
    \colhead{$\log$(sSFR)\tablenotemark{b}} &
    \colhead{$\log$(age)\tablenotemark{b}} &
    \colhead{$\chi^2_r$\tablenotemark{c}} &
    \colhead{$\log(M)$\tablenotemark{d}} &
    \colhead{$\log$(SFR)\tablenotemark{d}} &
    \colhead{$\log$(age)\tablenotemark{d}} &
    \colhead{$\chi^2$\tablenotemark{e}}
  }

  \startdata
679 & 1237656238472888902 & $10.08_{-0.06}^{+0.07}$ & $-10.33_{-0.07}^{+0.13}$ & $ 5.45_{-1.59}^{+1.41}$ & 0.33 & $10.10_{-0.16}^{+0.06}$ & $ 0.57_{-0.13}^{+0.17}$ & 9.0 & 5.03 \\
682 & 1237678617405227407 & $10.27_{-0.09}^{+0.11}$ & $-11.61_{-5.77}^{+0.83}$ & $ 7.23_{-1.87}^{+2.12}$ & 0.13 & $10.09_{-0.02}^{+0.01}$ & $-9.00_{-9.00}^{+9.00}$ & 3.5 & 99.90 \\
688 & 1237656906347708594 & $10.11_{-0.05}^{+0.04}$ & $-10.23_{-0.22}^{+0.09}$ & $ 4.38_{-1.05}^{+1.58}$ & 0.20 & $ 9.93_{-0.08}^{+0.42}$ & $ 0.02_{-0.96}^{+0.47}$ & 3.0 & 51.05 \\
691 & 1237663542608986381 & $10.50_{-0.08}^{+0.06}$ & $-10.51_{-0.41}^{+0.29}$ & $ 5.35_{-1.37}^{+1.86}$ & 0.06 & $10.24_{-0.11}^{+0.17}$ & $ 0.16_{-0.06}^{+0.45}$ & 2.0 & 20.44 \\
694 & 1237663542609183155 & $11.10_{-0.05}^{+0.05}$ & $-10.24_{-0.20}^{+0.13}$ & $ 4.27_{-0.99}^{+2.18}$ & 0.19 & $10.96_{-0.18}^{+0.38}$ & $ 1.05_{-0.40}^{+0.47}$ & 3.0 & 136.10 \\
695 & 1237656906351968456 & $11.50_{-0.08}^{+0.04}$ & $-10.51_{-0.28}^{+0.18}$ & $ 6.48_{-1.72}^{+1.40}$ & 0.17 & $11.17_{-0.08}^{+0.38}$ & $-9.00_{-9.00}^{+9.00}$ & 4.0 & 839.30 \\
697 & 1237663542611280181 & $ 9.64_{-9.64}^{+0.06}$ & $-10.05_{-0.00}^{+0.00}$ & $ 2.65_{-0.05}^{+0.08}$ & 1.61 & $ 9.59_{-0.10}^{+0.18}$ & $ 0.69_{-0.41}^{+0.27}$ & 1.8 & 18.40 \\
700 & 1237663542611542434 & $10.74_{-0.12}^{+0.15}$ & $-9.96_{-0.07}^{+0.09}$ & $ 2.74_{-1.02}^{+0.67}$ & 5.60 & $10.45_{-0.33}^{+0.50}$ & $ 1.55_{-0.45}^{+0.30}$ & 1.8 & 33.07 \\
701 & 1237663544221761583 & $11.03_{-0.08}^{+0.07}$ & $-16.65_{-26.69}^{+4.11}$ & $ 7.79_{-2.60}^{+2.00}$ & 0.02 & $10.83_{-0.01}^{+0.02}$ & $-9.00_{-9.00}^{+9.00}$ & 3.5 & 40.46 \\
702 & 1237663542612001229 & $11.22_{-0.08}^{+0.08}$ & $-10.18_{-0.05}^{+0.08}$ & $ 4.77_{-0.92}^{+0.22}$ & 8.65 & $10.95_{-0.39}^{+0.58}$ & $ 1.55_{-0.25}^{+0.38}$ & 7.0 & 65.39 \\
703 & 1237663544222483004 & $ 9.96_{-0.12}^{+0.12}$ & $-10.17_{-0.20}^{+0.17}$ & $ 3.73_{-1.35}^{+1.84}$ & 0.15 & $ 9.86_{-0.29}^{+0.30}$ & $ 0.41_{-0.55}^{+0.20}$ & 1.8 & 0.68 \\
708 & 1237663544224186552 & $10.64_{-0.09}^{+0.01}$ & $-10.38_{-0.07}^{+0.04}$ & $ 6.41_{-1.78}^{+0.46}$ & 37.19 & $10.76_{-0.63}^{+0.26}$ & $ 1.01_{-1.30}^{+0.20}$ & 7.0 & 454.80 \\
710 & 1237663544224645704 & $10.94_{-0.05}^{+0.17}$ & $-9.79_{-0.15}^{+0.18}$ & $ 0.99_{-0.58}^{+1.48}$ & 4.51 & $11.35_{-0.44}^{+0.15}$ & $ 1.86_{-0.21}^{+0.03}$ & 6.0 & 11.01 \\
717 & 1237663462608535732 & $10.81_{-0.07}^{+0.06}$ & $-10.48_{-0.40}^{+0.25}$ & $ 6.16_{-1.91}^{+1.70}$ & 0.08 & $10.63_{-0.04}^{+0.26}$ & $ 0.32_{-0.60}^{+0.27}$ & 4.5 & 20.42 \\
719 & 1237663462608797948 & $10.71_{-0.08}^{+0.07}$ & $-9.97_{-0.09}^{+0.04}$ & $ 2.60_{-0.54}^{+1.20}$ & 7.20 & $10.50_{-0.39}^{+0.42}$ & $ 1.55_{-0.47}^{+0.27}$ & 2.0 & 60.31 \\
\nodata
  \enddata

 \tablenotetext{a}{This table is a portion of the full SN catalog, which is
   published in its entirety as Table \ref{tab:fullCatalog} in the electronic
   edition of The Astrophysical Journal.  Selected columns relating to host
   galaxy properties are shown here for guidance regarding the form and
   content of these columns.}

\tablenotetext{b}{In the electronic edition $log(M)$ is called
  \texttt{logMassFSPS} and the upper limit is \texttt{logMasshiFSPS} and the
  lower limit is \texttt{logMassloFSPS}.  $\log(sSFR)$ is called
  \texttt{logSSFRFSPS} and the upper and lower limits are
  \texttt{logSSFRhiFSPS} and \texttt{logSSFRloFSPS}, respectively, and
  \texttt{age} is called \texttt{ageFSPS} with upper and lower limits
  \texttt{agehiFSPS} and \texttt{ageloFSPS}.}

\tablenotetext{c}{The reduced $\chi^2$ value of the fit.  This column is
  called \texttt{minredchi2FSPS} in the electronic edition.}

\tablenotetext{d}{In the electronic edition $log(M)$ is called
  \texttt{logMassPEGASE} and the upper limit is \texttt{logMasshiPEGASE} and
  the lower limit is \texttt{logMassloPEGASE}.  $\log(SFR)$ is called
  \texttt{logSFRPEGASE} and the upper and lower limits are
  \texttt{logSFRhiPEGASE} and \texttt{logSFRloPEGASE}, respectively, and
  \texttt{age} is called \texttt{agePEGASE}.}

\tablenotetext{e}{The $\chi^2$ value of the fit.  This column is called
  \texttt{minchi2PEGASE} in the electronic edition.}

\end{deluxetable}
\clearpage

\end{document}